\documentclass{aa}

\usepackage{graphicx}
\usepackage{txfonts}
\usepackage[citecolor=blue, linkcolor=blue]{hyperref}
\hypersetup{colorlinks=True}

\begin{document}

\title{The VANDELS survey: global properties of CIII]$\lambda$1908\r{A} emitting star-forming galaxies at $z\sim 3$}

\author{M. Llerena\inst{1}\fnmsep\thanks{e-mail: mario.llerena@userena.cl}
\and R. Amor\'in\inst{1,2}
\and F. Cullen\inst{3}
\and L. Pentericci\inst{4}
\and A. Calabrò\inst{4}
\and R. McLure\inst{3}
\and A. Carnall\inst{3}
\and E. Pérez-Montero\inst{5}
\and F. Marchi\inst{4}
\and A. Bongiorno\inst{4}
\and M. Castellano\inst{4}
\and A. Fontana\inst{4}
\and D. J. McLeod\inst{3} 
\and M. Talia\inst{6,7} 
\and N. P. Hathi\inst{8}
\and P. Hibon\inst{9}
\and F. Mannucci\inst{10}
\and A. Saxena\inst{11}
\and D. Schaerer\inst{12}
\and G. Zamorani\inst{7}
          }

   \institute{Departamento de Astronomía, Universidad de La Serena, Av. Juan Cisternas 1200 Norte, La Serena, Chile
         \and
             Instituto de Investigación Multidisciplinar en Ciencia y Tecnología, Universidad de La Serena, Raúl Bitrán 1305, La Serena, Chile
             \and
             SUPA, Institute for Astronomy, University of Edinburgh, Royal Observatory, Edinburgh, EH9 3HJ, UK
             \and
             INAF - Osservatorio Astronomico di Roma, Via di Frascati 33, 00078, Monte Porzio Catone, Italy
             \and
             Instituto de Astrofísica de Andalucía, CSIC, Apartado de correos 3004, E-18080 Granada, Spain
             \and 
             Dipartimento di Fisica e Astronomia, Universita di Bologna, Via Gobetti 93/2, I-40129, Bologna, Italy
             \and
             INAF-Osservatorio di Astrofisica e Scienza dello Spazio di Bologna, via Gobetti 93/3, I-40129, Bologna, Italy
             \and 
             Space Telescope Science Institute, 3700 San Martin Drive, Baltimore, MD 21218, USA
             \and European Southern Observatory, Avenida Alonso de Córdova 3107, Vitacura, 19001 Casilla, Santiago de Chile, Chile
             \and 
             INAF - Osservatorio Astrofisico di Arcetri, Largo E. Fermi 5, I-50125, Firenze, Italy
             \and 
             Department of Physics and Astronomy, University College London, Gower Street, London WC1E 6BT, UK
             \and 
             Observatoire de Genève, Université de Genève, 51 Ch. des Maillettes, CH-1290 Versoix, Switzerland
             }

   \date{Received ; accepted }

 
  \abstract
   {Strong nebular emission is ubiquitous in galaxies contributing to cosmic reionization at redshift $z\gtrsim6$. High-ionization UV metal lines, such as  CIII]$\lambda$1908\r{A}, show high equivalent widths (EW) in these early galaxies, suggesting harder radiation fields at low metallicity than low-$z$ galaxies of similar stellar mass. Understanding the physical properties driving the observed UV nebular line emission at high-$z$ requires large and very deep spectroscopic surveys, which are now only accessible out to $z\sim$\,4.
   }   
   {We study the mean properties of a large representative sample of 217 galaxies showing CIII] emission at $2<z<4$, selected from a parent sample of $\sim$750 main-sequence star-forming galaxies in the VANDELS survey. These CIII] emitters have a broad range of UV luminosities, thus allowing a detailed stacking analysis to characterize their stellar mass, star formation rate (SFR) and metallicity, as a function of the UV emission line ratios, EWs, and the carbon-to-oxygen (C/O) abundance ratio.
   }
   {Stacking provides unprecedented high signal-to-noise (S/N)  spectra of CIII] emitters over more than 3 decades in luminosity, stellar mass, and SFR. This enables a full spectral fitting to derive stellar metallicities for each stack. Moreover, we use diagnostics based on photoionization models and UV line ratios   to constrain the galaxies’ ionization source and derive the C/O abundance. 
   }
   {Reliable CIII] detections (S/N$\geq$3) represent $\sim$30\% of the parent sample. However, stacked spectra of non-detections (S/N$<$3) show weak (EW$\lesssim$\,2\r{A}) CIII] emission, suggesting this line is common in normal star-forming galaxies at $z\sim$\,3. On the other hand, extreme CIII] emitters (EW(CIII])$\gtrsim$8\r{A}) are exceedingly rare ($\sim$3\%) in VANDELS. The UV line ratios of the sample suggest no ionization source other than massive stars. Stacks with larger EW(CIII]) show larger EW(Ly$\alpha$) and lower metallicity, but not all CIII] emitters are Ly$\alpha$ emitters. The stellar metallicities of CIII] emitters are not significantly different from that of the parent sample, increasing from $\sim$10\% to $\sim$40\% solar for stellar masses $\log$(M$_{\star}$/M$_{\odot})$\,$\sim$9-10.5.  The stellar mass-metallicity relation of the CIII] emitters is consistent with previous works showing strong evolution from $z=0$ to $z\sim3$.
The C/O abundances of the sample range 35\%-150\% solar, with a noticeable increase with FUV luminosity and a smooth decrease with the CIII] EW. 
We discuss the CIII] emitters in the C/O-Fe/H and the C/O-O/H planes and find they follow stellar and nebular abundance trends consistent with those of Milky Way halo and thick disc stars and local HII galaxies, respectively. A qualitative agreement is also found with chemical evolution models, which suggests that CIII] emitters at $z\sim$\,3 are experiencing an active phase of chemical enrichment.
   }
   {Our results provide new insight on the nature of UV line emitters at $z\sim$\,2-4, paving the way for future studies at higher-$z$ using the James Webb Space Telescope.}

\titlerunning{Global properties of CIII]$\lambda$1908\r{A} emitting star-forming galaxies at $z\sim 3$}
\authorrunning{M. Llerena et al.}

   \maketitle
%

\section{Introduction}

The reionization of the Universe is an outstanding problem that still remains unsolved. While the time scales over which reionization ended are well established around redshift $z\sim 6$  \citep[e.g.][]{Mason2019,Yang2020,Paoletti2020}, the dominant sources of photons responsible for the transformation of the dominant neutral hydrogen into a mostly ionized medium have yet to be determined. Faint low-mass star-forming galaxies are considered candidates to lead reionization in this era due to their large number density and weak gravitational potential, favouring the strong and effective feedback needed to open low HI density paths for photons to escape \citep[e.g.,][]{Wise2014,Robertson2015,Bouwens2016,Finkelstein2019}. 
However, the contribution of additional sources with higher ionizing photon efficiency, such as luminous, massive starburst galaxies \citep[e.g.][]{Naidu2020,Endsley2021} and AGNs \citep[e.g.][]{Grazian2018} might have a significant contribution \citep[e.g.][and references therein]{Dayal2020}. 

A detailed characterization of the rest-frame ultraviolet (UV) spectra of star-forming galaxies (SFG) at $z\gtrsim$\,6 is thus essential to understand their ionization properties and thus shed new light into the reionization process. In the last few years, deep near-infrared (NIR) observations of galaxies during the reionization era have reported the presence of UV emission lines with unusually high equivalent widths (EW), such as CIV$\lambda$1550\r{A}, HeII$\lambda$1640\r{A}, OIII]$\lambda$1663\r{A}, or CIII]$\lambda$1908\r{A} \citep[see][for a review]{Stark2016}. 

UV nebular lines encode precious information on the physical conditions of the ionized gas in galaxies.  Different photoionization models, used to understand the role of age, ionization parameter, metallicity, and dust on the emergent UV nebular lines, struggle to explain their origin and strength \citep[e.g.][]{Jaskot2016,Gutkin_2016, Feltre_2016,Nakajima_2018,byler2018, Hirschmann2019}. However, so far, all models require the presence of hard radiation fields able to reproduce the observed UV emission lines with high ionization potentials, what also leads to more extreme ionization conditions in the interstellar medium (ISM) (e.g. high EWs and line ratios). Constraining available models with large and representative samples of emission line galaxies is therefore needed to improve our understanding of the physical mechanisms producing UV emission lines, thus paving the way for future extensive studies of galaxies at $z>6$. 

Emission lines are relevant not only to understand the physical conditions governing these early galaxies but also to provide a tool for their spectroscopic redshift identification. 
This is especially relevant at $z>$ 6, where  absorption features are weak and a significant drop in the number of galaxies with Ly$\alpha$ emission, often the strongest emission line in the UV, is observed due to the sharp increase of absorption by a predominantly neutral intergalactic medium (IGM)   \citep[e.g.,][]{Fan2006,Pentericci2014,Cassata2015,Fuller2020}.

One of the best alternatives to Ly$\alpha$ is the CIII] doublet --a combination of [CIII]$\lambda$1906.68\r{A}, a forbidden magnetic quadrupole transition and CIII]$\lambda$1908.73\r{A}, a semiforbidden electro-dipole transition (here, we are referring to vacuum wavelength), with an ionizing potential of 24.4 eV. This doublet (hereafter cited as CIII]$\lambda$1908 or CIII], for simplicity) is typically the brightest UV metal line in star-forming galaxies at intermediate redshift  \citep[$z\sim$\,3, e.g.][]{Shapley_2003} and has been proposed as an alternative to search for galaxies in the reionization era \citep{Stark2014}. Some searches have been successful, reporting high EW(CIII]) in the observed spectra of  galaxies at $z>6$ \citep{Stark2015,Stark2017,Mainali2017,Laporte2017,Hutchison2019}, but other studies failed in detecting the line in strongly star-forming systems \citep{Sobral2015,Schmidt2016}.

At low-$z$, studying UV emission lines requires space-based spectroscopy. Studies using {\it Hubble Space Telescope} (HST) observations of relatively small samples showed that strong CIII] emission is generally present in the spectra of local  low-metallicity galaxies  \citep[e.g.,][]{Garnett1995,Leitherer2011,Berg2016,Berg2018,Senchyna2017,Ravindranath2020}. However, the characterization of larger samples spanning a wider range of properties (e.g. stellar masses, star formation rates (SFR), metallicities) requires a stronger observational effort that has precluded studies with statistical significance. 
This is different at $2<z<4$, where both Ly$\alpha$ and CIII] are redshifted into the optical and can be probed over larger samples with ground-based 8-10m-class telescopes. Also, galaxies at $z>2$ are likely to be more similar to those at $z>6$ \citep[see][for a review]{Shapley2011}. Several studies now routinely report CIII] emission (along with other strong emission and absorption lines) in galaxies at cosmic noon, either from small samples of relatively bright galaxies \citep{Steidel1996,Erb2010,Steidel2014,Amorin_2017,Du2020}, fainter gravitationally lensed galaxies \citep{Pettini2000,Christensen2012,Stark2014,Rigby2015,Berg2018,Vanzella2021}, or in high signal-to-noise (S/N) stacks from larger galaxy samples \citep{Steidel2001, Shapley_2003,LeFevre_2019,meanNakajima_2018, Feltre2020}.

Deep surveys such as the VIMOS Ultra Deep Survey  \citep[VUDS][]{Lefevre2015} or the MUSE Hubble Ultra Deep Field Survey \citep[HDFS, ][]{Bacon2017} have recently studied large samples of CIII] emitters. 
\cite{LeFevre_2019} showed that only 24\% of the VUDS galaxies at $2.4<z<3.5$ shows CIII] emission, but only in $\sim$1\% this emission is as intense as the values found at $z>6$, i.e. EW(CIII])$\gtrsim$10-20\r{A}.  \citet{Amorin_2017} showed that extreme CIII] emitters at $2<z<4$ in VUDS are very strong Ly$a$ emitters (LAEs) characterized by very blue UV spectra with weak absorption features and bright nebular emission lines. These galaxies present high excitation, low metallicities, and low carbon-to-oxygen (C/O) abundances ratios, similar to the values expected to be common in most of the galaxies during the first 500 Myr of cosmic time. 

Using stacking of a large sample of LAEs from the MUSE HDFS, \cite{Feltre2020} found  that the mean spectra of LAEs with larger Ly$\alpha$ EW, fainter UV magnitudes, bluer UV spectral slopes, and lower stellar masses show the strongest nebular emission. 
 \citet{Maseda2017} arrived at similar conclusions for a sample of 17 CIII] emitters at $z\sim 2$ in the MUSE HDFS. For these galaxies, they found a correlation between EW(CIII]) and EW([OIII]$\lambda$5007), linking the properties of the stronger CIII] emitters to those of the so-called Extreme Emission Line Galaxies \citep[EELGs, e.g.][]{Maseda2014,Amorin2015}. These are low-metallicity starbursts defined by their unusually high EW([OIII]$\lambda$5007)$\gtrsim$200 \r{A} \citep[see also,][]{Tang2021, Matthee2021}. 
At lower-$z$,  the Green Pea galaxies \citep[][]{Cardamone2009, Amorin2010,Amorin2012}, are EELGs and they include  objects for which Lyman continuum leakage has been directly measured \citep{Izotov2016,Guseva2020,Wang2021}. These galaxies show prominent UV nebular lines, including high EW CIII]  \citep{Schaerer2018,Ravindranath2020}.  

While EELGs are likely analogs of the bright-end of reionization galaxies, the more common population of normal, main-sequence SFGs showing moderate or low EW(CIII]) still needs to be fully characterized at $z>2$. This requires large very deep samples achieving sufficiently high S/N spectra to detect and study the fainter CIII]-emitters. 
This work is the first of two papers aimed at exploiting the unprecedented ultra deep spectra provided by the VANDELS survey \citep{McLure_2018, Pentericci_2018} to assemble a large unbiased sample of main-sequence star-forming galaxies CIII] emitters at $2<z<4$ and characterize their main physical  properties as a function  of their UV line emission and chemical abundances. 

The metal content of galaxies does not only have a crucial role in the production and strength of nebular lines, but it is also sensitive to their star formation activity and to the presence of outflows, gas stripping, and dilution resulting from inflow of pristine gas \citep{Maiolino2019}. 
Scaling relations, such as the mass-metallicity relation (MZR), provide key insights into the physical mechanisms involved in the growth and evolution of galaxies. 
At $z\sim 2-4$, the MZR has been studied  using both the gas-phase metallicity \citep[e.g.][]{Erb2006b,Troncoso2014,Sanders2020} and the stellar metallicity \citep[e.g.][]{Sommariva2012,Cullen_2019,Calabro2020} metallicities, finding a strong redshift evolution towards lower metallicities at a given stellar mass. 

Moreover, as different chemical elements are produced by stellar populations at different timescales, the relative abundance of elements enables us to obtain constraints on the star formation history (SFH) of galaxies \citep{Maiolino2019}. The C/O abundance ratio is a powerful indicator because most of the oxygen is synthesized in massive stars ($>$10M$_{\odot}$), while carbon is produced in massive and intermediate-mass stars. Thus, a time delay in the production of carbon and its ejection to the ISM makes C/O a measurable "chemical clock" for the relative ages of the stellar populations in galaxies \citep{Garnett1995} and an important indicator to constraint chemical evolution models \citep{Vincenzo2018}. 

The C/O ratio has been studied in local dwarf galaxies and HII regions of disk galaxies  \citep[e.g.][]{Garnett1995,Chiappini2003,Esteban2014,Pena-Guerrero2017,Berg_2019}. A continuous increase of C/O with O/H is found above one fifth solar metallicity, but the relation flattens at lower metallicities (12+log(O/H)$<$8) showing a significant scatter of C/O values for a given metallicity \citep{Garnett1995,Berg2016,Perez-Montero_2017}. A detailed comparison with models led \citet{Berg_2019} to conclude that the C/O ratio is very sensitive to the assumed SFH, in such a way that longer and lower star formation efficiency bursts lead to low C/O ratios. Chemical evolution models with different prescriptions have been developed to understand the evolution of C/O with metallicity \citep[e.g.,][]{Carigi1994,Henry2000,Chiappini2003,Mattsson2010,Carigi2011,Molla2015,Vincenzo2018} but several variables remain unconstrained, especially at high redshifts. Therefore, measurements of C/O for galaxies at different metallicities remain crucial to study the properties of CIII] emitters and SFGs in general given that they can lead to variations in their observed CIII] emission \citep[e.g.][]{Jaskot2016,Nakajima_2018}. 

In this work, we focus on the average properties of CIII] emitters using the spectral stacking technique. One key goal is to study, for the first time at this redshift,  the relation between the mean stellar metallicity and C/O abundances of galaxies, which is discussed in terms of other physical properties of the sample. This will be useful to interpret future observations with the \textit{James Webb Space Telescope} (JWST) at higher redshifts where only the rest-frame UV spectral lines would be accessible. In a forthcoming paper (Llerena et al., in prep.) we will present a second study based on individual galaxies.

The paper is organized as follows. In Section \ref{sec:method}, we present the sample selection, the basic properties of the sample, and our stacking method. 
In Section~\ref{sec:results}, we present a qualitative and quantitative description of the emission and absorption lines detections via different emission-line diagnostics, the estimation of metallicities, C/O abundances, and different correlations found for our sample. In Section~\ref{sec:discussions},  we discuss our results, focusing on the stellar mass-metallicity relation and the  C/O-metallicity relation. Finally,  Section~\ref{sec:conclu} presents our conclusions.  

Throughout the paper we assume the following cosmology: $\Omega_{M}=0.3$, $\Omega_{\Lambda}=0.7$, H$_0=70$km s$^{-1}$ Mpc$^{-1}$. We adopt a \cite{Chabrier2003} initial mass function (IMF). We consider the solar metallicity Z$_{\odot}=0.0142$, log(O/H)$_{\odot}=8.69$, and log(C/O)$_{\odot}=-0.26$ \citep{Asplund2009}.  We assume by convention a positive EW for emission lines and rest-frame EW are reported. We use the following notation for metallicity for consistency with local conventions:
[X/Y] = $\log$(X/Y) - $\log$(X/Y)$_{\odot}$.

\begin{figure*}[t]
    \centering
    \includegraphics[width=0.8\textwidth]{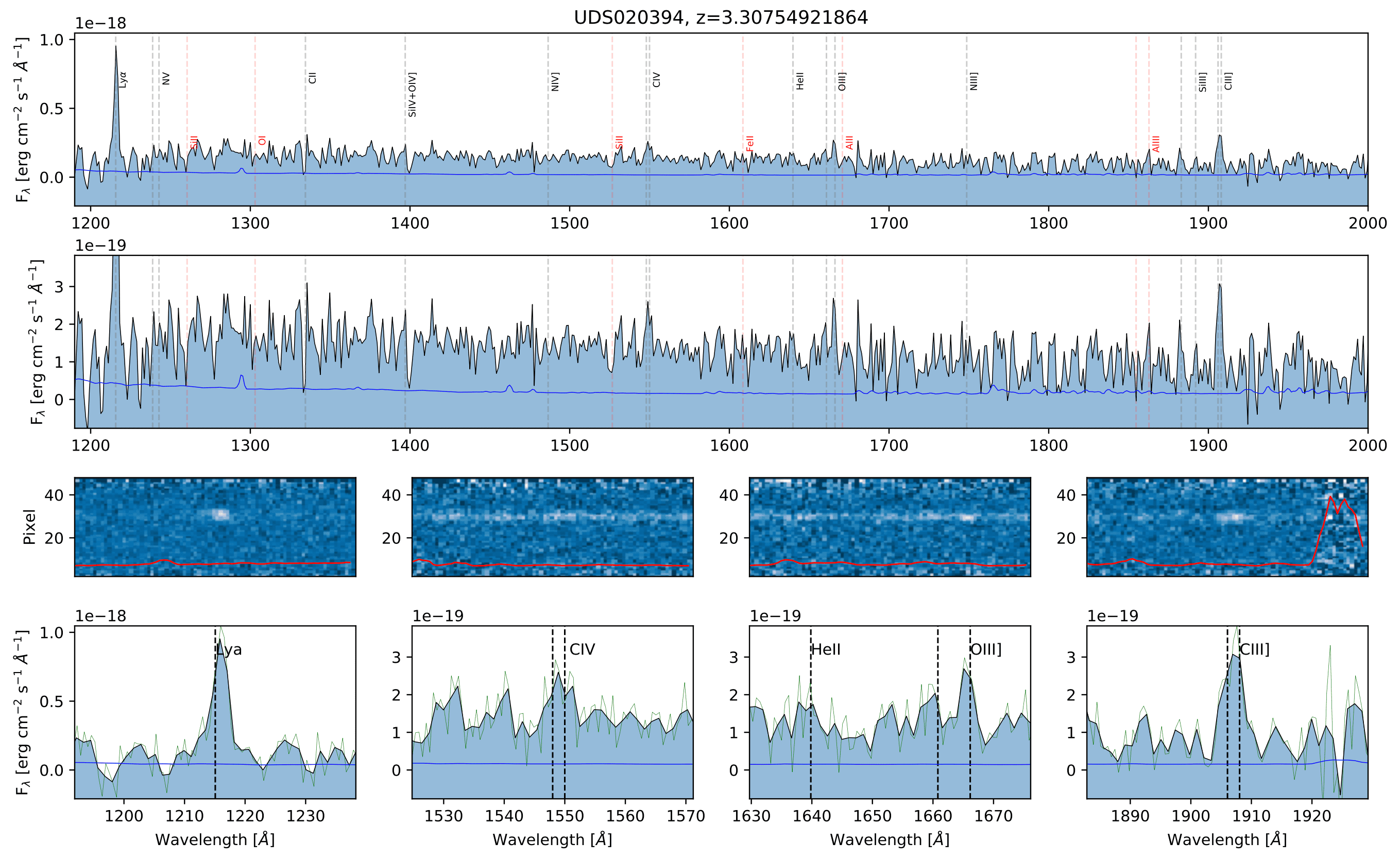}
    \caption{Spectrum of UDS20394, one of the more intense CIII] emitters in the C3 sample, whose estimated parameters are log(M$_*$/M$_{\odot})$=9.32, SFR=4.01 M$_{\odot}$ yr$^{-1}$, M$_{FUV}=-20$, M$_{K_s}=-20.46$,  EW(Ly$\alpha$)=19.2\r{A}, and EW(CIII])=12.3\r{A}. The green faint line is the de-redshifted VANDELS spectrum and the black line is the same but resampled by a factor of 2. The blue line in the upper panels is the error spectrum. The red line in the intermediate panels is the scaled sky spectrum. }
    \label{singlespectrum}
\end{figure*}

\section{Methodology}\label{sec:method}
\subsection{Sample selection}

In this work, we use spectra from VANDELS \citep{McLure_2018, Pentericci_2018}, an ESO public spectroscopic survey conducted with VIMOS at the \textit{Very Large Telescope}. VANDELS obtained unprecedented high S/N optical spectra of $\sim$2100 galaxies at redshift 1.0 $\leq z \leq$ 7.0 in the UKIDSS Ultra Deep Survey (UDS: 02:17:38, -05:11:55) and the Chandra Deep Field South (CDFS: 03:32:30, -27:48:28) fields. Ultra deep spectra for every single galaxy has a minimum (maximum) total exposure time of 20h (80h), and mean spectral resolution  $R \sim$ 580 and dispersion of 2.5 \AA/pixel in the wavelength range $\sim$4800-10000\AA. The survey strategy and design, including target selection and data reduction is fully described in a series of papers \citep{McLure_2018, Pentericci_2018,Garilli2021}. In short, VANDELS targets can be classified according to their selection criteria as bright SFGs in the range 2.4 $\leq z \leq$ 5.5 and Lyman-break galaxies (LBGs) in the range 3.0 $\leq z \leq$ 7.0, and a smaller sample of passive galaxies (1.0 $\leq z \leq$ 2.5) and AGN candidates. In this work, we only select galaxies from the SFGs and LBGs targets.  

Our sample is drawn from VANDELS DR3, which consists of 1774 galaxies --a subset of the 2087 galaxies included in the VANDELS final data release \citep{Garilli2021}. We select galaxies with spectroscopic redshift quality flag 3 or 4, which means 95\% and 100\% of confidence in their spectroscopic redshift \citep{McLure_2018}. We select galaxies at $2<z<4$ to ensure that the CIII] emission lines is included in the spectral range provided by the VANDELS spectra. Detection of CIII], typically the strongest nebular emission line in the sample, at S/N$>$3 is required to ensure a proper measurement of the systemic redshift. With this constraint, from a parent sample of 746 galaxies with the above redshift range and quality flags, a first sample of 225 galaxies are selected by their CIII] emission (130 in the CDFS field and 95 in the UDS field).  

{We cross-match the sample of CIII] emitters with the 7Ms CDF-S catalogue \citep{Luo_2016} and the $\sim$200-600Ks X-UDS catalogue \citep{Kocevski_2018} in order to discard galaxies with X-ray emission within 3 arcsec of separation. We also discard galaxies with spectral features consistent with AGNs or with strong sky residuals. A total of 8 galaxies were excluded from the sample. A more detailed analysis on the AGN sample in VANDELS will be presented in Bongiorno et al., in prep.} Our final sample of CIII] emitting galaxies is made of 217 galaxies (hereinafter C3 sample), which represents $\sim$30\% of the parent sample. Figure~\ref{singlespectrum} shows the rest-frame spectrum of one of the galaxies in the C3 sample. Some of the expected UV absorption and emission lines are marked by vertical lines. Detected emission lines that are relevant to our study are shown in both in 1D and 2D spectra and marked in different zoom-in panels. 

\subsection{Systemic redshift and basic properties of the sample}\label{section_props}

\begin{figure}[t]
    \centering
    \includegraphics[width=0.7\hsize]{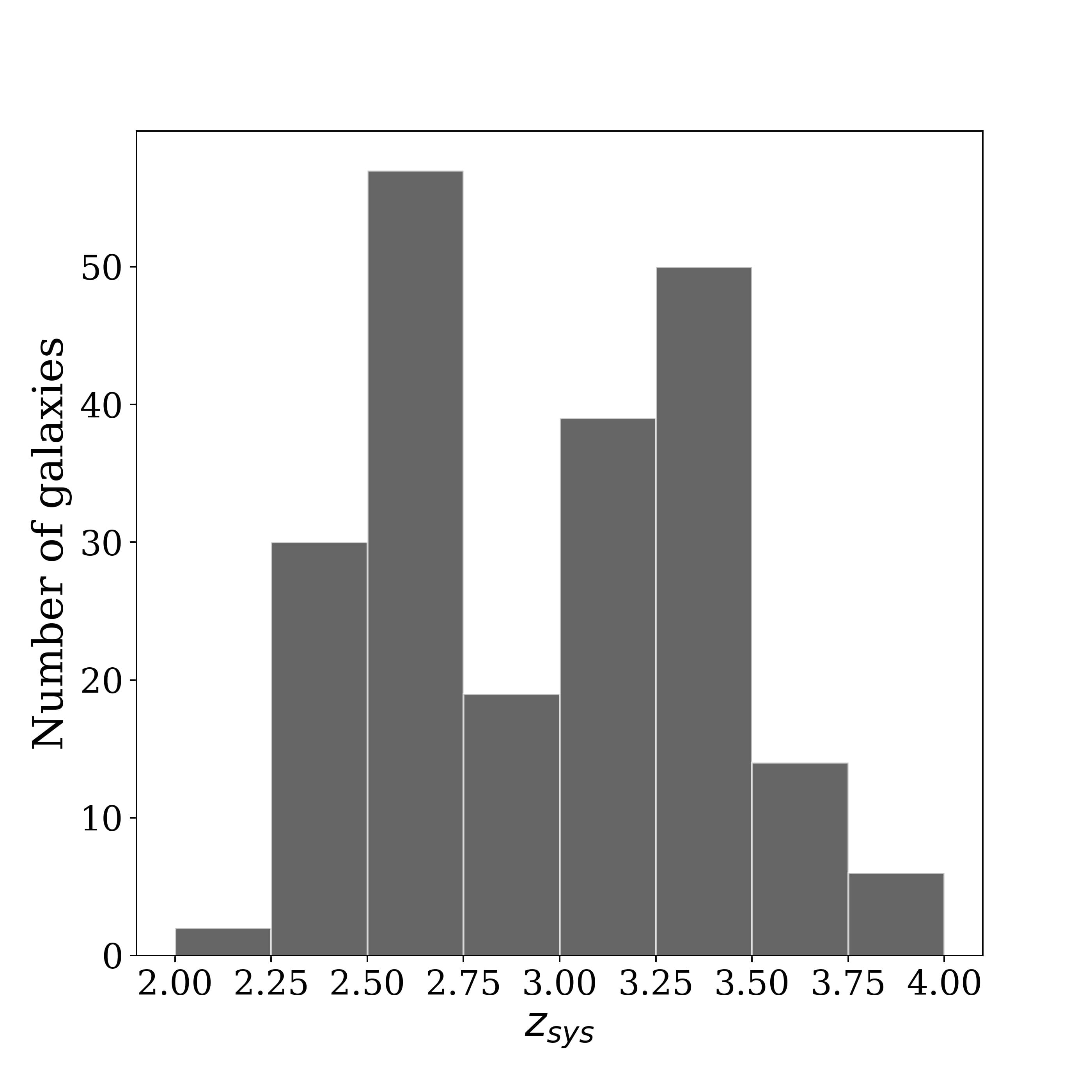}
    \caption{Systemic redshift distribution of the CIII] emitting galaxies in the C3 sample.}
    \label{samplezDR3}
\end{figure}

In order to prepare the C3 sample for the stacking procedure, we follow the methodology described in \citet{Marchi_2019} to derive accurate systemic redshifts ($z_{sys}$) using the nebular CIII] line. In Figure \ref{samplezDR3} we present the resulting $z_{sys}$ distribution for the C3 sample, which spans the range of $2.17-3.82$ ($\langle z_{sys} \rangle=2.98, \sigma=0.43$). {Compared with the spectroscopic redshifts ($z_{spec}$) of the sample reported in \cite{McLure_2018}, the systemic redshifts are slighly larger with a mean difference $\Delta (z_{sys}-z_{spec})=0.002$, that corresponds to $\sim$ 4\r{A} at the rest-wavelength of CIII].}

The physical properties of the C3 sample and the remaining galaxies in the VANDELS DR3 parent sample at $2<z<4$ are obtained from the Spectral Energy Distribution (SED) fitting using the Bayesian Analysis of Galaxies for Physical Inference and Parameter EStimation (BAGPIPES \footnote{\url{https://bagpipes.readthedocs.io/en/latest/}}) code. BAGPIPES is a state of the art Python code for modelling galaxy spectra and fitting spectroscopic and photometric observations \citep{Carnall_2018}, which has been now applied to the VANDELS final data release \citep{Garilli2021}. For this paper, the BAGPIPES code is run fixing the redshift and using the 2016 updated version of the \citet{Bruzual_2003} models using the MILES stellar spectral library \citep{Falcon2011} and updated stellar evolutionary tracks of \cite{Bressan2012} and \cite{Marigo2013}. The stellar metallicity is fixed to 0.2 Solar and the nebular component is included in the model assuming an ionization parameter $\log(U)=-3$. We choose to fix these parameters to typical mean values found in SFGs at similar redshift \citep[e.g.][]{Cullen_2019,Runco2021} to minimize the effects of possible degeneracies affecting the models \citep[see, e.g.][]{Castellano2014}. We note, however, that the results presented in subsequent sections remain unchanged if we allow these parameters vary within typically observed ranges. Dust attenuation is modelled using the \cite{Salim2018} model. The SFH is parameterized using an exponentially increasing $\tau$-model. We obtain a mean value for the timescale $\tau=5.49$ Gyr for the C3 sample ($\tau=5.36$ Gyr for the parent sample), which essentially implies constant star-formation. On the other hand, we obtain a mean age, i.e. the time since SFH begins, of 228 Myr for the C3 sample (270 Myr for the parent sample). The ages obtained from the SED fitting are thus longer than the timescales ($<20$Myr) in which the EW(CIII]) changes with age, according to photoionization models assuming continuous star formation \citep{Jaskot2016}.

\begin{figure}[t]
    \centering
    \includegraphics[width=0.99\linewidth]{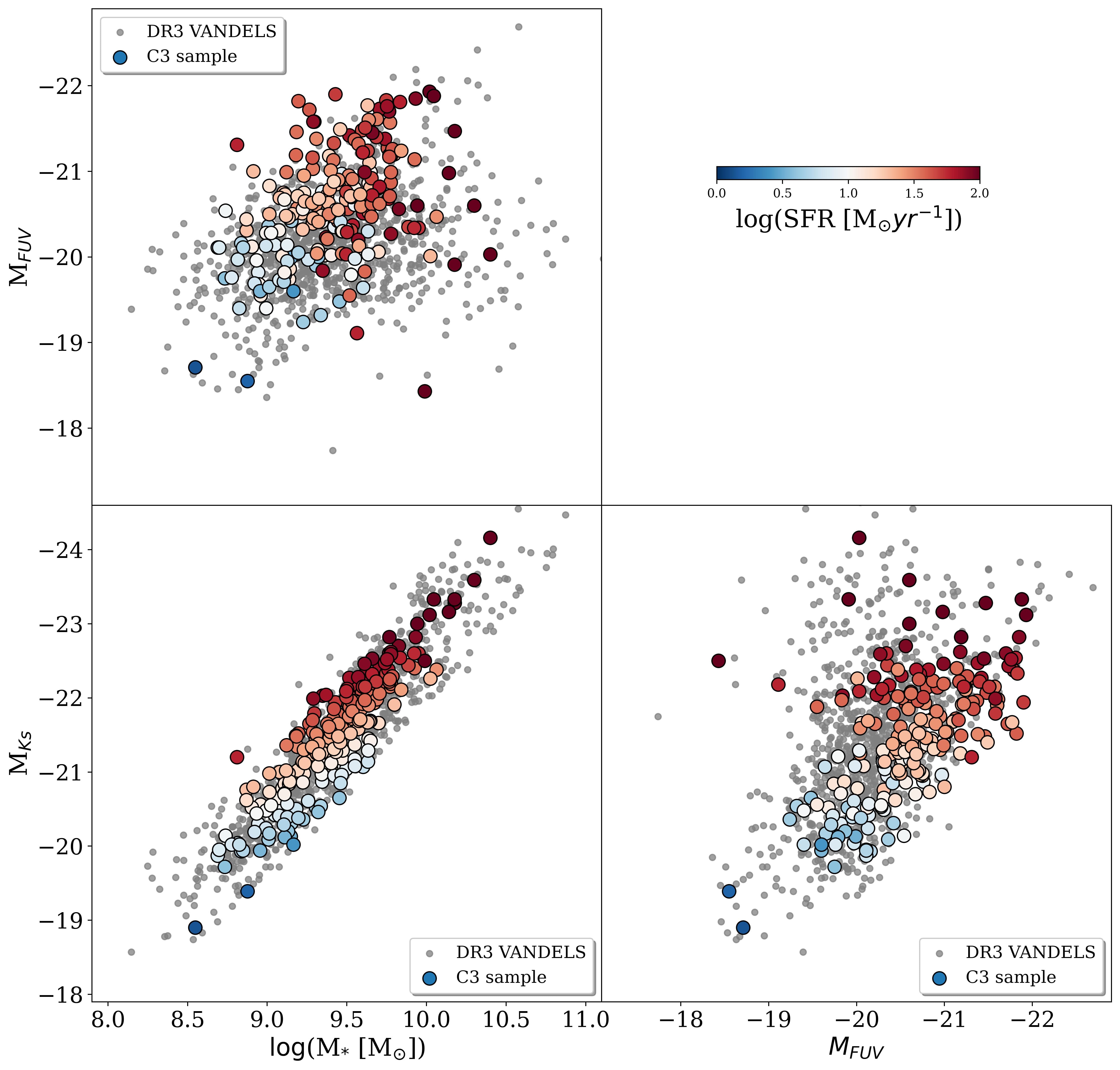}
    \caption{Relations between the resulting BAGPIPES parameters in the VANDELS main-sequence galaxies. The galaxies of the C3 sample are color-coded by star formation rate. Gray points are VANDELS galaxies of the parent sample that were not included in the C3 sample.}
    \label{sampleUMassbag}
\end{figure}

\begin{figure}[t]
    \centering
    \includegraphics[width=0.79\linewidth]{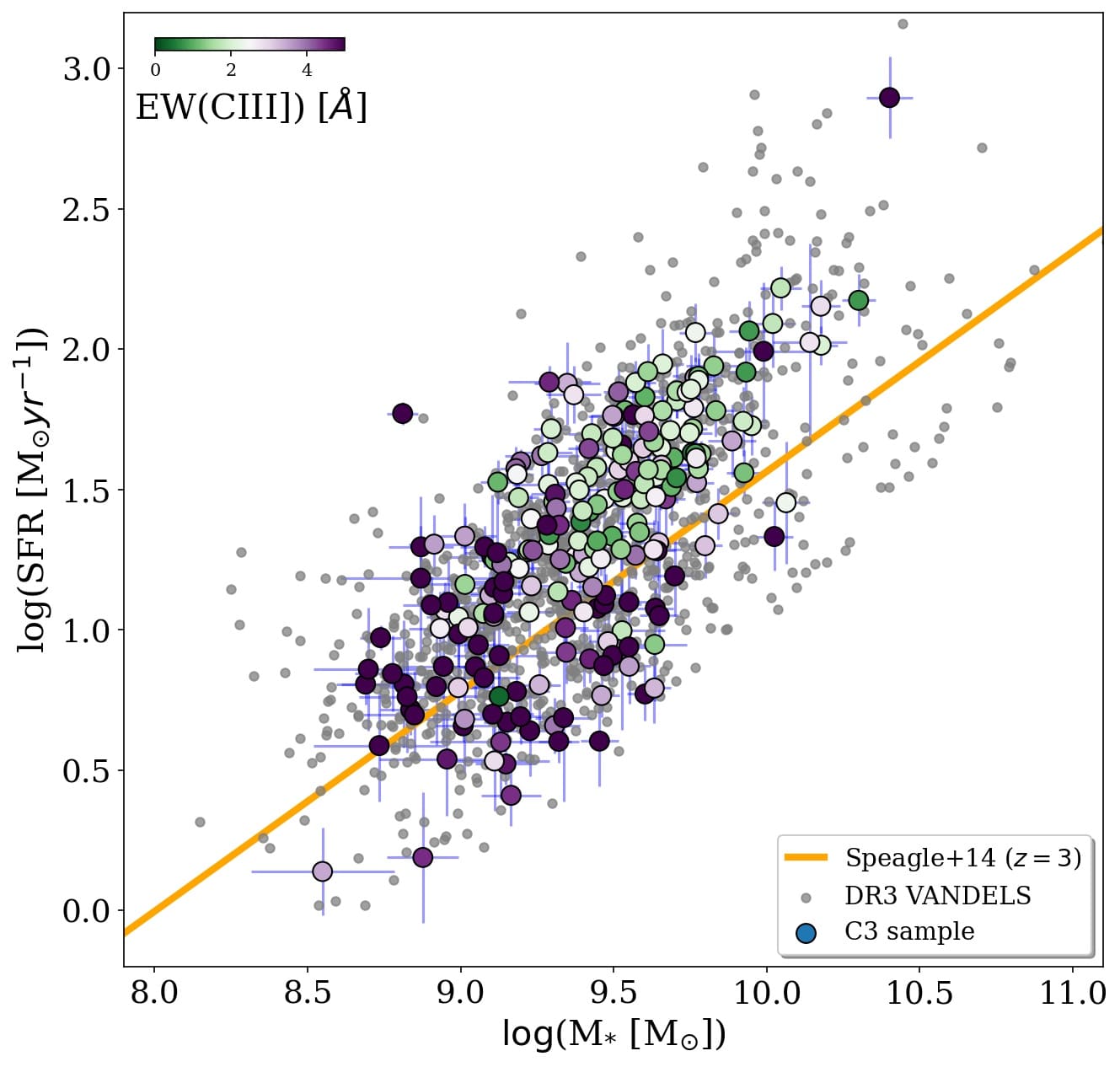}
    \caption{M$_{\star}$-SFR relation with our sample color-coded by EW (CIII]). The orange solid line is the main sequence at $z=3$ according to \cite{Speagle_2014}. Gray points are as in Fig.~\ref{sampleUMassbag}.}
    \label{sampleMSFR}
\end{figure}

\begin{figure*}[t]
    \centering
    \includegraphics[width=0.25\linewidth]{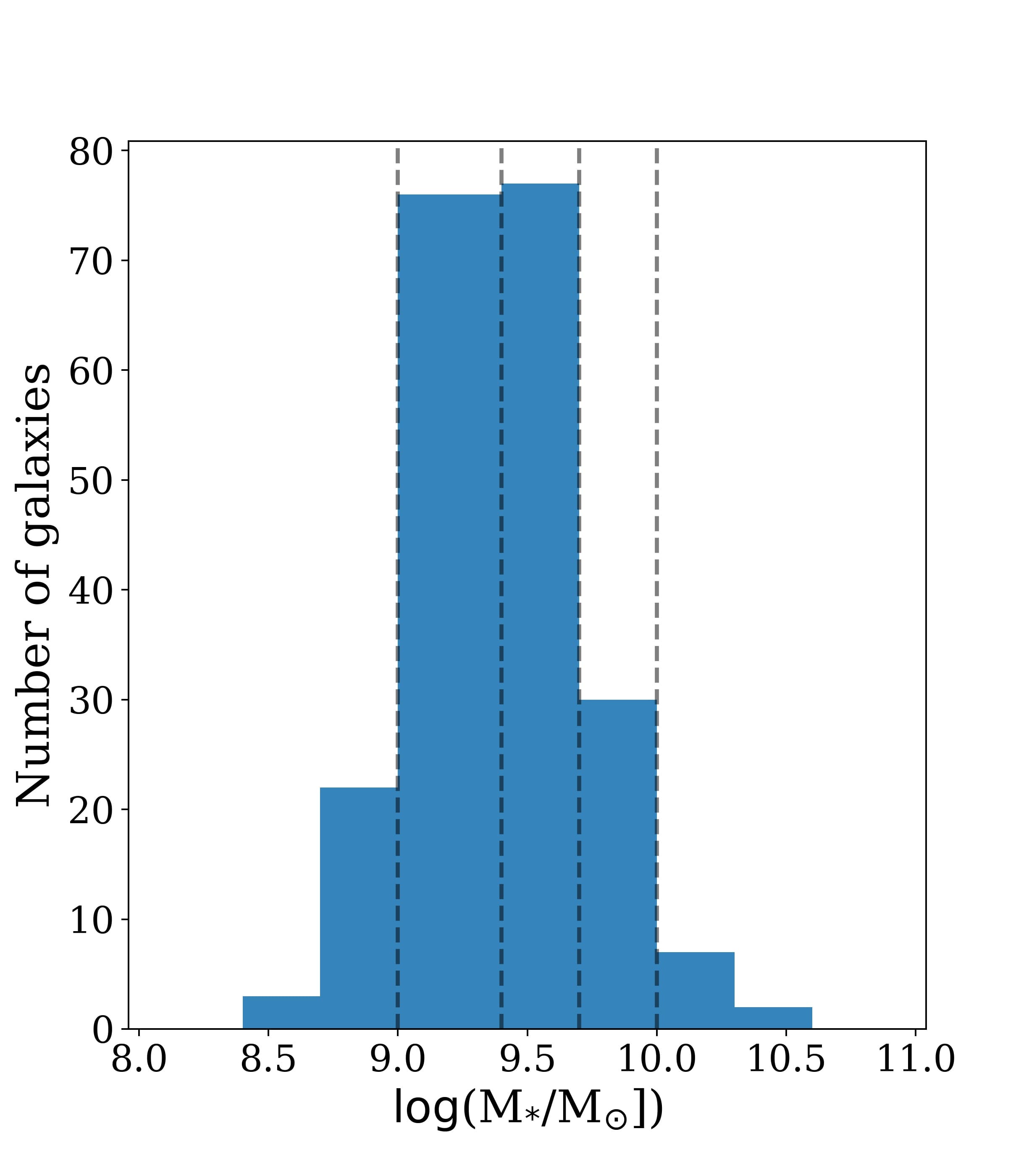}\,\includegraphics[width=0.25\linewidth]{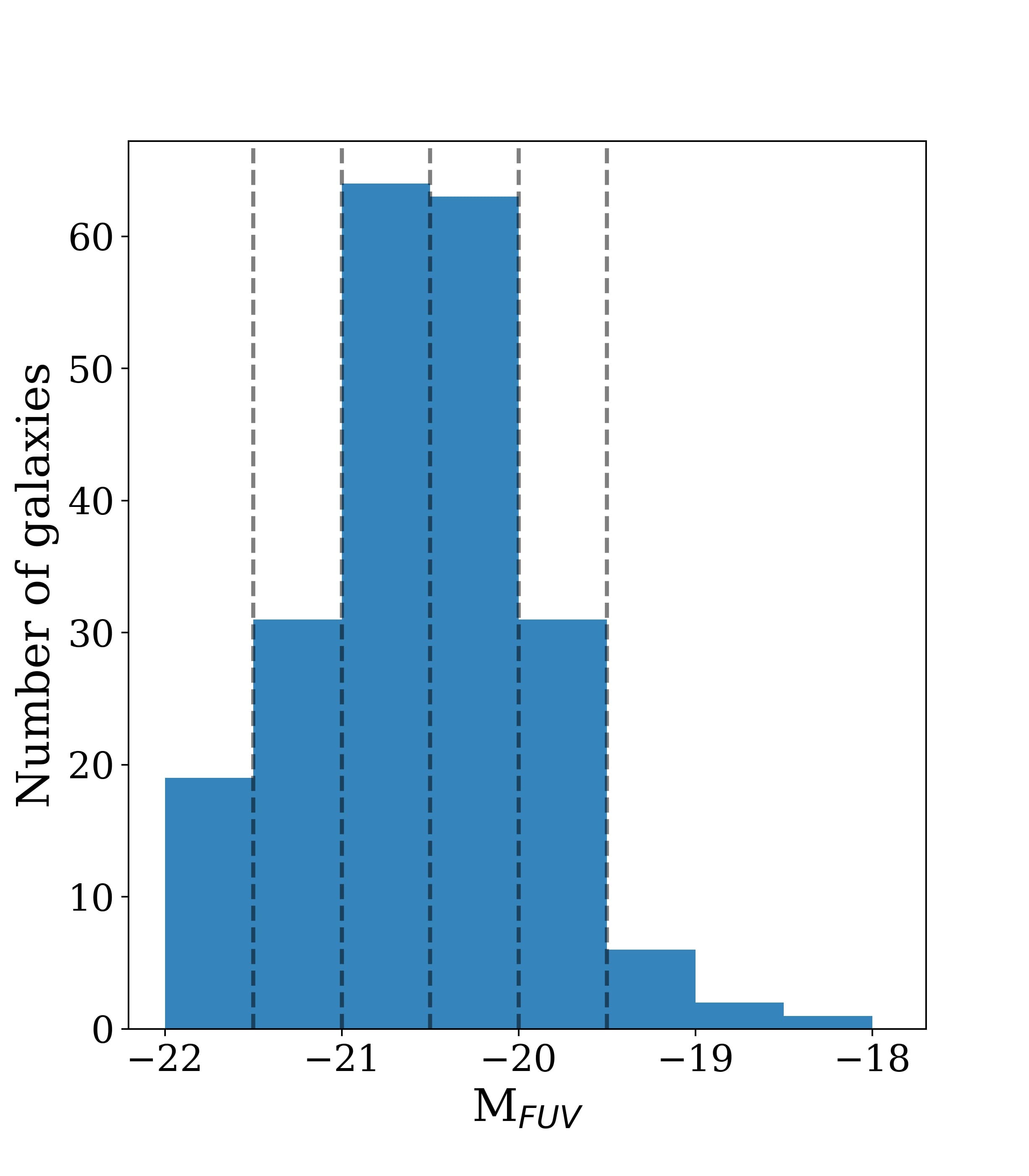}\,\includegraphics[width=0.25\linewidth]{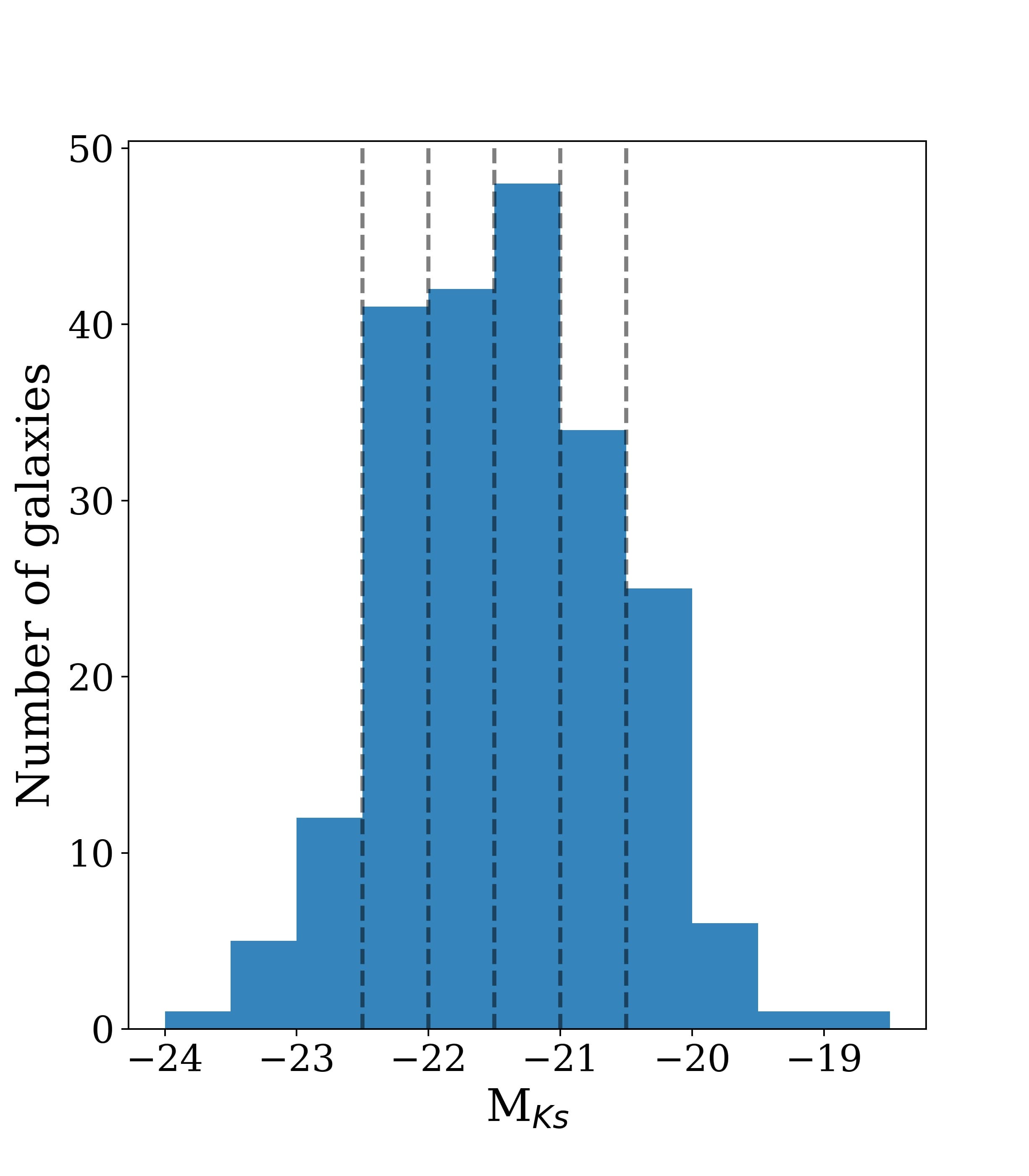}
    \caption{Distribution of the resulting BAGPIPES parameters in the C3 sample, \textit{from left to right}: stellar mass, FUV and Ks band luminosity. The vertical dashed lines represent the ranges of each bin for stacking according to Table ~\ref{tab:binsDR3}.}
    \label{samplemassesDR3}
\end{figure*}

In Figure \ref{sampleUMassbag}, we present some relations between the parameters extracted from the SED fitting, such as stellar mass, and {rest-frame} luminosity in different filters. The C3 sample is color-coded by SFR, which range from $\log \text{SFR} [M_{\odot} yr^{-1}]$=0.13-2.89  ($\langle \log \text{SFR} [M_{\odot} yr^{-1}]\rangle= 1.33$, $\sigma=0.42$). The stellar masses of the C3 sample span from $\log(M_{*}/M_{\odot})=8.54$ to 10.40 with a mean value of $\langle\log(M_{*}/M_{\odot})\rangle=9.41 \,(\sigma=0.33)$. The FUV(1500) luminosity, tracing the young stellar component of galaxies, span from $M_{FUV}=-21.93$ to $-18.43$ mag, with a mean value of $\langle M_{FUV}\rangle=-20.55\text{ mag}\,(\sigma=0.65)$. The K$_s$ band, which better traces the {evolved} stellar component, ranges between $-24.16$ and $-18.90$ mag, with a mean value of $\langle M_{K_s}\rangle=-21.44 \text{ mag}\,(\sigma=0.85)$. As expected, the rest K$_s$-band luminosity is a good tracer of the stellar mass of the galaxies, showing little scatter in Fig. \ref{sampleUMassbag}. Fig. \ref{sampleMSFR} demonstrates that the C3 sample is mainly located along the M$_{\star}$-SFR main-sequence followed by the parent sample. Only a few galaxies at the higher stellar mass end ($\gtrsim$  10$^{10}$M$_{\odot}$) appear offset to higher SFR. The C3 sample is therefore fairly representative of the VANDELS DR3 parent sample in this redshift range. The parameters shown in Fig.~\ref{sampleUMassbag} are thus used for the stacking analysis of the global properties of the sample and their distributions are displayed in the histograms in Figure \ref{samplemassesDR3}. 

\begin{table*}[t!]
    \centering
    \caption{Bin ranges used for the stack analysis with BAGPIPES parameters.}
    \begin{tabular}{c|c|c|c|c|c|c}
    \hline\hline
Bin&Bin range&N$_{gal}$ \tablefootmark{a} &Bin range&N$_{gal}$ \tablefootmark{a}&Bin range &N$_{gal}$ \tablefootmark{a}\\
&$\log M_{*}/M_{\odot}$&&FUV luminosity&&Ks luminosity&\\\hline
1&8.4 - 9.0 &25 &-22.0 : -21.5 &19 &-24.0 : -22.5 &19 \\
2&9.0 - 9.4&76 &-21.5 : -21.0 &31 &-22.5 : -22.0 & 41\\
3&9.4 - 9.7&77 & -21.0 : -20.5 & 64&-22.0 : -21.5 &42 \\
4& 9.7 - 10.0&30 &-20.5 : -20.0 & 63&-21.5 : -21.0 & 48\\
5&10.0 - 10.6&9 & -20.0 : -19.5&31 &-21.0 : -20.5 & 34\\
6&&&-19.5 : -18.0 & 9&  -20.5 : -18.5 &33 \\\hline\hline
    \end{tabular}
    \tablefoot{\tablefoottext{a}{Number of galaxies in each bin}
    }
    \label{tab:binsDR3}
\end{table*}{}

Besides the physical parameters obtained from BAGPIPES, we also measure the EW(CIII]) and EW(Ly$\alpha$) in all galaxies of the C3 sample. For CIII], we use \textsl{slinefit}\footnote{\url{https://github.com/cschreib/slinefit}} following a similar scheme to the one that will be explained in detail in Sec. \ref{sectionstack} for the stacked spectra. The EW(CIII]) distribution is presented in the left panel in Fig. \ref{EWdist}. The EW(CIII]) has a mean value of $\langle$EW(CIII])$\rangle=3.98$\AA\     ($\sigma=3.12$\AA). 
While most galaxies show low EW(CIII])$<$5\AA, we find a small number of strong CIII] emitters with EW(CIII]) up to $\sim$20\AA ($\sim$11\% of the C3 sample with EW(CIII])$>$8\AA). The EW(CIII]) values are shown in the color-code of Fig. \ref{sampleMSFR}, where the M$_{\star}$-SFR plane is shown.  It can be noticed that the intense and faint CIII]  emitters are above and below the main sequence, with some trend suggesting that the more intense CIII] emitter have lower stellar masses and then lower star formation rates.  

About half of our C3 sample are at $z>2.9$ with Ly$\alpha$ observable in the spectral range. For these 105 galaxies, we use the EW(Ly$\alpha$) obtained by \cite{Cullen_2020}. The distribution of such values is presented in Fig. ~\ref{EWdist}. We find that the EW(Ly$\alpha$) span a wide range from -48.37 to 99.79\r{A}, with a mean value of 12.69\r{A} ($\sigma=28.78$). {These values are significantly higher than the mean EW(Ly$\alpha)\sim 2$\r{A} of the parent sample}. Thus, the C3 sample includes both strong Ly$\alpha$ emitting galaxies and galaxies with weak or absent Ly$\alpha$ emission. About 34\% of the C3 sample with Ly$\alpha$ included in the spectral range are considered Ly$\alpha$ emitters galaxies (LAEs) (i.e. EW(Ly$\alpha$)$>$20\r{A}) consistent with what is commonly found for LGBs at these redshifts \citep[e.g.,][]{Cassata2015,Ouchi2020}. {A smaller fraction of LAEs ($\sim$23\%) is found for galaxies with non-detections (S/N$<$\,3) of CIII] in the parent sample.}

\subsection{Stacking procedure}\label{sectionstack}

\begin{figure}[ht]
    \centering
    \includegraphics[width=0.49\linewidth]{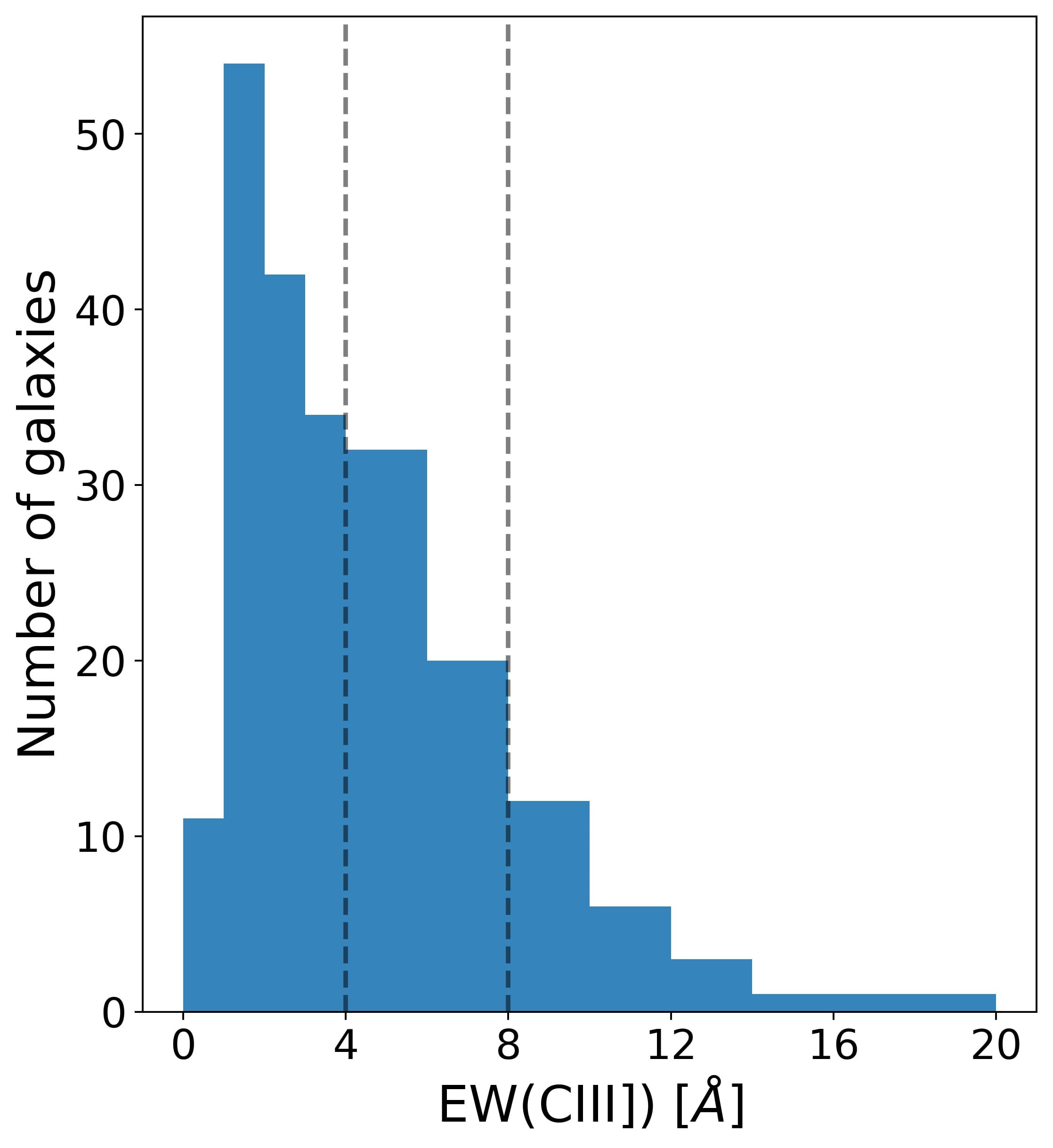}\includegraphics[width=0.49\linewidth]{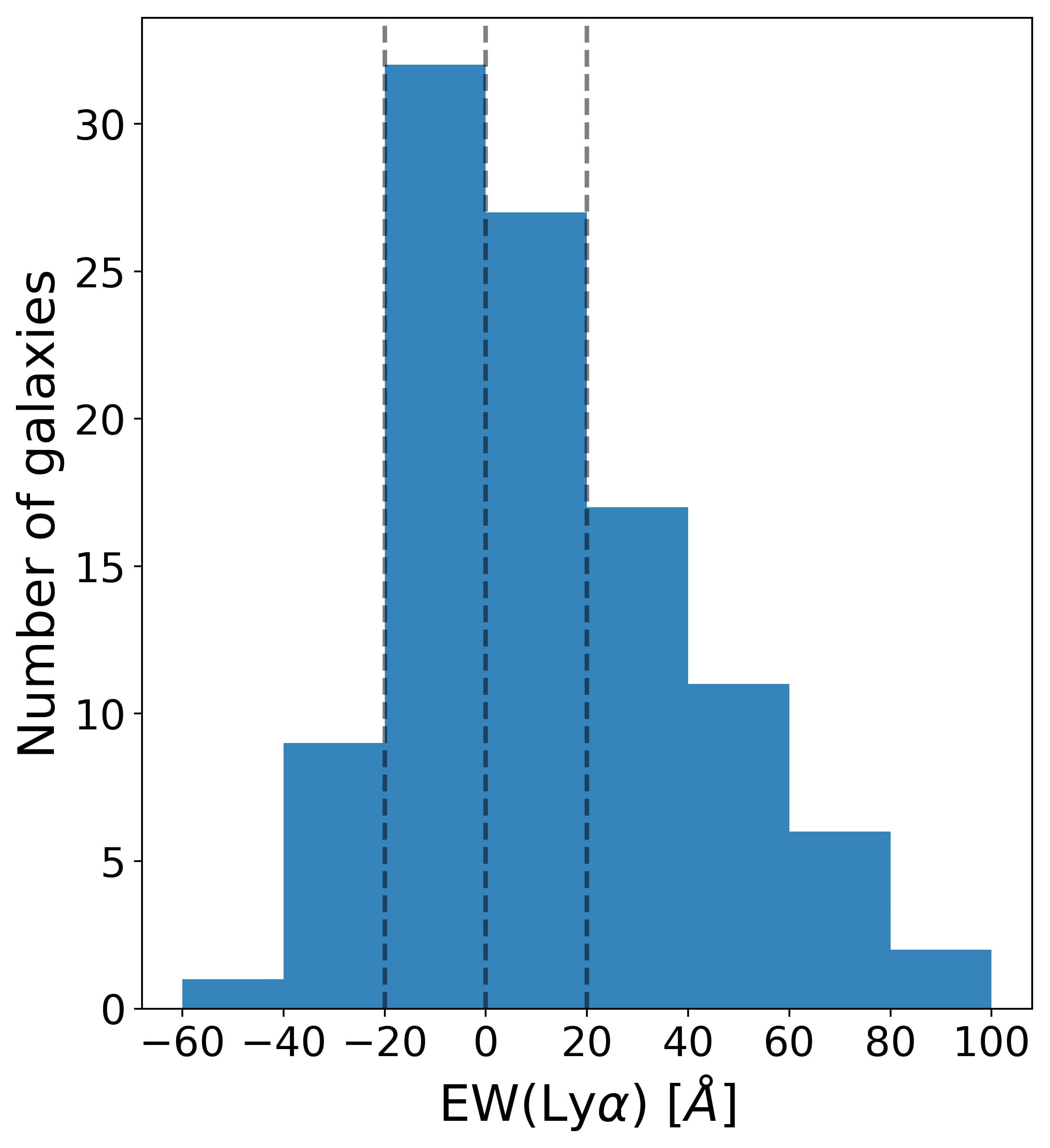}
    \caption{EW(CIII]) (\textit{left}) and EW(Ly$\alpha$) (\textit{right}) distributions of the galaxies in the C3 sample. The vertical dashed lines are the limits of the bins used for stacking according to Table \ref{tab:binsDR3EW} }
    \label{EWdist}
\end{figure}

\begin{figure*}[ht]
    \centering
    \includegraphics[trim=5 75 5 90, clip,width=0.33\linewidth]{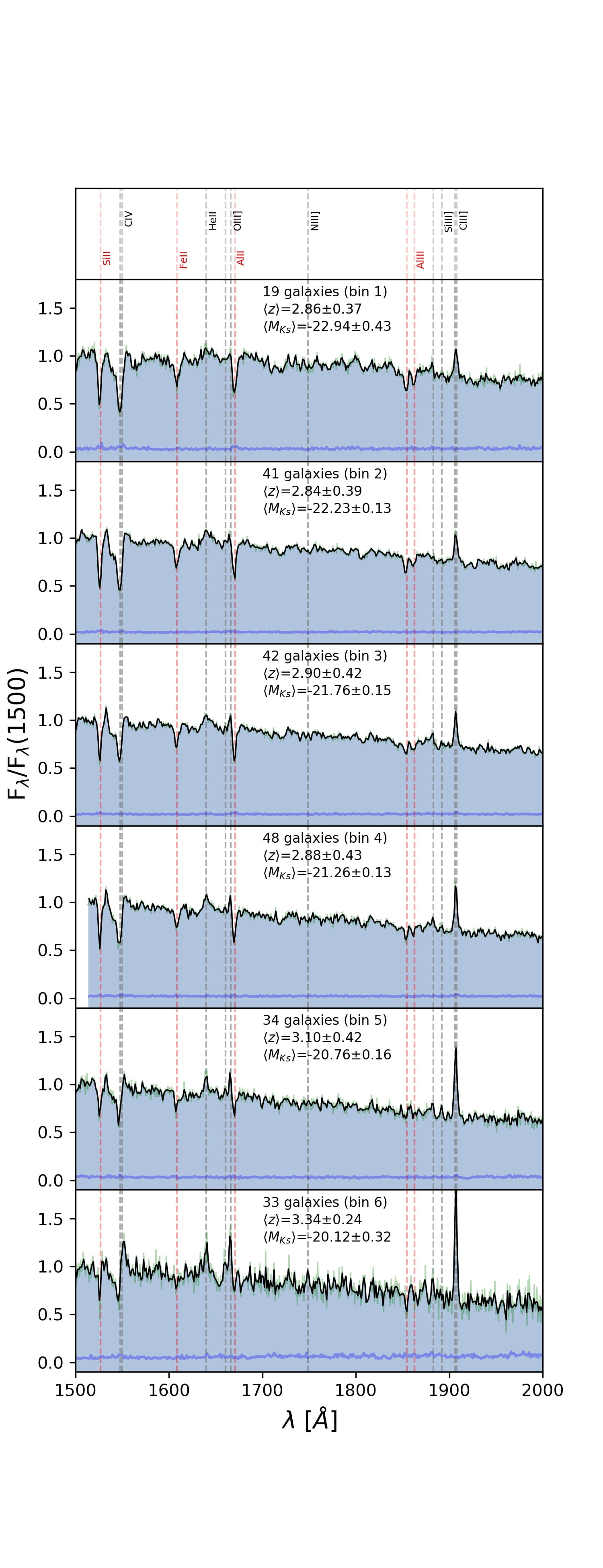}\,\includegraphics[trim=5 75 5 90, clip,width=0.33\linewidth]{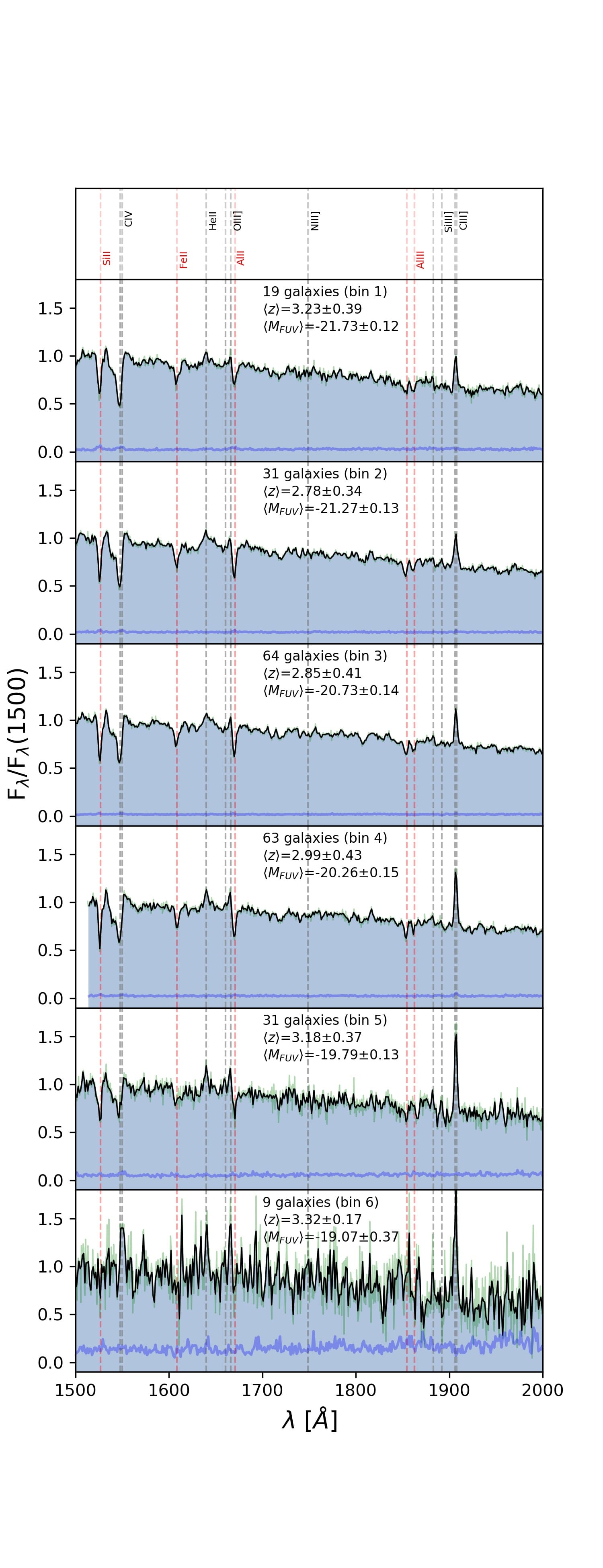}\,\includegraphics[trim=5 75 5 90, clip,width=0.33\linewidth]{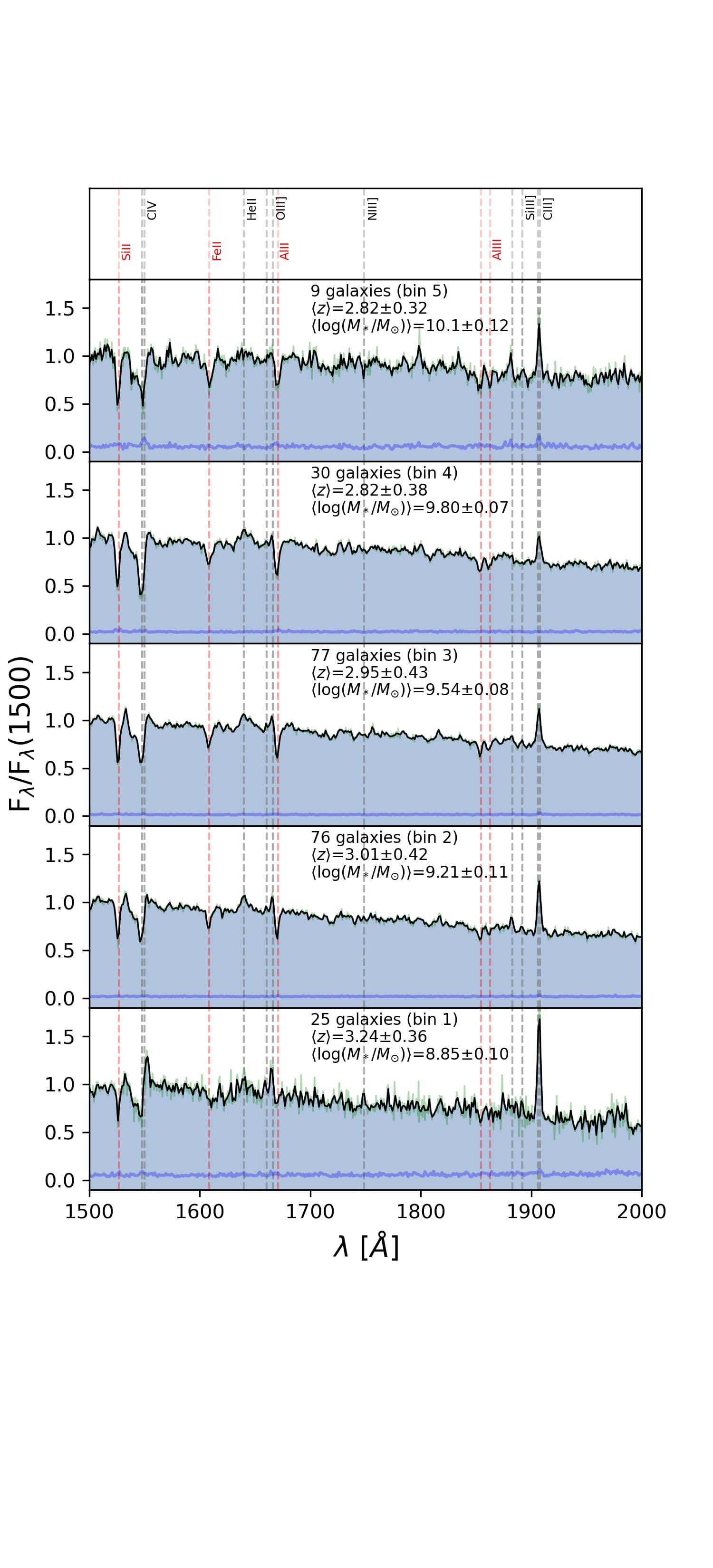}
    \caption{Resulting stacked spectra for each physical parameter with the C3 sample (what we call Stack A, see text for more details) . \textit{From left to right}: M$_{K_s}$, M$_{FUV}$, and stellar mass. In each panel, the green faint line is the stack spectrum with the $\sim$0.6\r{A}/pixel sampling, while the black one is with $\sim$1.2\r{A}/pixel. The blue line is the 1-$\sigma$ error spectrum. The vertical lines mark known UV lines (in black: emission lines, in red: ISM absorption lines). Information about the number of galaxies, the mean redshift and the mean parameter are included in each panel.}
    \label{StackA}
\end{figure*}

\begin{figure*}[ht]
    \centering
    \includegraphics[trim=5 775 5 50, clip, width=0.9\linewidth]{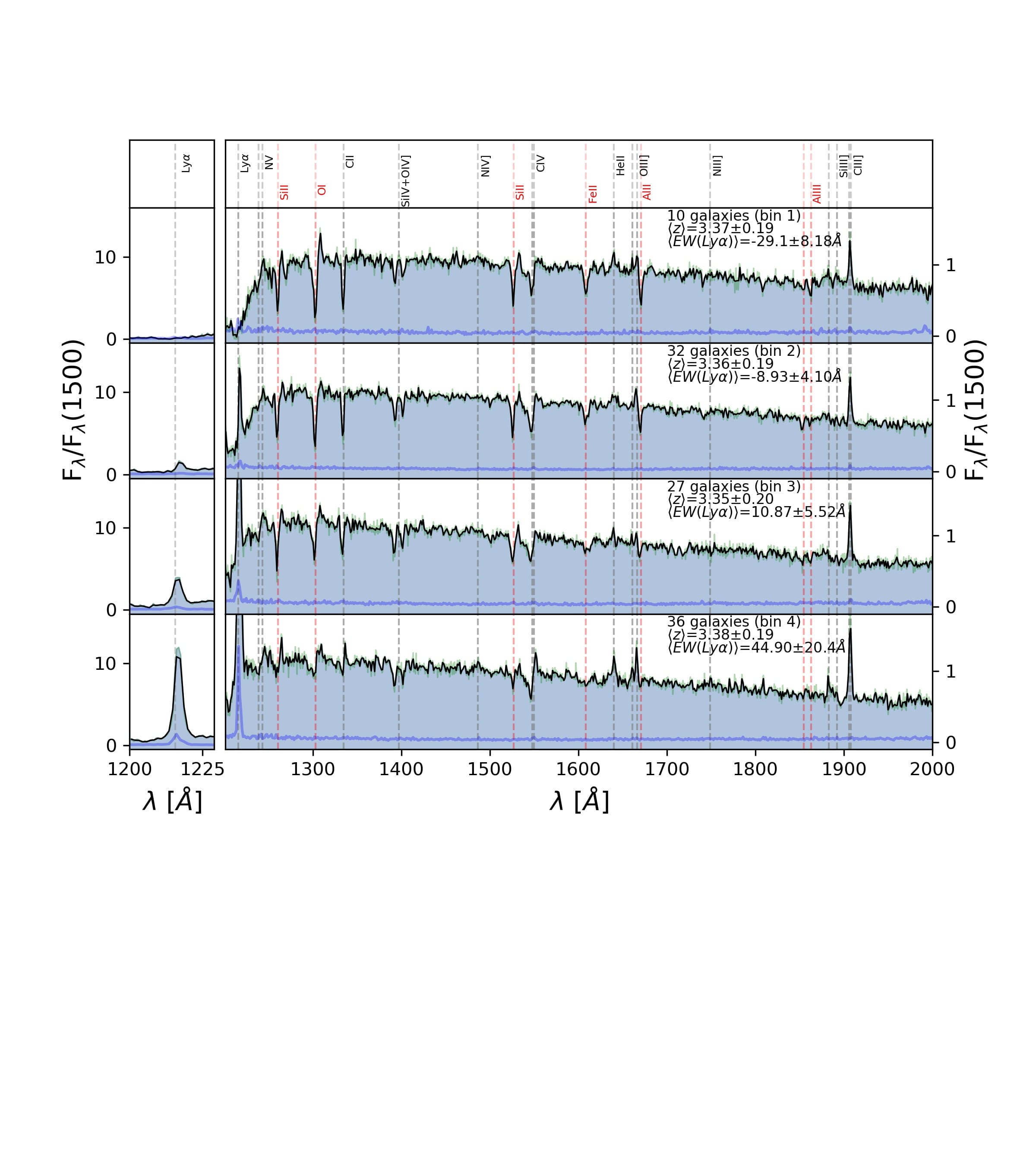}
    \caption{Resulting stacks by EW(Ly$\alpha$). Same as in Fig. \ref{StackA}. Information about the number of galaxies, the mean redshift and the mean EW(Ly$\alpha$) are included in each panel. The left panels show the relative Ly$\alpha$ strength in each bin.}
    \label{StackLya}
\end{figure*}

In this paper, we are interested in the characterization of the mean physical properties for the CIII] emitters in VANDELS. For this reason, we perform a stacking analysis, binning the C3 sample by stellar mass {and rest-frame luminosity and EW}. This allows us to increase the S/N of the data and probe properties, such as stellar metallicity, which would not be possible in individual objects. 

For this aim, we separate the selected galaxies in five stellar mass bins of width $\sim 0.3$ dex and six bins of luminosity of  $0.5$ dex each. This way, we have a significant number of galaxies per bin, as shown in Table ~\ref{tab:binsDR3} and marked by the vertical dashed lines in Fig. \ref{samplemassesDR3}. As shown in Fig. \ref{sampleUMassbag}, stellar mass and K$_s$ luminosity are correlated, as the latter is a good tracer of the  former. While we expect  stacks in these two quantities to produce similar results, we decided to use both of them to evaluate any possible difference due to the larger  dynamic range of the K$_s$ luminosity. 

We also separate the sample in bins of EW(CIII]) and EW(Ly$\alpha$). For the former, we separate them in 3 bins of 4\r{A}. And for Ly$\alpha$, the C3 sample is restricted to the subsample of galaxies with $z_{sys}>2.9$ where Ly$\alpha$ is observable, either in absorption or emission. We choose bins of $\sim$20\r{A} for the EW(Ly$\alpha$), which are presented in Table \ref{tab:binsDR3EW} and are marked by vertical dashed lines in Fig. \ref{EWdist}. 

\begin{table}
    \centering
    \caption{Bin ranges used for the stack analysis with EW.}
    \begin{tabular}{c|c|c|c|c}
    \hline\hline
Bin&Bin range&N$_{gal}$ \tablefootmark{a}&Bin range&N$_{gal}$ \tablefootmark{a}\\
&EW(CIII])[\AA]&&EW(Ly$\alpha$)[\AA]&\\\hline
1&0 : 4&141&-60 : -20&10\\\hline
2&4 : 8&52&-20 : 0&32\\\hline
3& 8 : 20&24&0 : 20&27\\\hline
4&&&20 : 100&36\\\hline\hline
    \end{tabular}
    \tablefoot{
    \tablefoottext{a}{Number of galaxies in each bin}}
    
    \label{tab:binsDR3EW}
\end{table}{}

For the stacking, we use a non-weighted scheme following \citet{Marchi2017}. All the individual spectra in the sample are first shifted into the rest-frame using the systemic redshift  and then they are resampled onto a common grid according to the mean systemic redshift of the sample ($z_{sys}\sim 2.98$) and normalized to the mean flux between 1460 and 1540\r{A}. The final flux at each wavelength was taken as the median of all the individual flux values after a 3-$\sigma$ clipping for rejecting outliers. The final wavelength range is where all spectra overlap and the spectral binning is 0.64\r{A}. The 1-$\sigma$ error spectrum is estimated by a bootstrap re-sampling of the individual fluxes for each wavelength and the standard deviation of the resulting median stacked spectra. {Changing the range of normalization to $\sim 1800$\r{A} does not affect the shape of the stacked spectra.} 

We also tested alternative weighted schemes for stacking, similar to the one presented in \citet{Marchi2017} and used in \cite{Saxena_2020} with a 1/$\sigma^2$ weight, where $\sigma$ is estimated as the flux error along the normalization range in each spectrum. The error spectra with the weighted scheme were larger compared  with the median stacking. For this reason, we consider the median stacking for this work, as they are a better representation of the global properties of the galaxies in each bin. 

Four different median stacking schemes were performed depending on the redshifts included in each bin. In the remaining of this work, they are named as follows:
\begin{itemize}
    \item Stack A: All the galaxies in the bin are stacked, considering the entire C3 sample. These stacks for each physical parameter are presented in Fig. ~\ref{StackA}. 
    \item Stack B: Only galaxies with $z>2.93$ are considered for stacking in each bin. These galaxies have Ly$\alpha$ included in the spectral range. The stacks for each physical parameter are presented in Fig. ~\ref{StackB}. 
    \item Stack C: Only galaxies with $z>2.93$ and with EW(Ly$\alpha$)$>$0 (i.e. Ly$\alpha$ in emission) are stacked. These stacks for each physical parameter are presented in Fig. ~\ref{StackC}. 
    \item Stack D: We stack only galaxies with $z<2.93$. In this subset, Ly$\alpha$ is not covered by the VANDELS spectra. Thus we ignore whether these galaxies are Ly$\alpha$ emitters or not. The stacks for each physical parameter are displayed in Fig. ~\ref{StackD}.
\end{itemize}
\begin{figure}[t]
    \centering
    \includegraphics[width=\linewidth]{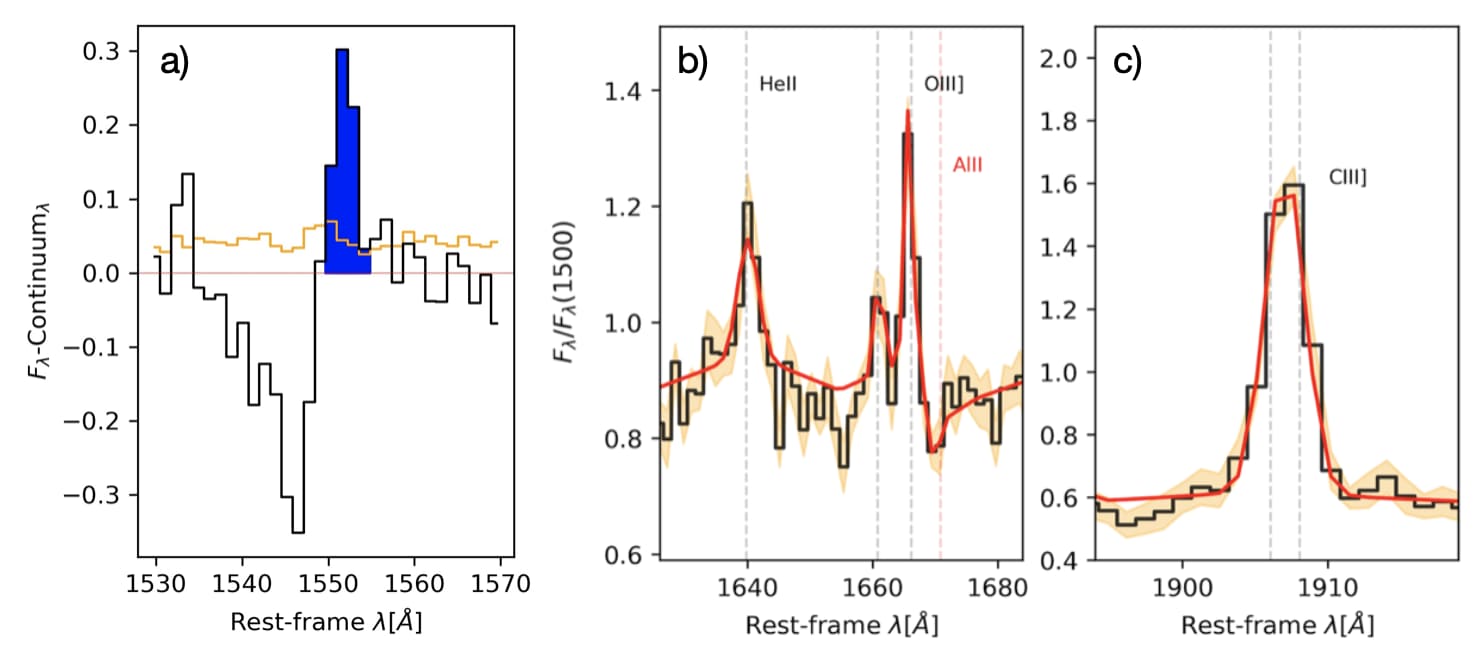}
    \caption{Emission-line flux and EW measurements in the stack by EW(Ly$\alpha$)$>$20. \textit{Panel a)} Continuum-subtracted spectrum. The orange line shows the error spectrum (1\,$\sigma$). The blue shaded region are the pixel integrated for the CIV line. \textit{b) and c)} The red line is the \textsl{slinefit} fitting for the spectrum. The orange shaded region shows 1\,$\sigma$ uncertainty of the stacked spectrum.}
    \label{example_slinefit}
\end{figure}
An additional subset of stacks A, B, C, and D by EW(CIII]) are presented in Fig. \ref{StackEWC3}. In the case of the stacks by EW(Ly$\alpha$), only Stack B is performed. The resulting stacks are presented in Fig. ~\ref{StackLya}. 

The above redshift dependence reduces the number of galaxies in each bin. In these cases, we adapt the binning for stacking to have at least four galaxies in each bin. This ensures the stack spectrum will gain at least a factor of 2 in the S/N ratio. The final number of galaxies for each stacked spectrum is included in labels in Fig. \ref{StackA},  \ref{StackB}, \ref{StackC}, \ref{StackD}, \ref{StackEWC3}, and \ref{StackLya}. 
 We find the composite spectrum of a bin to be representative of the median properties of the galaxies in each bin. Small changes in the bin sizes used for the stacking may change the error bars in the derived parameters in the least populated bins but they do not affect our results significantly. We discuss possible caveats related to the stacking analysis in Appendix \ref{apen0}.

\subsection{Line measurements}\label{sec:linemeas}
\begin{figure*}[t]
    \centering
    \includegraphics[width=\linewidth]{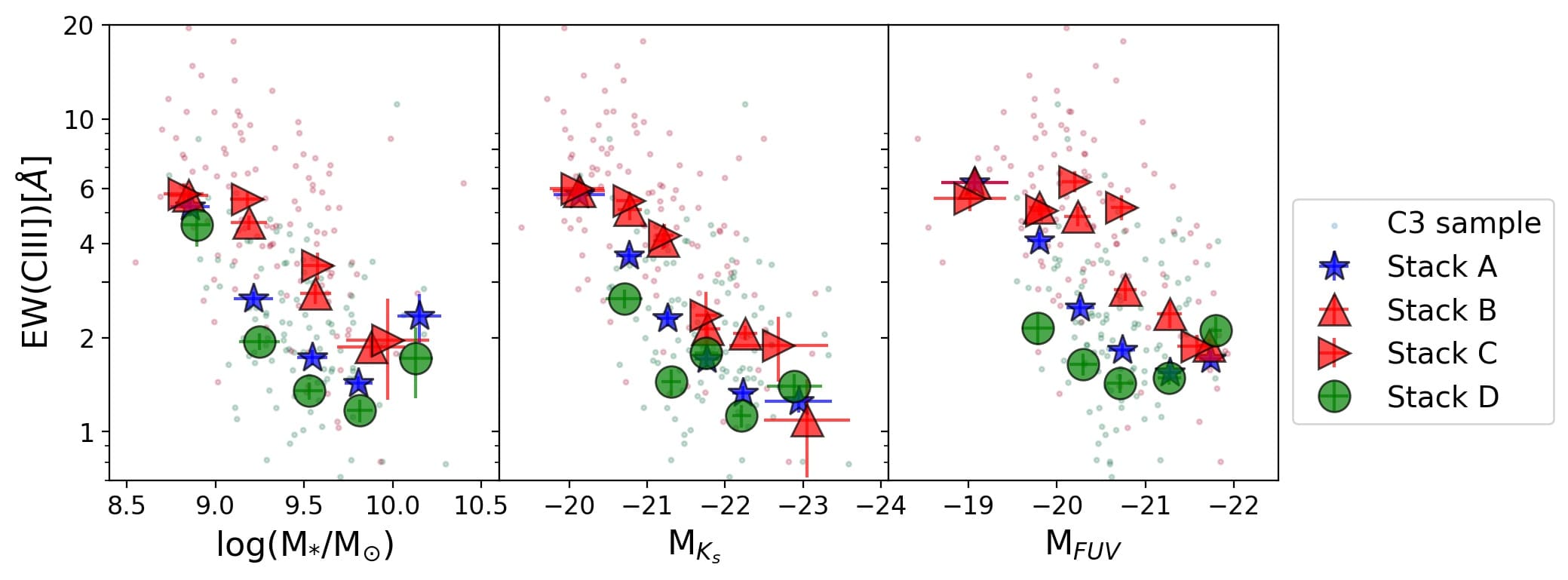}
    \caption{Relation of EW(CIII]) with (from left to right) stellar mass, K$_s$-band luminosity, and FUV luminosity, for the stacks A, B, C, and D for each parameter, and for the individual galaxies in the C3 sample (small circles. In green, galaxies at $z\lesssim 3$, and in red, galaxies at $z\gtrsim 3$). In the \textit{x}-axis, the parameters correspond to mean values for stacks, while for the C3 sample, the parameters correspond to the SED fitting values.}
    \label{c3relation}
\end{figure*}

Emission-line fluxes and EWs of the lines in each stacked spectrum were measured using \textsl{slinefit}, which is a software capable of simultaneously measuring emission and absorption lines and the UV-NIR continuum. For this purpose, \textsl{slinefit} uses templates built with the \citet{Bruzual_2003} stellar population models.

For our measurements, we include rest-frame UV emission and absorption lines at $\lambda_{\rm rest} >$\,1500\AA\ from \citet{Shapley_2003}. Rest-frame UV lines at $\lambda_{\rm rest} <$\,1500\AA\ were not included in the measurements. In particular, the Ly$\alpha$ line is instead measured following the same fitting technique presented in \cite{Cullen_2020}. In all the \textsl{slinefit} fitting runs, we allow that lines other than CIII] may have a small offset respect to the systemic velocity and the minimum width of the lines to be 100 km/s.  

In a first set of measurements, we only measure simultaneously CIII] and closer lines (AlIII$\lambda$1855\AA, SiIII$\lambda\lambda$1883,1892\AA, MgII$\lambda$2799\AA). We obtain the width of CIII] to be $\sim$300-350 km/s in all stacked spectra. We use this value to constrain the maximum width for the other emission lines.  

After that, in a second set of measurements, we measure HeII$\lambda$1640\AA, OIII]$\lambda\lambda$1666,1660\AA, and CIV$\lambda\lambda$1548,1550\AA\ (hereafter HeII, OIII], and CIV, respectively), with the constraint in the maximum width. In these cases, the maximum offset allowed is 100km/s, except for CIV for which a maximum of 1000km/s is allowed because larger offsets are observed. The P-Cygni profile of CIV is fitted assuming the same intensity for both components. Both components of OIII] are fitted with their ratio unconstrained.

Due to the complex CIV profile, the measurements with \textsl{slinefit} are found to slightly underestimate the continuum. For this reason, a more detailed continuum determination is performed. First, the continuum is fitted with a linear function between 1400 and 2000\r{A}, masking out regions with emission and absorption lines detected. Then, the spectrum is continuum-subtracted. Finally, the CIV flux is found by direct integration of the emission line profile after imposing a maximum base line width of 4\r{A} (or $\sim$ 390km/s), that is the typical value obtained for CIII]. The EW is estimated using the mean continuum flux in the same integrated range. 

All the above measurements are performed for all the stacked spectra with the 0.6\r{A}/pixel sampling, but the S/N in the case of faint emission lines is low and then the measurements are additionally performed in the resampled spectra by a factor of 2. For the resampling, we use \textit{SpectRes}\footnote{\url{https://github.com/ACCarnall/spectres}. More details on \citet{Carnall_2017}} that is a software that efficiently resamples spectra and their associated uncertainties, preserving the integrated flux. An example of the measurements can be seen in Fig. ~\ref{example_slinefit}. Hereafter, we consider the resampled spectra  measurements for the emission lines which are presented in Tables ~\ref{tablemassbin}, ~\ref{tableKsmagbin},  ~\ref{tableFUVmagbin}, \ref{tableEWLyabin}, and \ref{tableC3bin}.  

In the same tables, the mean color excess E(B-V) is reported for each stack. They are estimated from the individual E(B-V) values for each galaxy in each bin, which are obtained by BAGPIPES fitting. In the C3 sample, E(B-V) range from $\sim$0.01-0.43 mag, with a mean value of 0.098$\pm$0.014 mag. The mean E(B-V) of the stacks are used to compute the reddening correction\footnote{Using the \textit{extinction} code at \url{http://github.com/kbarbary/extinction}} using the \cite{Calzetti2000} extinction curve for simplicity and assuming that the color excess of the stellar continuum is the same that the color excess for the nebular gas emission lines. Despite the evidence that this assumption could not be true \citep[e.g.][]{Calzetti2000, Reddy2015} and the ionized gas E(B-V) could be larger than a factor of $2.2$ the stellar E(B-V) (in particular, galaxies with high SFR), we assume it for simplicity. However, we note that the results of this paper are not affected if we change this prescription to more extreme assumptions. Using the calibration presented in \cite{Sanders2020} to correct the gas extinction from the SED extinction, we obtain a factor up to $\sim 3$ of difference between gas and stellar extinction, but even with those values, the trends found in this paper are not altered significantly.  

Line fluxes are presented uncorrected by extinction in Tables ~\ref{tablemassbin}, ~\ref{tableKsmagbin},  ~\ref{tableFUVmagbin}, \ref{tableEWLyabin}, and \ref{tableC3bin}. However, results and figures shown in subsequent sections consider dereddened quantities.

\begin{figure*}[ht]
    \centering
    \includegraphics[width=0.9\linewidth]{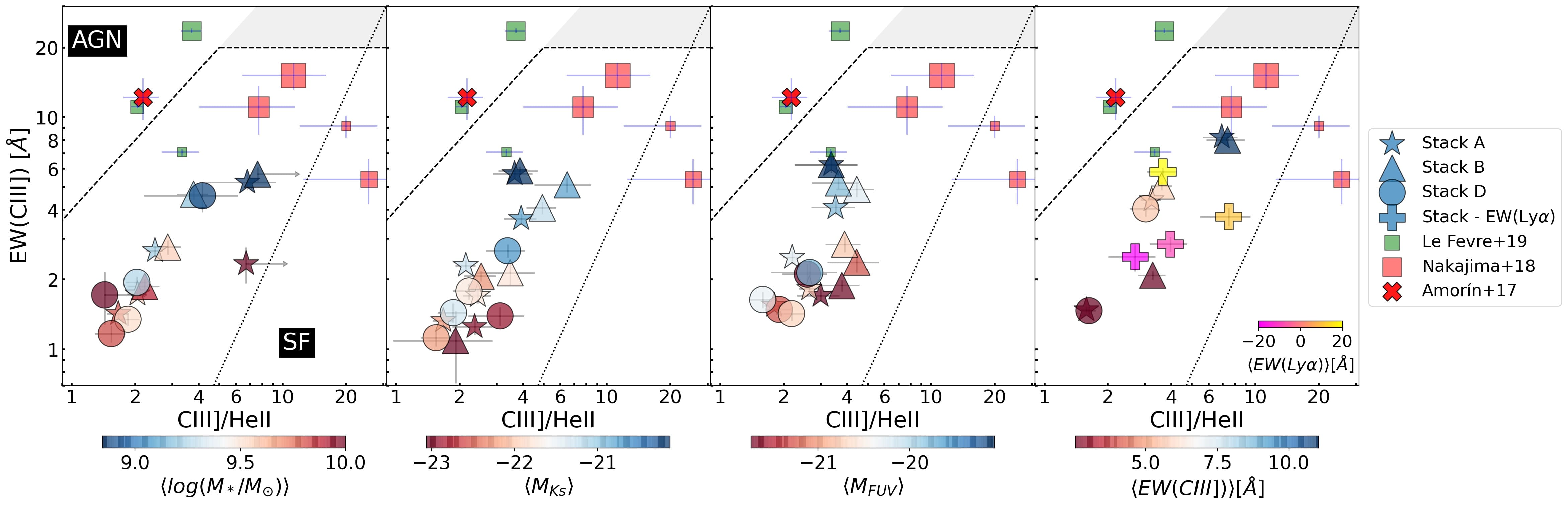}\\\includegraphics[width=0.9\linewidth]{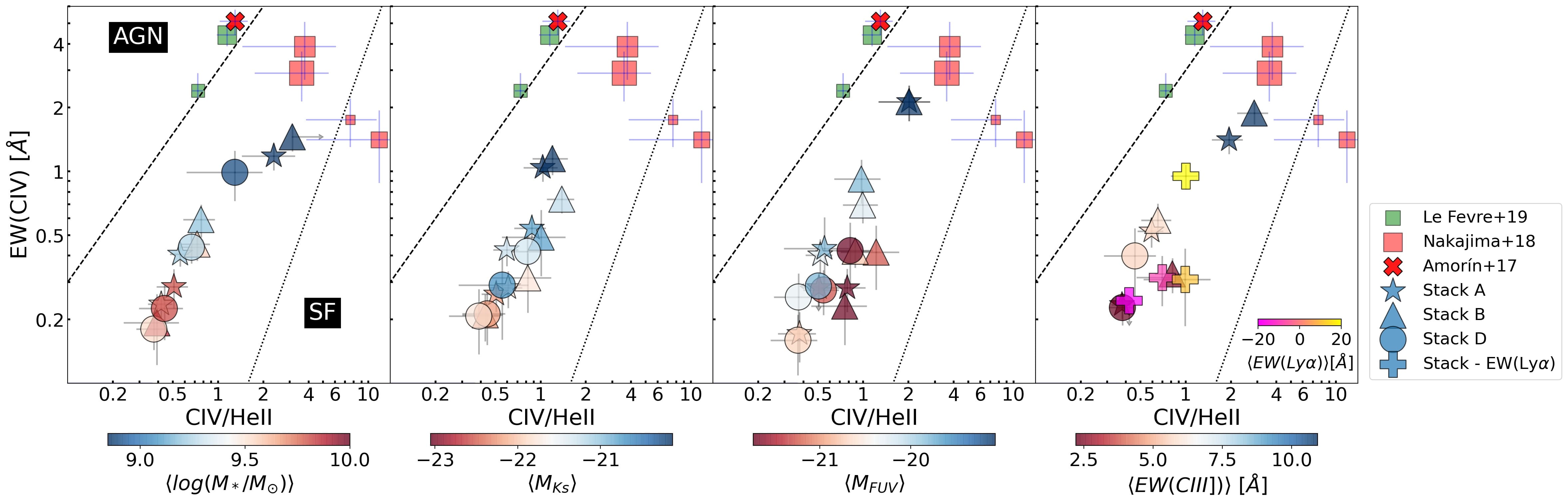}\\\includegraphics[width=0.9\linewidth]{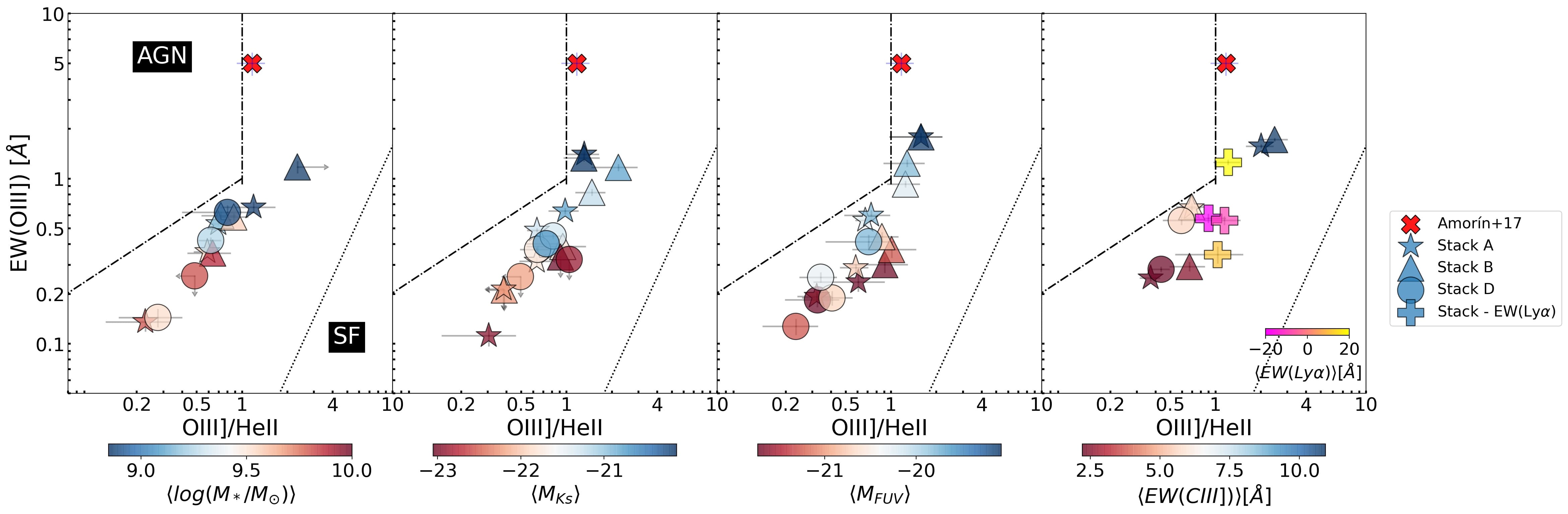}
    \caption{Diagnostic diagrams with EW of UV emission lines for the different stack sets, color-coded by the parameter used for stacking. The black dashed lines separate AGNs and star-forming galaxies as proposed in \cite{Nakajima_2018}. The dotted and dotted-dashed lines are from \cite{Hirschmann2019} to separate between SFGs and composite, and composite and AGNs, respectively. \textit{Top row}: Diagnostic of EW(CIII])-CIII]/HeII ratio. The gray shaded region is where the models overlap. \textit{Middle row}: Diagnosis of EW(CIV)-CIV/HeII ratio.  \textit{Bottom row}: Diagnosis of EW(OIII])-OIII]/HeII ratio. The green boxes are the composites from \cite{LeFevre_2019} and the higher the size, the higher the EW (CIII]). The red X mark is the composite from \cite{Amorin_2017}.
    The red rectangles are the stacks from \citep{meanNakajima_2018} with the brightest M$_{UV}$ (smallest rectangle), smaller EW(Ly$\alpha$), faintest M$_{UV}$, and larger EW(Ly$\alpha$) (largest rectangle).}
    \label{diagC}
\end{figure*}

\begin{figure*}[ht]
    \centering
    \includegraphics[width=0.85\linewidth]{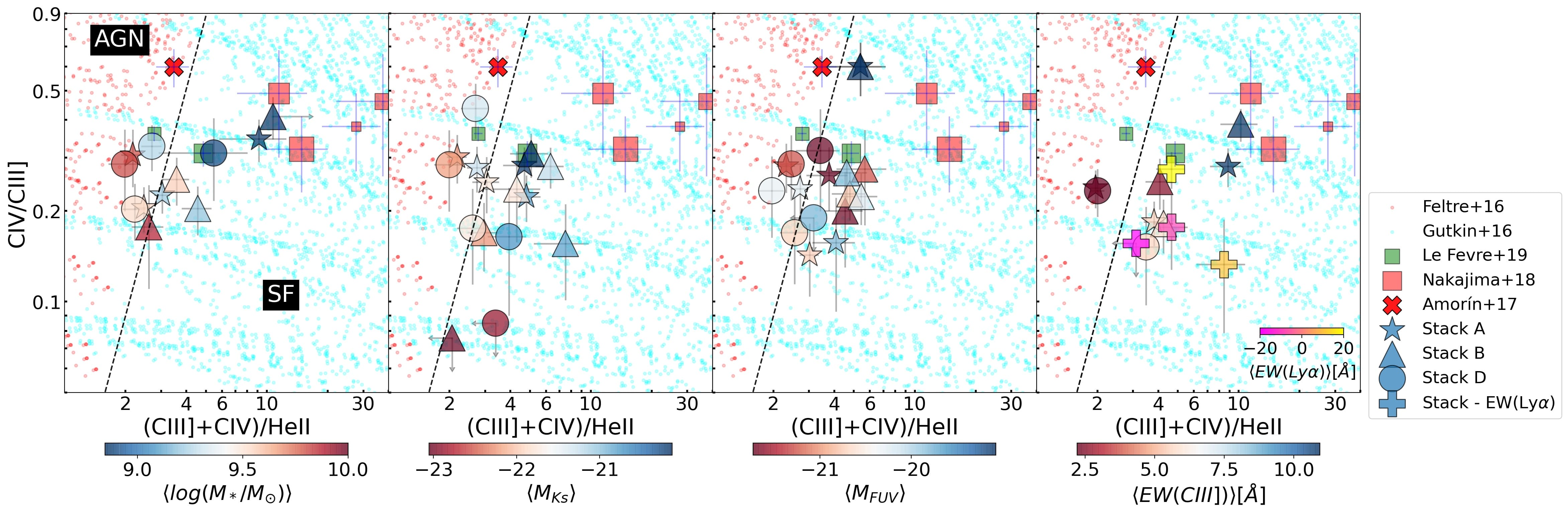}\\
    \includegraphics[width=0.85\linewidth]{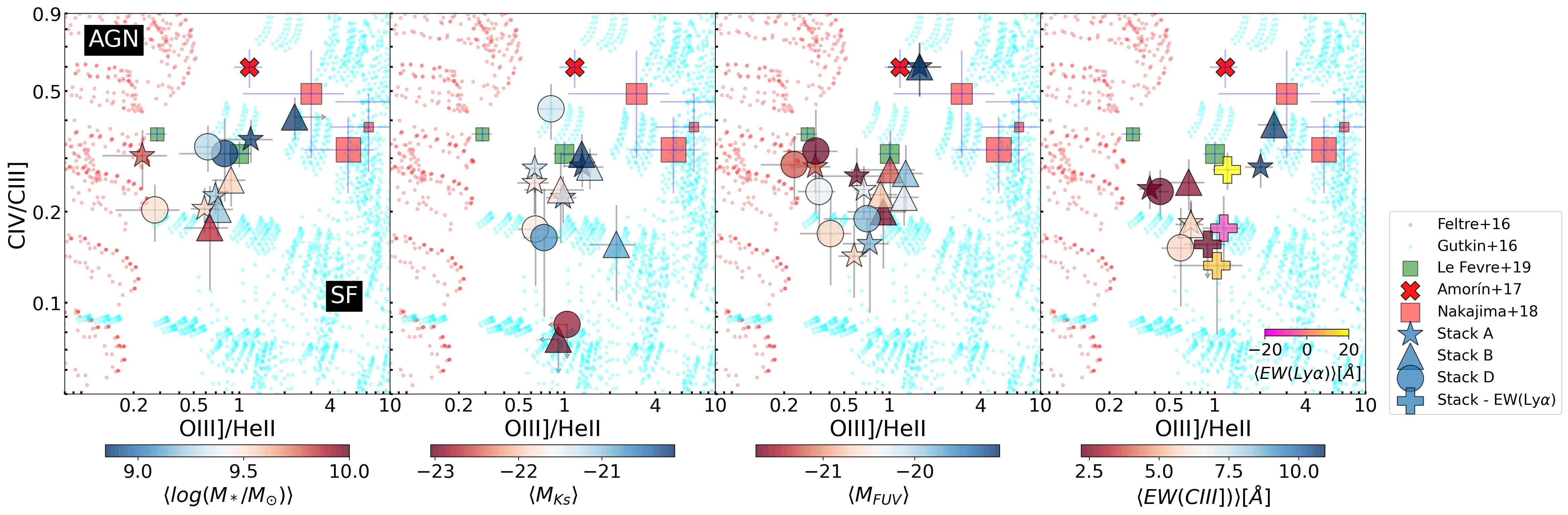}\\\includegraphics[width=0.85\linewidth]{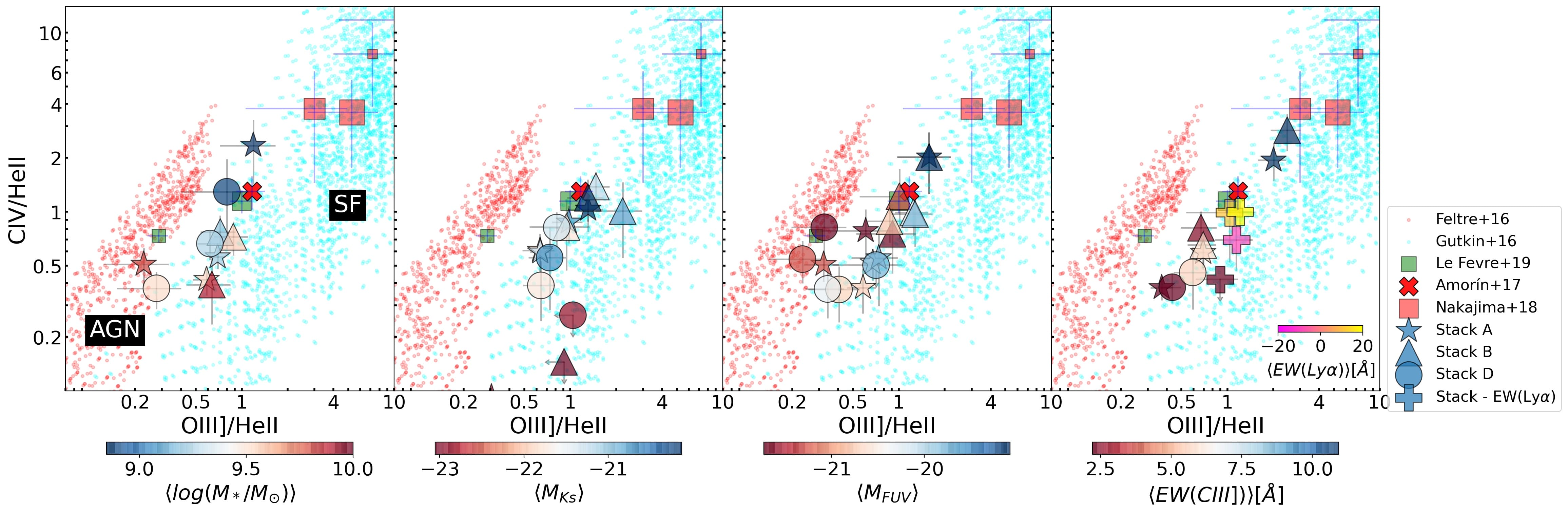}\\\includegraphics[width=0.85\linewidth]{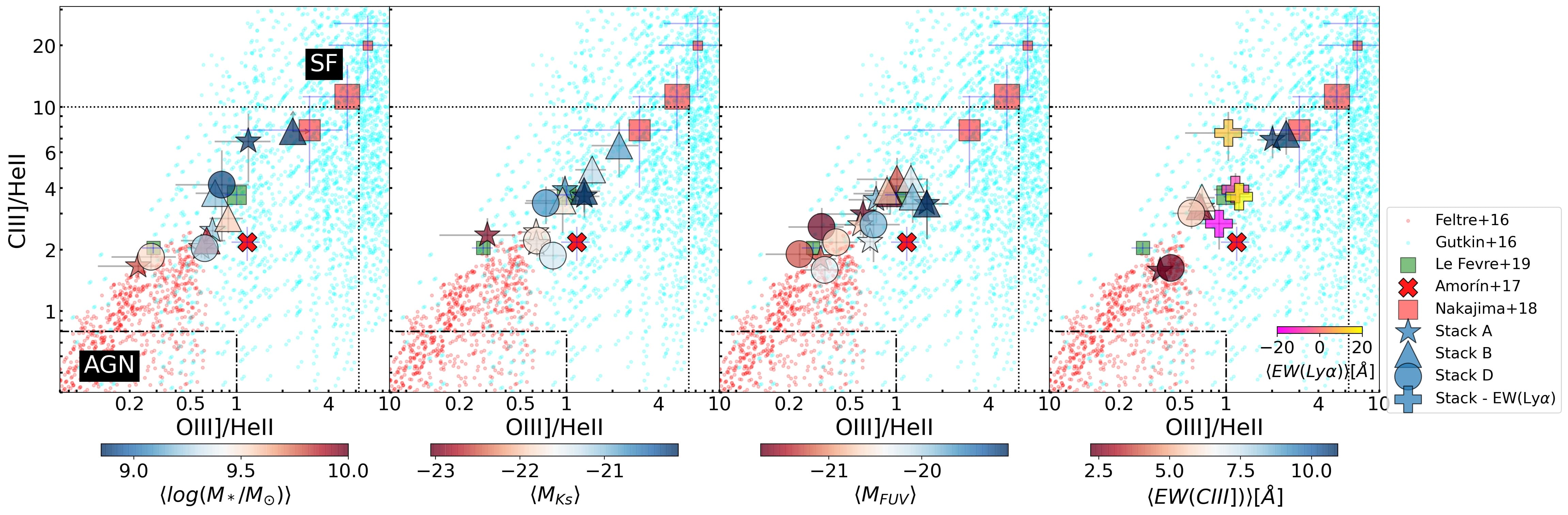}\\
    \caption{Diagnostic diagrams with fluxes of UV emission lines for the different stack sets, color-coded by the parameter used for stacking. Symbols are the same as in Fig. \ref{diagC}. The red and cyan points are the photoionization models for AGN \citep{Feltre_2016} and SF \citep{Gutkin_2016} for metallicity $Z$=0.0001, 0.0002, 0.0005, 0.001, 0.002, and for ionization parameter $\log(U)$=-4, -3, -2, -1. The dotted-dashed lines in the bottom panel separates composite and AGNs according to \cite{Hirschmann2019}.}
    \label{diagO}
\end{figure*}

\section{Results}\label{sec:results}
\subsection{Identification of UV absorption and emission lines} \label{subsec:lines}

The high S/N spectra of the stacks allows us to identify several interesting features both in absorption and emission in the rest-frame UV spectra. Among these features, low-ionization interstellar lines such us SiII$\lambda$1260\r{A}, OI+SiII$\lambda$1303\r{A}, CII$\lambda$1334\r{A}, SiII$\lambda$1526\r{A}, FeII$\lambda$1608\r{A}, and AlII$\lambda$1670\r{A} are found. In Fig. \ref{StackA}, where the stacks A (redshift-independent) by stellar mass and luminosities are shown, we find that the stronger ISM absorption lines are in the stacks built with higher stellar masses (or more luminous at a given broadband). {This is expected since these lines are saturated in low resolution spectra and then their width increases with dynamical mass.} The same trend is shown in Fig. \ref{StackB}, \ref{StackC}, \ref{StackD} for the stacks B, C, and D by stellar mass and luminosities.

We find a similar trend when considering the stacks by EW(Ly$\alpha$) in Fig. \ref{StackLya}. We find the stronger low-ionization ISM absorption lines in the stacks with smaller EW(Ly$\alpha$), i.e. when Ly$\alpha$ is in absorption, while the ISM absorption lines are barely identified in the stack with the larger EW(Ly$\alpha$). This is consistent with previous observations \citep[e.g.][]{Shapley_2003}. Regarding the stacks by EW(CIII]), we note the same trend as we show in Fig. \ref{StackEWC3}. The stronger low-ionization ISM absorption lines are found in the stacks with smaller EW(CIII]). 

In addition to the low-ionization features associated with neutral outflowing gas, we identify high-ionization interstellar absorption lines such as SiIV$\lambda\lambda$1393,1402\r{A}, CIV, and NV$\lambda\lambda$1238,1242\r{A}. While SiIV$\lambda\lambda$1393,1402\r{A} and NV$\lambda\lambda$1238,1242\r{A} are only identified in all the stacks B and C, and in the stacks by EW(Ly$\alpha$) due to the spectral range, CIV is identified in all the stacks. We note in Figs. \ref{StackA}, \ref{StackB}, \ref{StackC}, \ref{StackD}, \ref{StackLya}, \ref{StackEWC3} that the stronger absorption lines are in the stacks of higher stellar mass, brighter in any luminosity, lower EW(Ly$\alpha$), and lower EW(CIII]), i.e. similar to the trend observed in low-ionization ISM features. 

In our stacks we also identify fine structure emission lines of SiII that have been observed in the rest-UV spectrum of star-forming galaxies  \cite[e.g.,][]{Shapley_2003}. SiII*$\lambda$1533\r{A} is in the spectral range of all the stacks. This faint line is between two ISM absorption lines and is identified in most of the stacks. This line is particularly more intense in the more massive and luminous galaxies showing both lower EW(Ly$\alpha$) and lower EW(CIII]), but the trend is less clear than the one we find in the ISM absorption lines (see Fig. \ref{StackA}, \ref{StackB}, \ref{StackC}, \ref{StackD}, \ref{StackLya}, \ref{StackEWC3}). Other fine structure lines, such as  SiII*$\lambda$1265\r{A} and SiII*$\lambda$1309\r{A} are also identified in stacks B and C, and in the stack by EW(Ly$\alpha$) (see Fig. \ref{StackB}, \ref{StackC}, \ref{StackLya},\ref{StackEWC3}). These lines are identified in all the stacks, irrespective of stellar mass, luminosity or EW bin, suggesting they are a more common feature in the UV spectra of CIII] emitters.  

In addition to CIII], we identify nebular emission lines such as OIII], CIV, and HeII, which are central to the main goals of this paper. We observe that the strength of the nebular lines depends on the stellar mass and luminosity. We find that in general the less massive (and fainter in any band) stacks show the more intense Ly$\alpha$, CIV, HeII, OIII] and CIII] nebular lines. In the case of more massive (and brighter in any band), we find that the same set of nebular lines tend to be fainter. In particular for CIV and HeII, they show a stellar wind component as suggested by the P-Cygni profile (in the former) or broad profile (in the latter). In particular, CIV shows a P-Cygni type profile in all stacks with the emission being more intense in the less massive (or faintest) stacks and in the ones with lowest EWs, either CIII] or Ly$\alpha$ (see Fig. \ref{StackA}, \ref{StackB}, \ref{StackC}, \ref{StackD}, \ref{StackLya}, \ref{StackEWC3}). 

In Fig. \ref{StackLya} it is worth noticing that the nebular features appear less dependent of the EW(Ly$\alpha$) than the ISM absorption features, for which clearer differences are seen. We also note that, comparing stacks B and C, that differ by the inclusion of galaxies with Ly$\alpha$ in absorption or not, we find no strong difference in nebular emission lines or ISM absorption, but there are differences in the strength of Ly$\alpha$.  

\subsection{Relation of EW(CIII]) with luminosity and stellar mass}\label{subsec:c3relations}

In this section, we perform a qualitative description of the relation of EW(CIII]) with the physical parameters used for stacking, which are shown in Fig. \ref{c3relation}. We find a trend in which the stacks with more massive galaxies have lower EW(CIII]), while the more intense CIII] emitters correspond to the ones built with the lowest stellar mass, but the scatter is large when we consider individual objects (small circles) and the relation is weak, especially for stacks A and D, which are those containing galaxies at $z\lesssim$\,3 (small green circles). Something similar is observed with the broad-band luminosities, where the stacks of fainter objects tend to have higher EW(CIII]). The scatter follows the one observed for the luminosity of the individual galaxies of the C3 sample. We note that galaxies in the C3 sample at $z\lesssim$\,3 (included in stack D) show a mean EW(CIII])=2.4\r{A} and tend to have lower EW(CIII]) than galaxies at $z\gtrsim$\,3 (included in stack B and C) which show a mean EW(CIII])=5.3\r{A}. This is related to the overlap of VANDELS targets with slightly different selection criteria around $z\sim 3$. For instance, galaxies selected as LBGs tend to show lower stellar masses than those selected as bright SFGs. We refer to \cite{Garilli2021} for a more detailed discussion on the effect of the VANDELS selection criteria on the galaxies physical properties.

Comparing the different schemes of stacking, it can be noticed that the Stack B and C, i.e. the stacks that only include galaxies at $z\gtrsim 3$, have larger EW(CIII]) than Stacks A and D for a given stellar mass or luminosity. On the other hand,  Stack D, including only galaxies at $z\lesssim 3$, tend to have lower EW(CIII]) for a given mass or luminosity than all the other stacking schemes. Except for the stack with the lowest stellar mass, EW(CIII])$\gtrsim 3$\r{A} are not observed in Stack D. In addition, comparing Stack B and C, i.e. the effect of including or not galaxies with Ly$\alpha$ in absorption,  we note that slightly high EW(CIII]) are observed when only Ly$\alpha$ in emission is included.

A similar brief analysis is presented in Appendix \ref{appen1} for the non-detected CIII] emitters from the parent sample. We find they are consistent with these results and they are intrinsically faint CIII] emitters with EW(CIII])$\lesssim 2$\r{A} for any FUV luminosity.

\subsection{Diagnostic diagrams based on UV emission lines}\label{subsec:diag}

Different diagnostic diagrams using rest-frame EWs of CIII], CIV, OIII] and the line ratios of CIII], CIV, and HeII have been proposed to identify the main source of ionizing photons in distant galaxies \citep[e.g.,][]{Nakajima_2018,Hirschmann2019}. These diagnostics are useful for determining the nature of the dominant ionizing source of galaxies. However, they need to be constrained using large samples of galaxies where these lines are detected. Given that the C3 sample is draw from a parent sample of photometrically-selected  SFGs, they can be used to probe and constrain these models using nebular flux ratios and EWs of systems which are not particularly extreme in their properties.

In this subsection, we explore the ionization properties of the C3 sample using different diagnostic diagrams proposed by \citet{Nakajima_2018}, which are presented in Fig. \ref{diagC} and \ref{diagO}. These diagnostic diagrams consider equivalent widths and line ratios of UV nebular lines to classify star-forming and AGN-dominated galaxies, respectively. They are based on the prediction of photoionization models  showing that UV lines are sensitive to the shape of the incident radiation field and can be used to distinguish the nature of the dominant ionizing source --pure star formation or AGNs (cyan and red circles in Fig. \ref{diagO}, respectively). We include also the results from synthetic emission lines from \cite{Hirschmann2019} that take into account composite galaxies classified by BPT diagram \citep{Baldwin1981} for the limits in the diagnostic diagrams and the criteria for defining a composite galaxy depends on the ratio of black hole accretion rate and star-formation rate.  

In Fig. \ref{diagC}, we present the diagrams EW(CIII]) versus CIII]/HeII, EW(CIV) versus CIV/HeII, and EW(OIII]) versus OIII]/HeII. Our results are color-coded by the physical parameter used for stacking and with symbols representing each type of stack (see Fig. ~\ref{diagC}). As a reference for comparison, we include in Fig. \ref{diagC},\ref{diagO} similar results for stacks of CIII-emitters from the VUDS survey \citep{Lefevre2015}. The black dashed lines are the limits between star-forming galaxies (on the right of the lines) and AGNs (on the left of the lines). The gray shaded region in the first diagram is where the models overlap and the classification may be ambiguous. In stacks where one of the lines in the diagnostic ratios is not detected at 2$\sigma$, 2$\sigma$-limits are taken into account. Instead, if two lines involved in a given diagnostic are undetected, then the stack is not considered. 

Overall, the main result from Fig. \ref{diagC} is that all the stacks explored in {our VANDELS sample} lay within the region dominated by ionization driven by star-formation. In the upper panel, stacks with lower stellar mass and fainter broad-band luminosity tend to show higher EW(CIII]) and higher CIII]/HeII ratios. Similar results are found in the middle panel of Fig. \ref{diagC} which shows similar trend with EW(CIV) and CIV/HeII. Finally, the bottom panel of Fig. \ref{diagC} shows the EW(OIII]) as a function of the OIII]/HeII ratio. Similar trends are found but with few stacks closer to the demarcation lines. Besides being consistent with the region dominated by star-formation, all the stacks lay in the region of composite galaxies according to \cite{Hirschmann2019} in all diagnostic diagrams by EW.  

{We note that our stacks with stronger CIII] emission have consistent line ratios to the stack of  CIII] emitters of similar EW in VUDS  ($5<$EW(CIII])$<10$\r{A}, the smaller green rectangle in the top panel in Fig. \ref{diagC}) presented in \cite{LeFevre_2019}. Bigger green symbols show VUDS galaxies with EW(CIII])$>10$\r{A}, which are closer to the demarcation lines. Therefore, our stacks explore a region in the emission line parameter space of photoionization models that is poorly explored in previous surveys and is occupied by the weak CIII] emitters.}

In order to complement the above diagnostics with EW, we also explore diagnostic diagrams using only UV emission-line flux ratios. We compare them with the \citet{Feltre_2016} and \citet{Gutkin_2016} photoionization models to probe the dominant ionizing source in our C3 sample. For the OIII] fluxes, we consider the sum of the OIII]$\lambda$1661\r{A} and OIII]$\lambda$1666\r{A} lines by adopting a theoretical ratio of OIII]1661\r{A}/OIII]1666\r{A}=0.41 from \cite{Gutkin_2016} for a $\log U=-2$.   

In the panels of Fig. \ref{diagO} from top to bottom, we present  CIV/CIII] as a function of (CIV+CIII])/HeII ratio, and CIV/CIII], CIV/HeII and CIII]/HeII as a function of OIII]/HeII, respectively. Overall, the C3 stacks are fully consistent with diagnostics shown in Fig. \ref{diagC}, and point to pure star formation as the dominant source of ionization. {Moreover, given that the galaxies in the C3 sample are selected to be star-forming and X-ray sources were excluded, these results lead to constrain photoionization models at $z\sim 3$.}  

A few stacks in these diagnostic diagrams lie close to the demarcation lines. As shown in previous studies (\cite{Feltre_2016, Gutkin_2016,Nakajima_2018}) these diagnostics may have some overlapping region in which both star-forming and AGN driven ionization may coexist. Only a few points lie at low OIII]/HeII and low CIII]/HeII (or CIV/HeII), a region where some contribution from AGNs might be expected according to models. Also Fig. \ref{diagO} shows that the less massive galaxies tend to have higher OIII]/HeII ratios. In the bottom panel of Fig. \ref{diagO}, demarcation lines from \cite{Hirschmann2019} are shown. The stacks lay consistently in the region of composite galaxies. 

As shown in \cite{Hirschmann2019}, the demarcation lines between composite and SFGs are clearer at $z=0-1$. At higher redshifts the models overlap in this zone. This could be an effect of the evolution of the criteria for defining a composite galaxy with redshift, or the change of the demarcation lines in the BPT diagram depending on the ionization parameter, electron density or extreme UV ionization field \citep{Kewley2019} which are likely to change due to the evolution of mass-metallicity relation with redshift. The higher electron temperatures at low metallicity can enhance the strengths of collisionally excited lines. It is also possible to find some pure SFGs with higher C/O in the composite region,  because the demarcation lines are based on fixed C/O at a given metallicity.
 
\subsection{An apparent EW(Ly\texorpdfstring{$\alpha$}{a})-EW(CIII]) correlation}\label{subsec:corr}

Previous studies have reported a positive correlation between EW(Ly$\alpha$) and EW(CIII]) \citep[e.g.][]{Shapley_2003,Stark2014,LeFevre_2019,Cullen_2020}. Such  relation is potentially useful for using CIII] to identify galaxies in the epoch of reionization where Ly$\alpha$ is strongly attenuated by the IGM. 

\begin{figure}[t]
    \centering
    \includegraphics[width=\columnwidth]{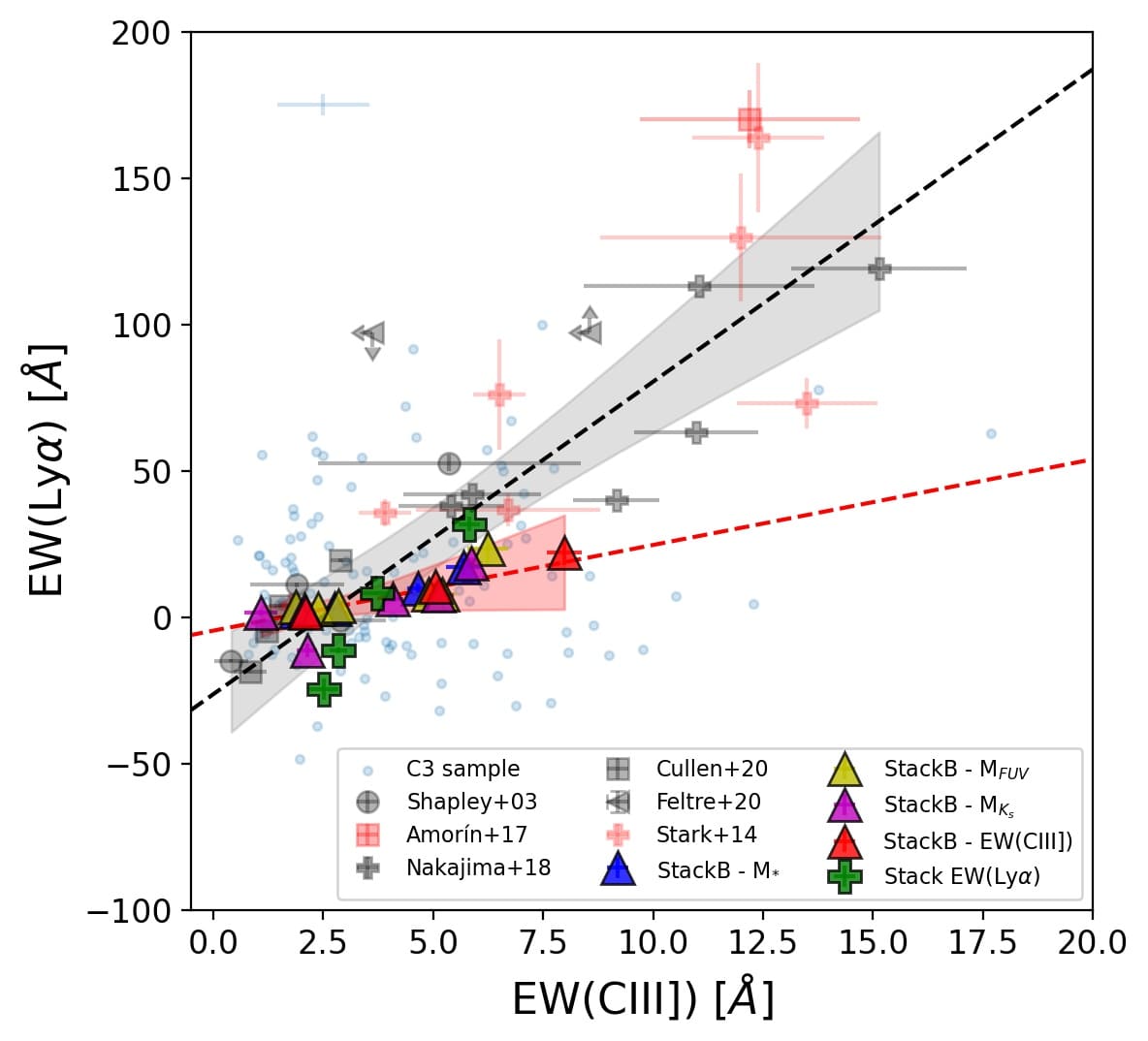}
    \caption{$EW(Ly\alpha)$-EW(CIII]) relation. The results from stacks B are represented by the triangle symbols, with different color depending on the parameter used for stacking: blue for stellar mass, yellow for FUV luminosity, magenta for K$_s$ luminosity, red for EW(CIII]), and green for EW(Ly$\alpha$). Individual galaxies with measured Ly$\alpha$ from the C3 sample are the small blue circles with their typical errors on the upper left.  Previous results from literature at similar redshift are also displayed (Stacking: \cite{Shapley_2003,Amorin_2017,meanNakajima_2018,Cullen_2020,Feltre2020}. Individual objects: \cite{Stark2014}). The dashed black line is the best fit in Eq. \ref{EWeq} including our stacks and the sample from literature. The red dashed line is the best fit in Eq. \ref{EWeqred} including only our stacks.}
    \label{corrEW}
\end{figure}

In Figure~\ref{corrEW}, we explore the possible correlation between EW(Ly$\alpha$) and EW(CIII]) for our C3 sample. We consider spectra from Stack B by stellar mass, FUV, K$_s$ luminosity, EW(CIII]), and EW(Ly$\alpha$). We also include data from literature, both stacks \citep{Shapley_2003,Amorin_2017,meanNakajima_2018,Cullen_2020,Feltre2020} and individual galaxies at similar redshifts \citep{Stark2014}, for which the existence of such correlation has been confirmed. We fit a linear regression to all the above data, which gives
\begin{equation}
    EW(Ly\alpha)=(10.67 \pm 0.89) \times EW(CIII]) - (26.38\pm 5.72).
    \label{EWeq}
\end{equation}

{The linear fit is performed using \textsc{LMFIT} \citep{LMFIT} with a least-squared method weighted by 1/$\sigma$, where $\sigma$ is the uncertainty in the parameter used for fitting. We use the same method for linear fitting in the following sections. When the linear fit is displayed in figures, the shaded region is the 3-$\sigma$ uncertainty band of the fitting and it covers the range where the fit is performed and valid.}

The relation in Eq. \ref{EWeq} fits the data with a relatively large scatter, which is shown at 3$\sigma$ level by the grey band in Fig. \ref{corrEW}. If we only include our stacks in the linear fit, we find that the best fit is
\begin{equation}
    EW(Ly\alpha)=(2.92 \pm 0.85) \times EW(CIII]) - (4.65\pm 2.36),
    \label{EWeqred}
\end{equation}
which has a lower slope. {The shallower relation can be an effect for the low EW(CIII]) of our stacks and because most of the literature sample is selected by $Ly\alpha$ or by strong CIII] emission. If we compare our stacks built by EW(Ly$\alpha$) (green symbols in Fig. \ref{corrEW}), we find that they are in better agreement with Eq. \ref{EWeq}.}

Our VANDELS data allow us to probe the low EW end of this relation. Fig.~\ref{corrEW} shows that there is a clear trend which seems to hold even in the absence of very strong CIII] emitters in our stacks. 
We caution, however, that the functional form in Eq.~\ref{EWeq} and Eq.~\ref{EWeqred} is representative for the average population of star-forming galaxies at $z\sim2-4$ and that strong deviations for individual galaxies can exist, especially at low EW. This is illustrated in Fig. \ref{corrEW} where we show the distribution of individual galaxies in the C3 sample. We note that these objects are not included in the linear fit in Eq.~\ref{EWeq} and Eq.~\ref{EWeqred}. The scatter shown by the individual galaxies is larger than the typical uncertainties. A similar diagram was presented in \cite{Marchi_2019} for few individual VANDELS sources. 

Based on Cloudy photoionization models, \cite{Jaskot2016} show that at a given EW(Ly$\alpha$), the scatter in EW(CIII]) can be as high as 10-20\r{A}, comparable to the observed level of scatter among galaxies, depending on the different metallicities, ionization parameters, and ages considered for the models. In general, higher EW(CIII]) for a given EW(Ly$\alpha$) indicates higher ionization parameter, younger ages and lower stellar metallicity. In a future work, we will address this point by using our large sample of CIII] emitters to constrain different photoionization models with their individual measurements.  

Overall, the correlation found in Fig.~\ref{corrEW} suggests that CIII] emitters are good markers of LAEs, especially for galaxies with low stellar mass, low luminosity, high star formation rates. This confirms the potential use of CIII] to identify and study galaxies at the epoch of reionization, for which Ly$\alpha$ emission is strongly attenuated due to IGM opacity. However, this could be  challenging due to the lower EWs of CIII] compared to that of Ly$\alpha$ in SFGs. A similar stacking approach will be useful with large enough samples at $z>6$ for studying their global properties, but CIII] may be the only robust and high S/N emission that may be observed to have individual detections. 

\subsection{A relation between stellar metallicity and the CIII] and   Ly\texorpdfstring{$\alpha$}{a} equivalent widths}\label{subsec:metall}

\begin{figure}[t]
    \centering
    \includegraphics[width=\columnwidth]{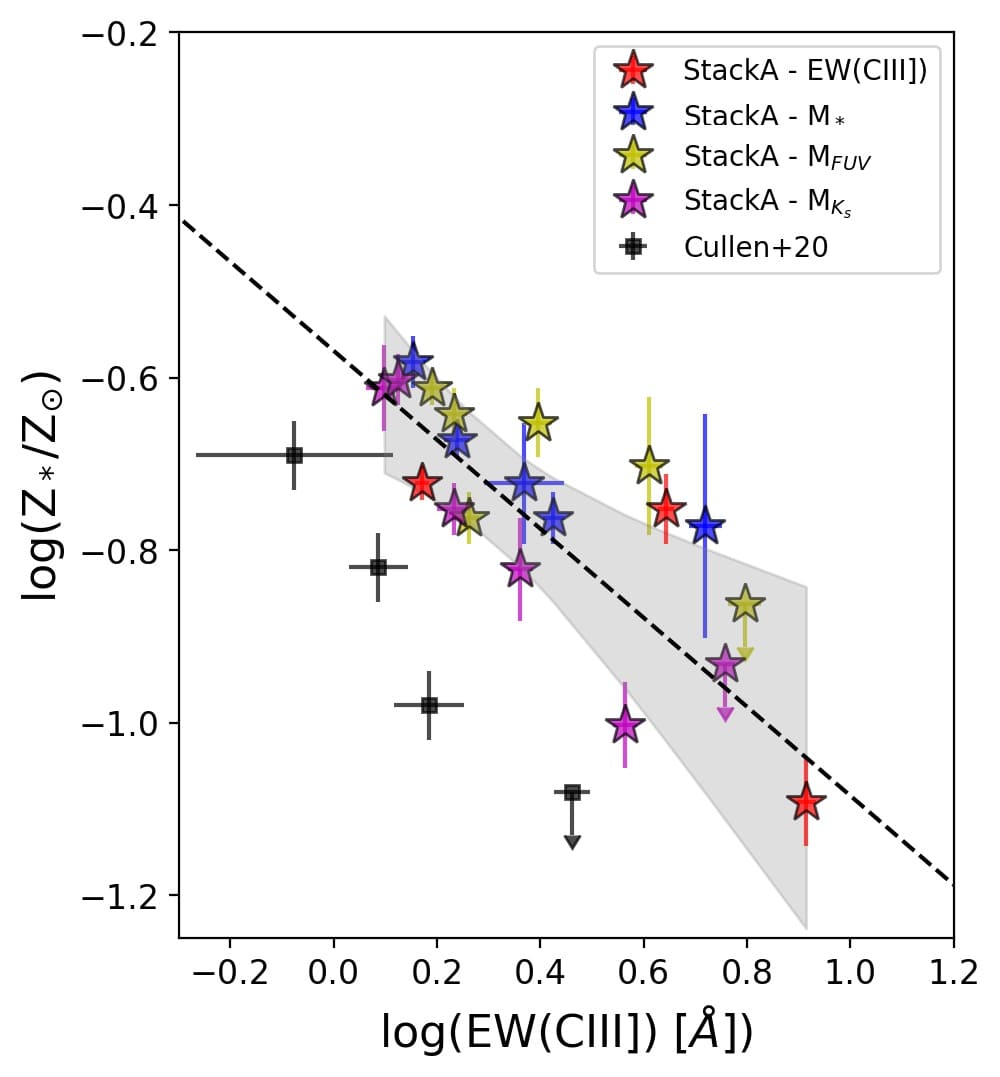}
    \caption{EW(CIII])-stellar metallicity relation. The results from our stacks A are represented by star symbols. Colors for our stacks are the same as in Fig. \ref{corrEW}. Black dashed line is the best fit in Eq. \ref{ZEWc3eq} and the gray shaded region is the 3-$\sigma$, band uncertainty.}
    \label{corrZEW}
\end{figure}

The rest-frame UV spectrum is dominated by the continuum light from young, massive stars that contains features of the chemistry of stellar photospheres and expanding stellar winds. The strength of these features have a strong dependence on the total photospheric metallicity. The full UV-spectrum fitting uses the strength of these faint photospheric features to estimate the stellar metallicity.  

We study the stellar metallicity (tracing the Fe/H abundance) of the C3 sample following two complementary approaches. First, we follow the full spectrum fitting method described in \cite{Cullen_2019}. In short, the high S/N stacked spectra are fitted using a Bayesian approach with Starburst99 (SB99) high-resolution WM-Basic theoretical models with constant star-formation rates \citep{Leitherer2014}.   
As a result, for some stacked spectra the stellar metallicity is an upper limit because the model parameter space does not extend below Z$_{\star}$=0.001 (0.07 Z$_{\odot}$). In these cases, a 2-$\sigma$ upper limit is reported.  

A second set of stellar metallicity estimates is performed using the method presented in \cite{Calabro2020}. This method is based on stellar photospheric absorption features at 1501\r{A} and 1719\r{A}, which are calibrated with SB99 models and are largely unaffected by stellar age, dust, IMF, nebular continuum or interstellar absorption. 
These estimations were only possible in stacks A and B, which are the ones with the highest S/N ($\sim$20-30). Using this method, we find consistent stellar metallicity values with those obtained with the full spectral fitting. However, the stellar metallicities based on the two photospheric indices show larger uncertainties (up to $\sim 0.6$ dex). Hereafter, we use the results from the first approach. 

The stellar metallicity of the C3 sample ranges from $\log(Z_{*}/Z_{\odot})=-1.09$ ($\sim$8\% solar) to -0.38 ($\sim$40\% solar), with a mean value of $\log(Z_{*}/Z_{\odot})=-0.8$ ($\sim$16\% solar). All stellar metallicities for stacks are reported in Tables \ref{tablemassbin}, \ref{tableKsmagbin}, \ref{tableFUVmagbin}, \ref{tableEWLyabin}, and \ref{tableC3bin}.

We explore the relation of the stellar metallicity with the EW(CIII]), that is shown in Fig. \ref{corrZEW}. For this, we use the spectra from Stack A, which include the entire C3 sample of galaxies, irrespective of redshift. We include the stacks by stellar mass, FUV luminosity, K$_s$ luminosity, and EW(CIII]) for the fitting. The best linear fit to data is  
    
\begin{equation}
    \log(Z_{*}/Z_{\odot})=(-0.51 \pm 0.08) \times \log(EW(CIII])) - (0.57 \pm 0.03),
    \label{ZEWc3eq}
\end{equation}
with EW(CIII]) in \r{A}. We find a decrease of EW(CIII]) with stellar metallicity. Comparing these results with \cite{Cullen_2020}, we find an offset towards higher EW(CIII]) for a given stellar metallicity. {We find three  reasons for such offset.  Firstly, the two samples only overlap in a narrow redshift range, as the  \cite{Cullen_2020} sample include galaxies between $z=3$ and $z=5$, thus excluding galaxies at $z<3$ which are numerous in our sample. Secondly, the selection criteria for our sample is based on CIII], leading to stacks with higher EW(CIII]). Finally, the use of accurate systemic redshifts based on CIII] line profile fitting lead to stacks with higher EW(CIII]) than stacks for the entire parent sample using the VANDELS spectroscopic redshift. We find the latter can explain up to a difference of $\log$(EW(CIII]))$\sim 0.2$ dex in the stacks of lower stellar mass and luminosity. }

\begin{figure}[t]
    \centering
    \includegraphics[width=\columnwidth]{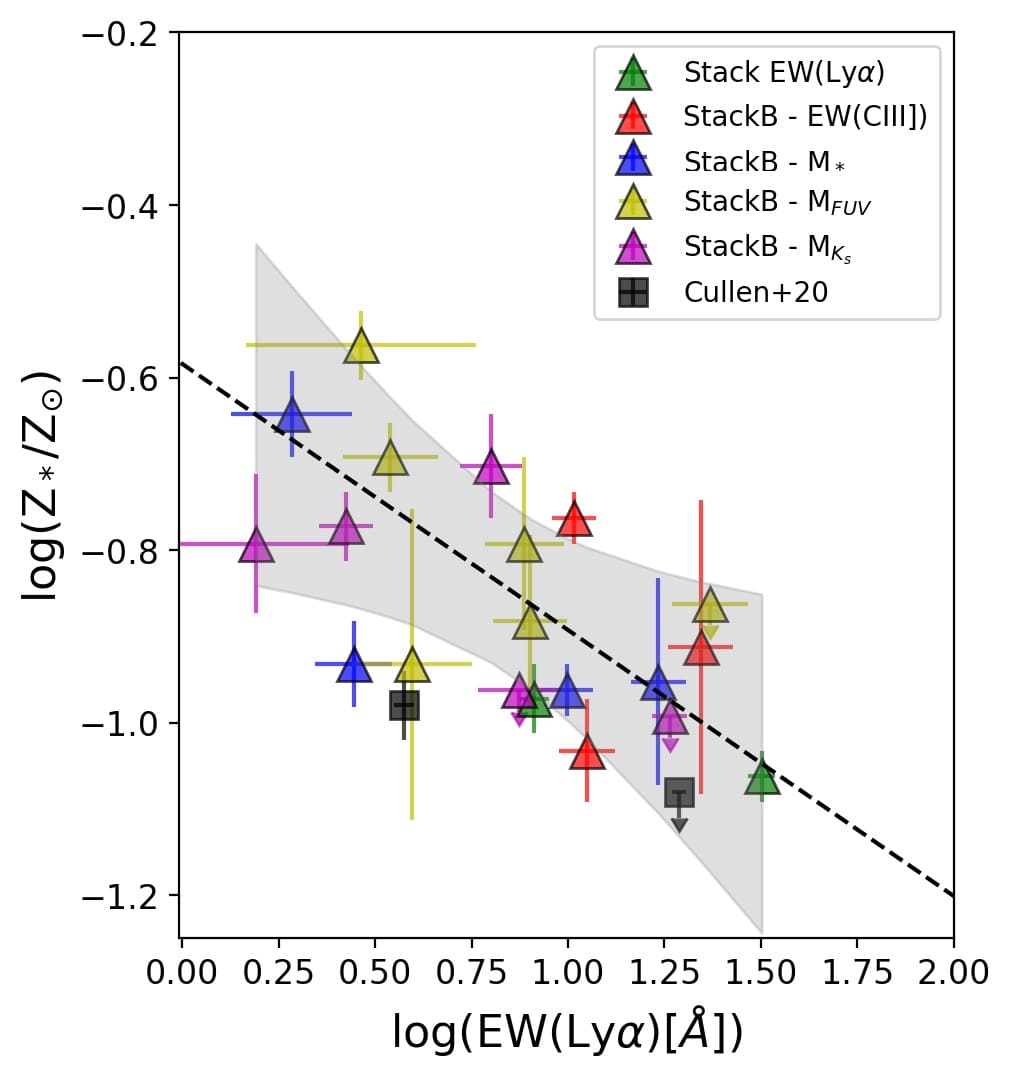}
    \caption{EW(Ly$\alpha$)-stellar metallicity relation. The symbols and colors for our stacks are the same as in Fig. \ref{corrEW}. Black dashed line is the best fit to the stacked data (triangles) presented in Eq. \ref{ZEWLyaeqlog}, with the gray shaded regions at 3-$\sigma$, respectively. The black symbols are the stacks in \cite{Cullen_2020} with Ly$\alpha$ in emission.}
    \label{corrZEWlyalog}
\end{figure}

In Figure~\ref{corrZEWlyalog}, we show stellar metallicities as a function of EW(Ly$\alpha$) for the set of spectra from Stack B using different colours for stacks by stellar mass, EW(CIII]), and FUV, K$_s$ luminosity, and EW(Ly$\alpha$). We perform a linear fitting to only our stack data, excluding two bins for which the EW(Ly$\alpha$) is negative (i.e. Ly$\alpha$ is in absorption). Our best fit is
\begin{equation}
    \log(Z_{*}/Z_{\odot}) = (-0.30 \pm 0.07)\times\log(EW(Ly\alpha))-(0.58\pm 0.06 )
    \label{ZEWLyaeqlog}
\end{equation}
with EW(Ly$\alpha$) in \r{A}. We find a decrease of EW(Ly$\alpha$) with stellar metallicity. The relation in Eq. \ref{ZEWLyaeqlog} provide larger metallicities at fixed EW(Ly$\alpha$) compared to the one presented by \cite{Cullen_2020} {which is based on stacking of galaxies at $z>$\,3 and include spectra with Ly$\alpha$ in absorption. In our fit, instead, we only consider stacks with Ly$\alpha$ in emission.
We note, however, that despite the difference in redshift, the two stacks from \cite{Cullen_2020} (black rectangles in Fig. \ref{corrZEWlyalog} and built by binning in  EW(Ly$\alpha$)) with Ly$\alpha$ in emission show a trend that is are  consistent with our stacks based on EW(Ly$\alpha$) (green triangles).}

Overall, the relation found in Fig.~\ref{corrZEWlyalog} confirms and adds robustness to the anticorrelation found for VANDELS galaxies out to $z \sim$5 in \cite{Cullen_2020}. We demonstrate in Fig. \ref{corrZEW} and  Fig.~\ref{corrZEWlyalog} that galaxies with stronger CIII] emission show larger Ly$\alpha$ EW and lower stellar metallicities of $\lesssim$10\% solar.
In correlations involving Ly$\alpha$, it is important to note that the Ly$\alpha$ emission is resonantly scattered and the correlations are not easily interpreted as for the CIII] or other nebular emission lines. As shown in \cite{Cullen_2020}, Eq. \ref{ZEWLyaeqlog} indicates that harder ionizing continuum spectra emitted by low metallicity stellar populations plays a role in modulating the Ly$\alpha$ emission in star-forming galaxies.

\subsection{C/O ratio and its relation with EW and physical parameters}\label{subsec:co}

\begin{figure}[t]
    \centering
    \includegraphics[width=0.95\linewidth]{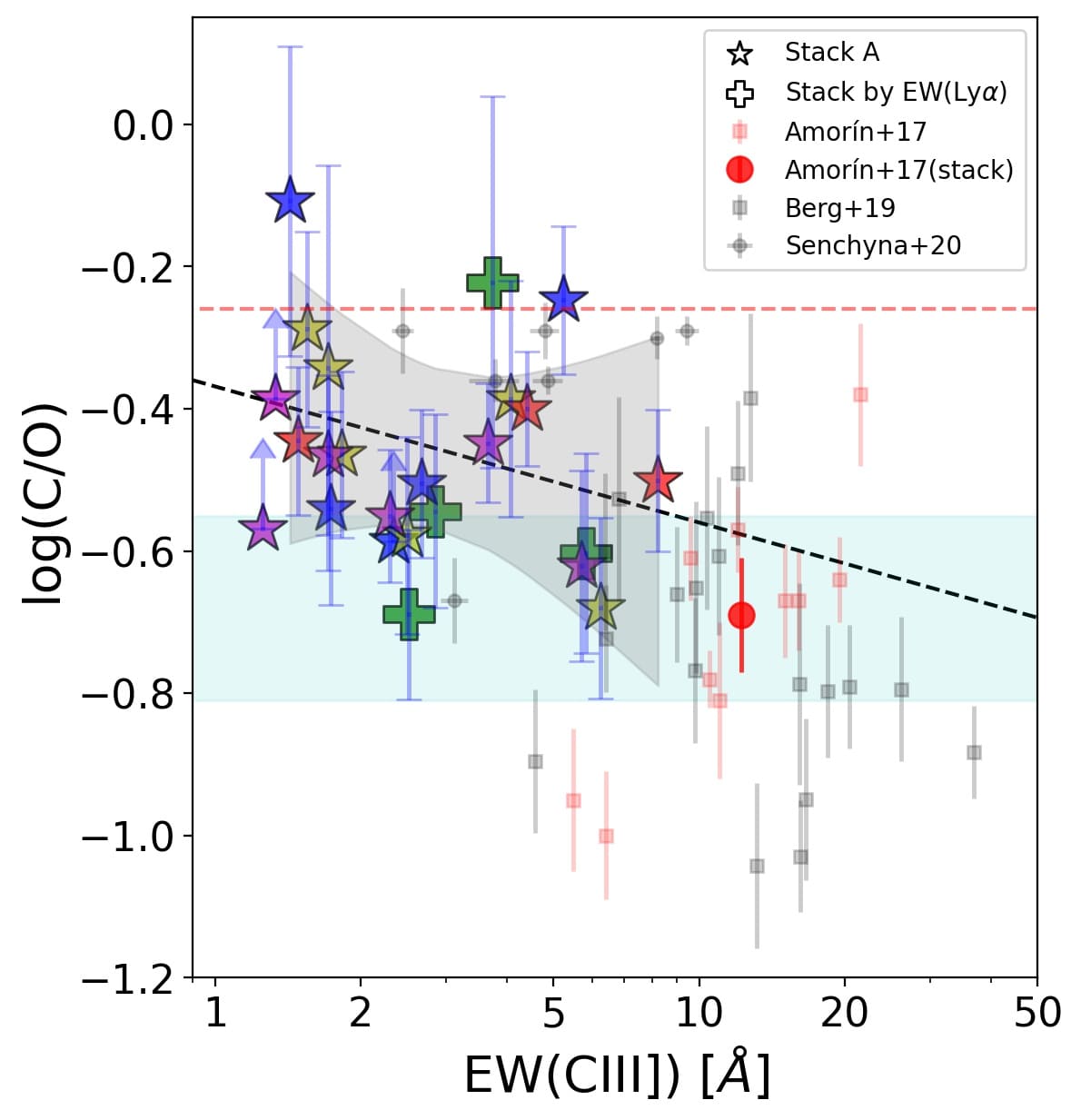}
    \caption{Relation between the EW(CIII]) and C/O ratio as derived using \textsc{HCm-UV}. The red dashed line corresponds to the solar value. The cyan shaded region is the average C/O for Lyman-break galaxies in \cite{Shapley_2003} and the red circle is the result for a composite spectrum of CIII]-emitters in \cite{Amorin_2017}. 
    Individual CIII] emitters at high$-z$ \citep[(red squares),][]{Amorin_2017} and low$-z$ \citep[(black squares and circles),][]{Berg_2019,Senchyna2020} are also included. 
    Colors for our stacks are the same as in Fig. \ref{corrEW}. The black dashed line is the best linear fit to our stacks, as displayed in Eq. \ref{COc3}.}
    \label{C/O_EW}
\end{figure}

The relative abundances of carbon, nitrogen and other alpha elements to oxygen may provide insight not only on the origin of carbon in  galaxies but also in their chemical evolution. However, constraining the C/O abundance is often difficult. For local galaxies, the emission lines often used to derive C/O are exceedingly faint carbon recombination lines \citep[e.g][]{Esteban2014} or the CIII] collisionally excited line, which is accessible for low-metallicity objects only from space \citep[e.g.][]{Senchyna2017,Berg_2019}. At $z\gtrsim$\,1, the required emission lines lie in the optical range but even for low-metallicity objects their faintness require very deep observations or stacking  \citep[e.g.][]{Shapley_2003,Amorin_2017}.  
While this makes the C/O difficult to constrain, this abundance ratio is essential to understand different emission-line diagnostics \citep[e.g.][]{Feltre_2016,Jaskot2016,Nakajima_2018,byler2018} and more generally the origin of carbon and the chemical evolution of star-forming galaxies. Here we explore the C/O ratio, which can be derived from the observed C and O lines in the UV.

We estimate the C/O abundance using the code HII-CHI-mistry in its version for the UV \citep{Perez-Montero_2017} (hereafter \textsc{HCm-UV}\footnote{We used the version 3.2 publicly available at \url{https://www.iaa.csic.es/~epm/HII-CHI-mistry-UV.html}}) considering PopStar stellar atmospheres \citep{Molla2009}, for the photoionization models used by the code. This \textsc{python} code derives the carbon-to-oxygen ratio (i.e. log(C/O)) from a set of observed UV emission-line intensities, which are also used to estimate ionization parameter and gas metallicity in a consistent framework with results provided by the direct Te-method. More details on this methodology can be found in \citet{Perez-Montero_2017}. For C/O, we use as an input the CIV, CIII], OIII] fluxes (and their errors) which are reported in Tables ~\ref{tablemassbin}, ~\ref{tableKsmagbin} and  ~\ref{tableFUVmagbin}, after extinction correction ({as explained in Sec. \ref{sec:linemeas}}). In most of the stacks, the OIII]$\lambda$1660\r{A} is not detected at 3$\sigma$. For this reason, we consider the theoretical ratio of  OIII]$\lambda$1660\r{A}/OIII]$\lambda$1666\r{A}$\sim$0.4 from photoionization models with an ionization parameter of -3 and -2 \citep{Gutkin_2016}. In the code, 25 Monte Carlo iterations are performed to estimate the uncertainties in the C/O calculations. In the cases where OIII]$\lambda$1666\r{A} is detected with a $<$2$\sigma$ level or when the C/O uncertainties are larger than 0.9 dex, C/O is estimated as a lower limit. Results are reported in Tables \ref{tablemassbin}, \ref{tableKsmagbin}, \ref{tableFUVmagbin}, \ref{tableEWLyabin}, and \ref{tableC3bin}. We find log(C/O) values ranging from -0.68 (38\% solar) to -0.06 ($\sim$150\% solar) with a mean value of -0.50 (60\% solar).

Alternatively to \textsc{HCm-UV}, we used the empirical calibration between C3O3 ($\equiv$log((CIII]+CIV)/OIII])) and C/O found by \cite{Perez-Montero_2017} using a control sample with C/O and metallicities obtained from UV and optical lines. Using this calibration {$\log($C/O$) = -1.07 + 0.80\times$C3O3}, which essentially provides an accurate fit to models predictions for the C3O3 index, we find consistent results, within the typical error of $\sim 0.2$ dex, with those of \textsc{HCm-UV}. Small differences can be attributed to slight changes in the ionization parameter \citep[see Fig. 2 in][]{Perez-Montero_2017}, which is constrained by \textsc{HCm-UV} using CIV and CIII]. All C/O results presented in subsequent analysis and figures are derived with  \textsc{HCm-UV} but they are fully consistent with the C3O3  calibration.

\begin{figure*}[t]
    \centering
    \includegraphics[width=0.9\linewidth]{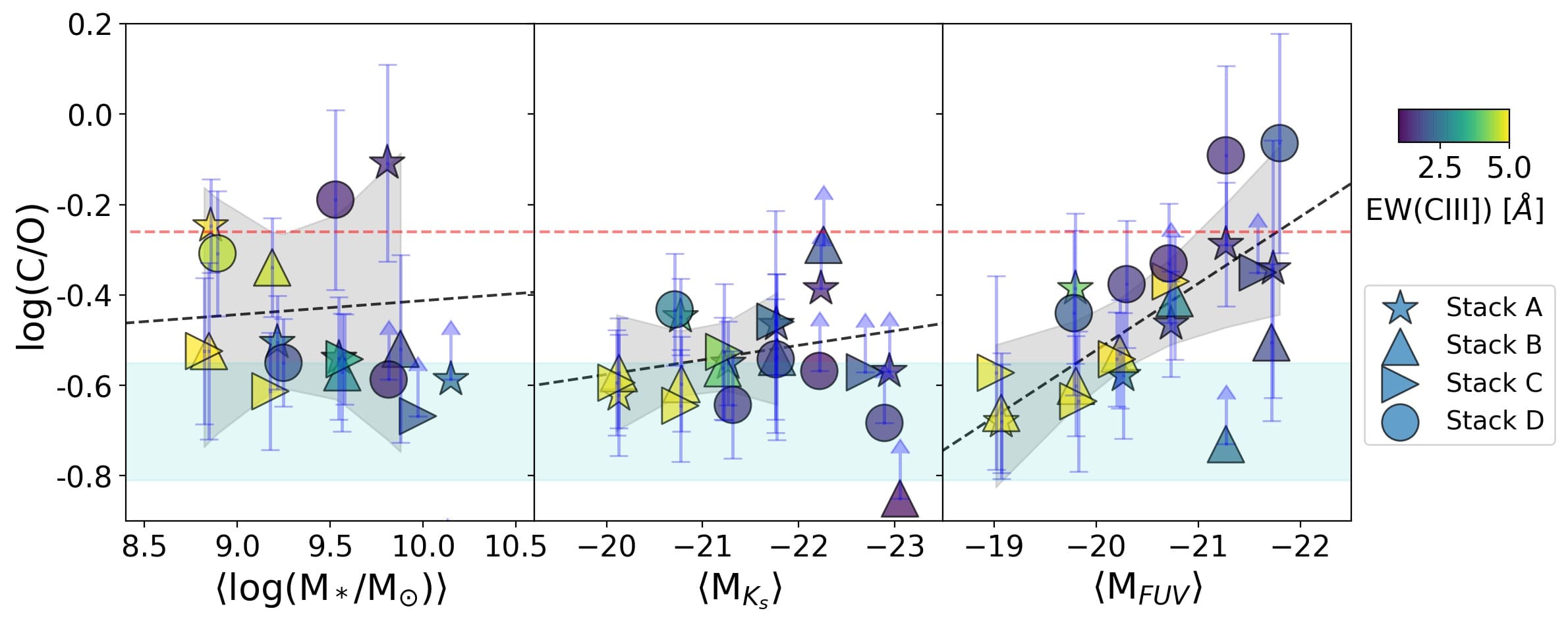}
    \caption{Relation between the different physical properties used for the stacks and C/O ratio using as derived using \textsc{HCm-UV}. The red dashed line corresponds to the solar value. The cyan shaded region is from \cite{Shapley_2003}. The results of the stacks are color coded by EW(CIII]). The black dashed line is the best linear fit for each parameter in Eq. \ref{COmass}, \ref{COKs}, and \ref{COFUV}, respectively. }
    \label{C/OHII}
\end{figure*}

We explore the relation of log(C/O) with EW(CIII]) in Fig. \ref{C/O_EW}. We only include the results from stack A by stellar mass, luminosities and EW(CIII]), and the stacks by EW(Ly$\alpha$). A linear regression to data gives 
\begin{equation}
    \log(C/O)=-(0.19\pm 0.14)\times\log(EW(CIII]))-(0.36\pm 0.07),
    \label{COc3}
\end{equation}
with a Pearson correlation coefficient $r_p=-0.30$. This relation suggests a decrease of C/O abundance ratio with EW(CIII]), i.e. the more extreme CIII] emitters tend to have lower log(C/O). We showed in Eq. \ref{ZEWc3eq} that strong CIII] emitters also have low stellar metallicities, which lead to  less cooling and higher nebular temperatures that enhance the CIII] emission. Therefore, Eq. \ref{COc3} suggests a change in stellar metallicity. The relation between C/O and stellar and gas  metallicities will be discussed in Section \ref{subsec:discCO}.

We also observe a weak relation ($r_p=-0.29$) between C/O with EW(Ly$\alpha$), where LAEs tend to have lower C/O, thus suggesting higher EW(CIII]) and lower metallicity, i.e. a younger chemical age. For EW(Ly$\alpha$)=20\r{A}, a log(C/O) $\sim-0.5$ is found ($\sim$60\% solar) and a corresponding EW(CIII])$\sim10$\r{A} (from Eq. \ref{COc3}). We stress, however,  that the relation between both C/O and EW(Ly$\alpha$) could not be physically motivated and it relies on previous correlations found between EWs in Eq. \ref{EWeqred}. 

On the other hand, in Figure~\ref{C/OHII} we present the results of C/O for the C3 sample as a function of the physical parameter used for stacking and color-coded by EW(CIII]). We find an apparent mild increase of C/O with stellar mass (left panel on Fig. \ref{C/OHII}). A linear fit to data after excluding lower limits gives 

\begin{equation}
   \log(C/O)=(0.03\pm 0.12)\times \log(M_{*}/M_{\odot})-(0.72\pm 1.20),
   \label{COmass}
\end{equation}

with $r_p=0.07$. This  relation is weak and of limited use due to the lack of reliable C/O estimations in the high mass end for which OIII] is barely detected in our stacks. This makes the dynamical range of stellar mass too small to provide a more robust relation. 

In the $K_s$ band (middle panel on Fig. \ref{C/OHII}), we perform a linear fitting excluding the lower limits, which gives 
\begin{equation}
   \log(C/O)=-(0.03\pm 0.03 )\times M_{K_s}-(1.21\pm 0.72),
   \label{COKs}
\end{equation}

with {$r_p=-0.37$. We find an increase of C/O with K$_s$ luminosity}, but again the relation is weak mostly due to the lower limits for high-luminosity stacks. 

On tthe other hand, the FUV luminosity (right panel of Fig. \ref{C/OHII}) has a stronger correlation with C/O. A linear regression excluding lower limits gives 
\begin{equation}
   \log(C/O)=-(0.14\pm 0.03)\times M_{FUV}-(3.47\pm 0.63),
   \label{COFUV}
\end{equation}
with {$r_p=-0.74$}. This correlation clearly shows an increase of C/O in galaxies with higher FUV luminosity and can be used to estimate a mean C/O value from a galaxy luminosity. {Assuming  the FUV luminosity is a tracer of  the recent SFR and that more evolved stellar populations may have a larger contribution in the C/O relation with stellar mass and K$_s$ luminosity, the above differences might be explained invoking a strong dependence of C/O with star formation histories. However, larger samples and deeper spectra, especially  for high mass galaxies,  would be needed to provide better insight.}

We note that our results suggest  that stacking by FUV luminosity  results  in a more homogeneous distribution of C/O within the bins, compared to the stellar mass and K$_s$ luminosity. In the short dynamical range for stellar mass ($\sim$1 dex where C/O was estimated) there is no  clear trend for this sample.

In summary, the C/O abundance of CIII] emitters increases from less than half solar for fainter (low-mass) galaxies to about solar abundance for our brighter (high-mass) objects. The stronger CIII] emitters, i.e. higher EWs, are found in low luminosity, low C/O galaxies.

\subsection{On the relation of SFR with EW(CIII]) and C/O}
\begin{figure}[t]
    \centering
    \includegraphics[width=0.9\columnwidth]{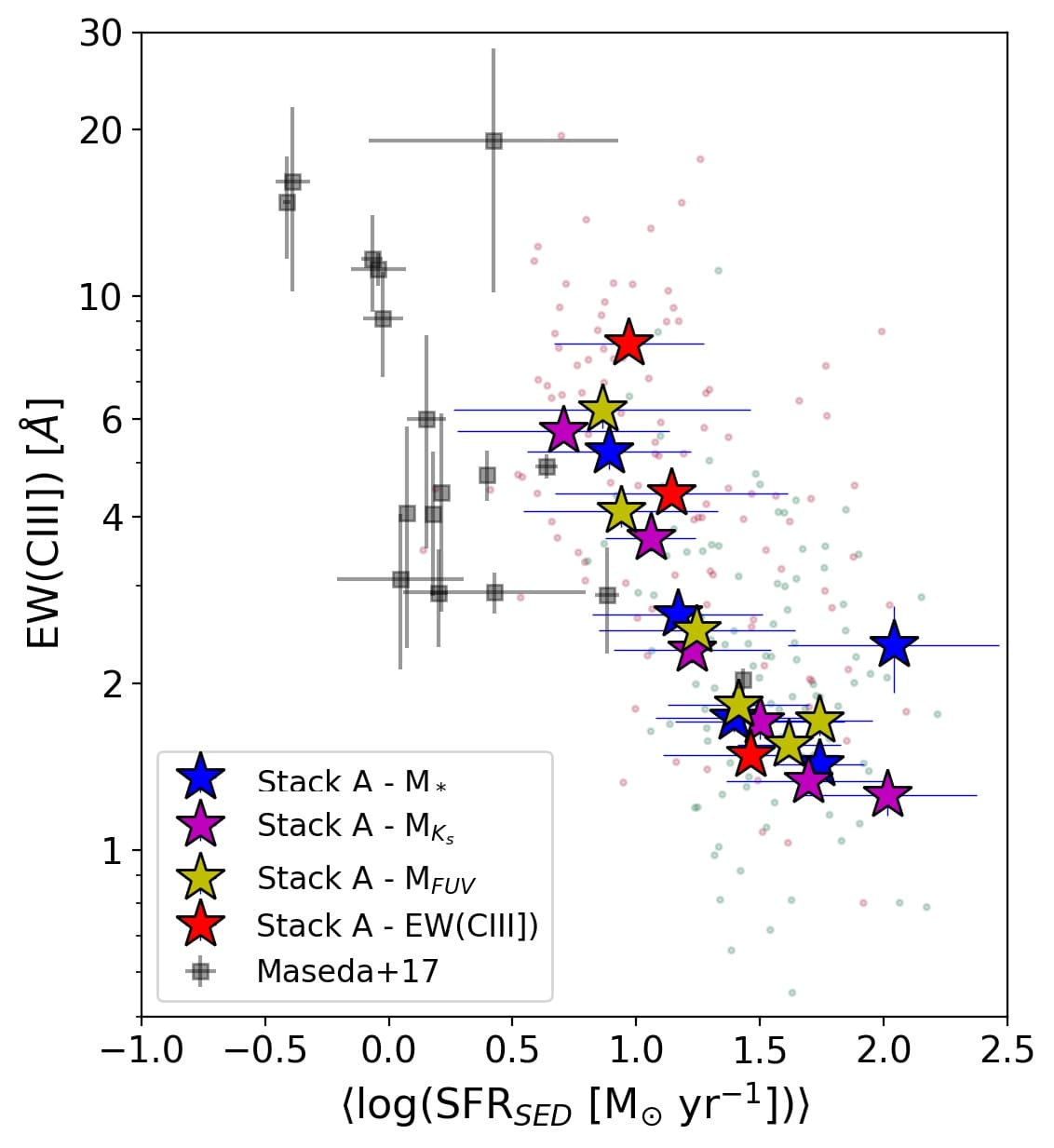}\\\includegraphics[width=0.9\columnwidth]{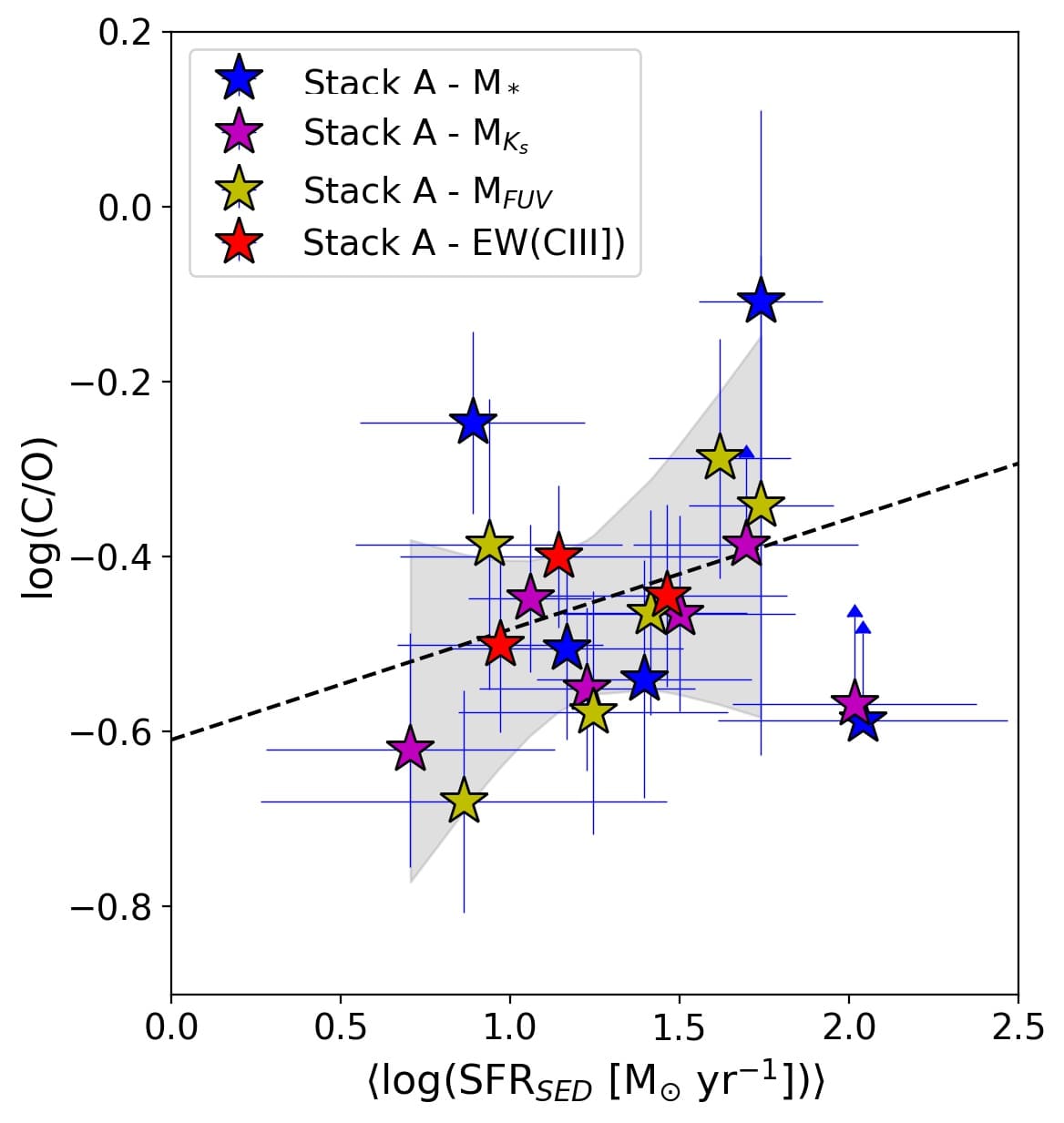}
    \caption{Relation between SFR-EW(CIII]) (\textit{top panel}) and SFR-C/O (\textit{bottom panel}). Stars symbols are Stack A color-coded depending on the parameter used for stacking, according to legend, as in Fig. \ref{corrEW}. \textit{On the top panel:} The black squares are results from CIII] emitters in \cite{Maseda2017}. The small dots are the C3 sample with same colors as in Fig. \ref{c3relation}. \textit{On the bottom panel:} The dashed black line is the best fit for the SFR-CO relation in Eq. \ref{eqSFR-CO}.}
    \label{SFR-relation}
\end{figure}

We explore the relation between EW(CIII]) and C/O abundance ratio with SFR. In both cases, we illustrate these findings using the sample of Stack A by stellar mass, FUV and $K_{s}$ luminosities, and EW(CIII]). We estimate the SFR with the mean value of the SED-based SFR of individual galaxies in each bin. In Fig. \ref{SFR-relation} we present these results. Overall, we find that EW(CIII]) decreases with SFR, as we also find for individual galaxies in the C3 sample. This is consistent with the trend observed in \cite{Maseda2017} for a smaller sample of strong CIII] emitters. Our stacks show that galaxies with SFR$\gtrsim 30$ M$_{\odot}$/yr have EW(CIII])$\lesssim$\,3 \r{A}. Instead, for galaxies with SFR$\lesssim 30$ M$_{\odot}$/yr, the relation becomes steeper and shows more dispersion on the EW(CIII]) values, with the more intense CIII] emitters having lower SFRs. Stacks with average SFR$\lesssim$\,10 M$_{\odot}$/yr show all  EW(CIII]) larger than $\sim$ 3 \r{A}. We note that the result in Fig. \ref{SFR-relation} does not imply that higher SFR tends to suppress CIII] emission, as this relation hide an underlying metallicity dependence that is key for interpreting the CIII] emission (see Section \ref{subsec:disMZR} for a discussion on stellar mass-metallicity relation). Higher SFR implies a larger number of ionizing photons, which tend to enhance CIII] emission. However, since our sample lie in the star-forming main sequence, galaxies with higher SFR have higher stellar mass and also higher metallicity, which imply more efficient cooling and weaker collisionally-excited lines such as CIII].

Finally, we do not find a correlation of EW(CIII]) with specific SFR (sSFR$=$SFR/M$_{*}$). For the stacks shown in Fig. \ref{SFR-relation} and based on the mean values of stellar mass and SFR from the SED fitting, we find values for $\log$(sSFR [yr$^{-1}$]) ranging between $-8.3$ and $-7.9$ with no clear correlation with EW(CIII]). Note, however, that the dynamical range in sSFR probed by the C3 sample appears too small  compared with the uncertainties to probe  possible correlations with other observables.

On the other hand, Fig. \ref{SFR-relation}  shows a trend between SFR and C/O abundance. The C/O tends to increase with the average SFR. This relation shows a large scatter and appears mainly driven by the stacks by M$_{FUV}$. This result shows consistency with the relation found between C/O and M$_{FUV}$  in Fig.~\ref{C/OHII}, as M$_{FUV}$ is a good tracer of SFR. A linear fit to this relation, excluding upper limits, gives 
\begin{equation}
    \log(C/O)=(0.12\pm 0.11)\times\log(SFR_{SED})-(0.61\pm0.13), 
    \label{eqSFR-CO}
\end{equation}
where SFR is in M$_{\odot}$/yr ({$r_p=0.51$}).
Considering that our C3 sample is representative of main-sequence galaxies, this relation shows the limitation of Eq. \ref{COmass} because faint OIII] lines are not detected in stacks with high-mass galaxies. 

It is worth noting that in Fig. \ref{corrZEW}, \ref{corrZEWlyalog}, \ref{C/O_EW}, and \ref{SFR-relation}, different dynamic ranges in stellar metallicity and C/O are obtained depending on the parameter used for stacking. Overall, the trends found in these and other relations are mainly driven by the stacks binned by luminosity, which are more uniformly populated. In Appendix \ref{apen0}, we show that our results remain unchanged if we restrict the fitting to only those stacks binned by luminosity.

\section{Discussions}\label{sec:discussions}
\subsection{On the stellar mass-metallicity relation of CIII] emitters at \texorpdfstring{$z\sim3$}{z-3}}\label{subsec:disMZR}
\begin{figure*}[t]
    \centering
    \includegraphics[width=0.45\textwidth]{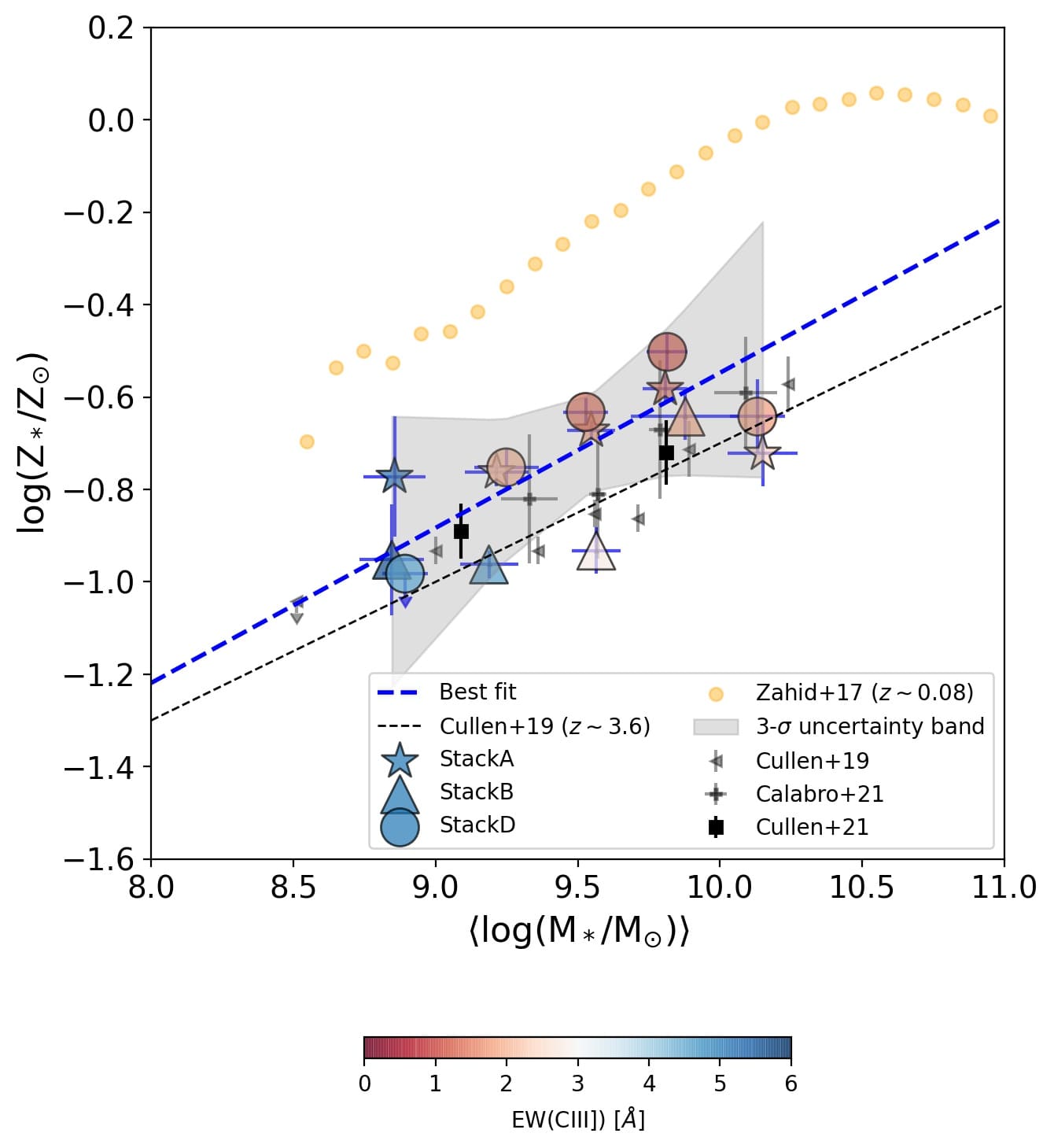}\,\includegraphics[width=0.45\textwidth]{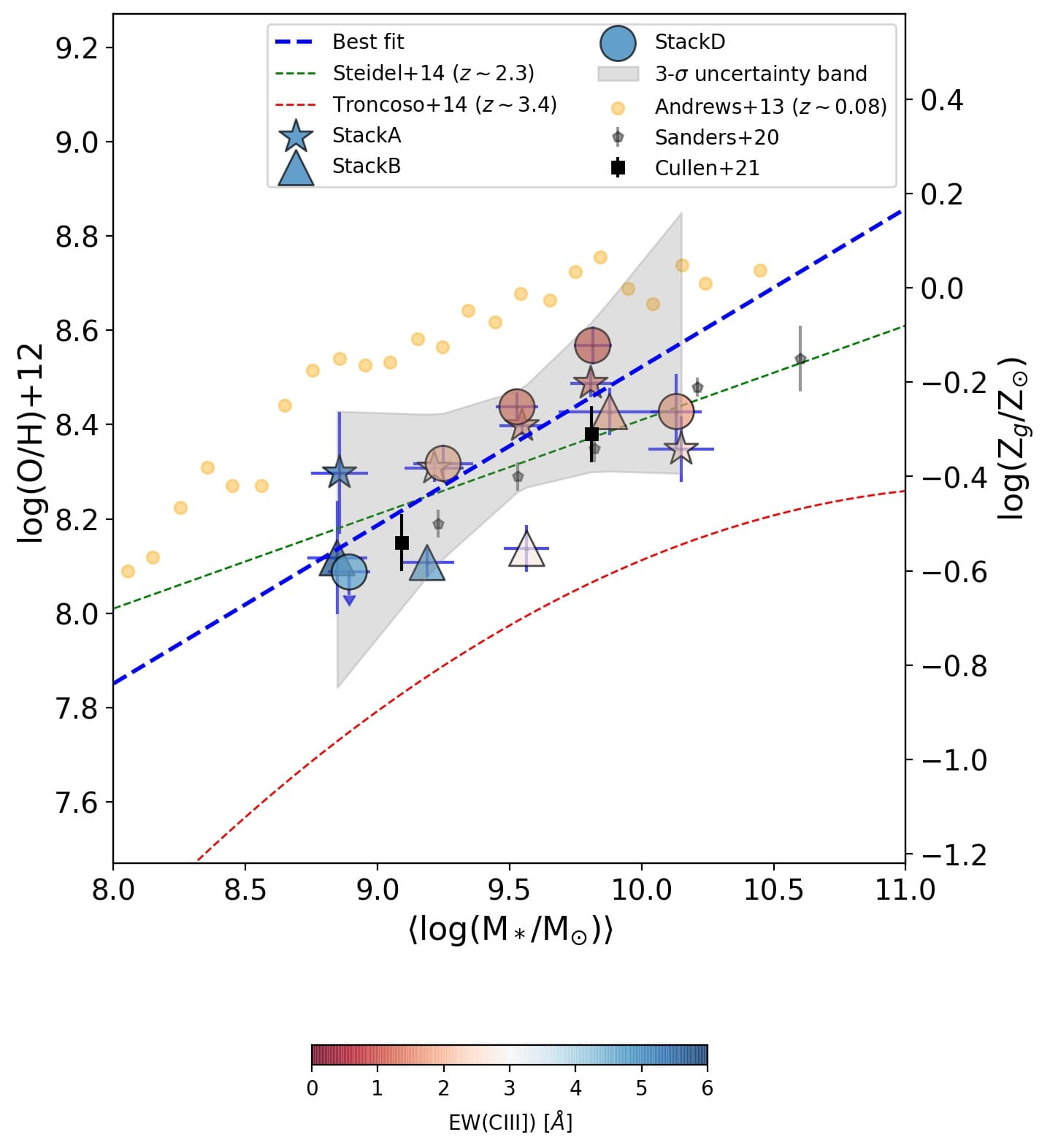}
    \caption{\textit{Left panel}: MZRs with the stacks A, B, D by stellar mass with star, triangle and circle symbols, respectively, and color-coded by EW(CIII]). The black symbols are stacks from \cite{Cullen_2019, Calabro2020,Cullen2021}.  The orange circles is the local MZRs from \cite{Zahid2017}. The blue dashed line is the best fit with the shaded region at 3-$\sigma$ in Eq. \ref{mzeq}, and the black dashed line is the MZRs from \cite{Cullen_2019}. \textit{Right panel}: MZRg with the same stacks as left panel but re-scaled with the $\alpha$-enhancement. Green dashed line from \cite{Steidel2014}. Red dashed line from \cite{Troncoso2014}. Black symbols are values from literature \citep{Sanders2020,Cullen2021}. The orange points are the local MZRg from \cite{Andrews2013}. The $Z_g$ scale in the y-axes corresponds to the gas-phase metallicity obtained from Eq. \ref{eq_alpha} (see text for more details). }
    \label{MZrelation}
\end{figure*}

Scaling relations, such as the stellar mass-metallicity relation (MZR), are important diagnostics to understand the evolution of galaxies. In particular, the MZR is shaped by different physical processes such as strong outflows produced by stellar feedback, infall of metal-poor gas,  the so-called stellar mass “downsizing” for which high-mass galaxies evolve more rapidly and at higher redshifts than low-mass ones, or by the shape of the high-mass end of the initial mass function (IMF) \citep[see][for a review]{Maiolino2019}. Thus, the MZR of a galaxy population in a determined redshift may provide clues on the dominant processes that affect its evolution in that period. 

The gas-phase MZR (hereafter MZRg) has been explored in the local universe, exploiting the methods based on optical emission lines applied to large samples of galaxies \citep[e.g.][]{Tremonti2004, Andrews2013}. In the same way, the MZRg has been explored at higher redshifts \citep[e.g.][]{Erb2006b,Mannucci2010,PM2013,Troncoso2014,Lian2015,Sanders2020} thus providing a clear evolutionary picture up to $z\sim 3.5$ \citep[e.g.][]{Maiolino2008,Mannucci2010}. The  MZRg is found to evolve with redshift, with metallicity declining with redshift at a given stellar mass.  In one of the latest works, \cite{Sanders2020} use $T_e$-consistent metallicity calibrations to derive O/H for star-forming galaxies out to $z \sim$\,3.3 and explore the MZRg evolution. They find similar slopes at all redshifts for $\log(M_{*}/M_{\odot})\lesssim$\,10 and a nearly constant offset of about 0.2-0.3 dex towards lower metallicities compared to local galaxies at a given stellar mass. After comparing with chemical evolution models, the authors argue that this is driven by both higher gas fraction (leading to stronger dilution of ISM metals) and higher metal removal efficiency, e.g. by feedback. 

On the other hand, studying the redshift evolution of the stellar MZR (hereafter MZRs) has been historically more challenging due to the required high S/N continuum spectra. 
In the local universe, first studies of the MZRs by  \citet{Gallazzi2005} and subsequent work  via stacking of SDSS optical spectra for statistical samples of galaxies  \citep{Zahid2017}, found stellar metallicity increasing over a large range of  $\log$(M$_{\star}$/M$_{\odot})\sim$\,9-11. 
At higher redshifts, however, the lack of high S/N optical spectra for statistical samples precludes similar analyses. Estimates of stellar metallicity are found, instead, from metallicity-sensitive indices or full spectral modelling of deep  rest-frame UV spectra sampling  young, massive stars \citep[e.g.][]{Sommariva2012,Cullen_2019,Calabro2020}. Therefore,  they provide $Z_{*}$ values expected to be similar to those derived for the ISM out of which young stars have recently formed. Recent studies have shown  an evolving MZRs  up to $z\sim 3$, decreasing $Z_{*}$ at fixed M$_{\star}$ by more than a factor of 2 from $z=0$ to $z=3.5$, similarly to what is found for the MZRg \citep[e.g.][]{Cullen_2019,Cullen_2020,Calabro2020}. 

Here, we explore the position of normal galaxies with detected CIII] emission in the MZR at $z\sim$\,2-4, while comparing with some of the above results from the literature. 
Using the results presented in previous sections, we probe the stellar mass-metallicity relation MZRs on the left panel in Fig. \ref{MZrelation}. In this figure, we use stacks A, B, and D computed by stellar mass. Our best linear fit to these points gives the following relation,

\begin{equation}
    \log(Z_{*}/Z_{\odot})=(0.33 \pm 0.10) \times [\log(M_*/M_{\odot})-10] - (0.54 \pm 0.05),
    \label{mzeq}
\end{equation}

which is represented by the blue dashed line in Fig.~\ref{MZrelation}. The MZRs of CIII] emitters at $2<z<4$ have a nearly constant offset of $\sim$ 0.4 dex compared with the local MZRs from \cite{Zahid2017} over more than one decade in $M_*$. 
Compared to other MZRs at similar redshift from \citet{Cullen_2019, Cullen_2020} and \citet{Calabro2020}, we find a relatively good agreement in slope and normalisation thus suggesting that CIII] emitters are not different from their parent sample of star-forming galaxies in the MZRs. While stacks A and D are slightly shifted to high $Z_*$, stacks B appear more consistent with the MZRs derived in the above previous VANDELS studies. These small differences may arise from the different selection criteria. In particular, redshift selection used in these works only include galaxies at $z>3$, while our stacks D include only galaxies at $z<$\,2.9 and stacks A include a mix of galaxies below and above $z=2.9$. {We also find that for a given stellar mass, the intense CIII] emitters tend to lie below the MZRs, while the faint emitters, tend to lie above the MZRs (see Fig. \ref{MZrelation}).} {On the other hand, the offset between Eq. \ref{mzeq} and the MZRs reported in \cite{Cullen_2020} is explained in part due to the difference in the redshift range covered by the samples and  differences in the assumptions leading to  stellar mass derivations. In \cite{Cullen_2020},  stellar masses are derived from SED fitting assuming solar metallicity and models that do not account for nebular emission. These assumptions lead to an offset of around $\sim0.2$ dex towards higher stellar masses compared with our updated  catalogue.}

If we assume that stellar and gas-phase metallicity are the same, our results are consistent with the MZRg derived from \cite{Troncoso2014} (at 1-$\sigma$) for the range of stellar mass covered by our stacks. However, this assumption is not necessarily true, especially at high-$z$.

\cite{Sommariva2012} found a small difference ($\sim0.16$dex)  between $Z_g$ and $Z_*$, but their conclusion was not robust given their large reported uncertainties. One might expect small differences because we are measuring stellar metallicity using UV absorption features driven  by massive stars with short lifetimes and similar properties of the interstellar gas where they were formed. However, larger differences can be found for galaxies whose ISM has been enriched primarily by core collapse supernovae with highly super-solar O/Fe, as discussed in \cite{Steidel2016,Topping2020b}. Such conclusion has been recently reached in a work by \cite{Cullen2021} where it was found that a subset of galaxies in VANDELS at $z\sim3.4$ are $\alpha$-enhanced (i.e., their O/Fe ratios are more than two times solar) from a direct comparison of their stellar and gas metallicities.

Studies of the MZRg using exclusively the rest-UV spectrum has strong limitations due to the lack of hydrogen lines besides Ly$\alpha$. While this line has been used in galaxies with extremely high EWs \citep{Amorin_2017}, we avoid the use of Ly$\alpha$ in our VANDELS sample because it is generally affected by resonant scattering and absorption by the IGM. An alternative method to constrain gas-phase metallicity is using nebular HeII instead of Ly$\alpha$. This possibility has been implemented, for instance, in \textsc{HCm-UV} (version 4) and in \cite{Byler2020} using a He2-O3C3 calibration. However, HeII is generally weak in most stacks and it may include both nebular and stellar origin, which being  difficult to disentangle is thus an additional source of error. 

We estimate gas-phase metallicities  for our stacks using the above two methods. When we compare them with the derived stellar metallicities we find a mean difference of $\log Z_{*} - \log Z_{g} \sim$\,0.16 dex for the stacks by stellar mass, FUV luminosity, and EW(CIII]). However, the dispersion is larger ($\Delta(\log Z_{*} - \log Z_{g}) \sim$\,0.25 dex) and since the uncertainties for the metallicities are also larger (up to $\sim 0.4$ dex), the above comparison does not provide a robust assessment of the true difference between stellar and gas-phase metallicities, which are proxies  of the Fe/H and O/H abundances, respectively. 

For this reason, we follow an alternative approach to estimate gas-phase metallicities in our stacks. Following \cite{Cullen2021}, we consider that 
\begin{equation}
    \log(Z_g/Z_{\odot})=\log(Z_g/Z_*)+\log(Z_*/Z_{\odot}), 
    \label{eq_alpha}
\end{equation}
where $\log(Z_g/Z_*)\sim$[O/Fe] is a proxy of the  $\alpha$-enhancement, which depends on stellar mass. Then, adopting the difference found by \cite{Cullen2021} for MZRs and MZRg, we infer that [O/Fe]$\sim$0.37-0.40 dex for the range of stellar masses of our sample. Thus, we use [O/Fe]$\sim$0.38, the value corresponding to the mean stellar mass, to convert our $Z_{\star}$ into $Z_g$ values. In the right panel of Fig. \ref{MZrelation} we show the MZRg obtained following the above approach. Despite our C3 sample is a subsample of the one used by \cite{Cullen2021}, the assumed [O/Fe] appears reasonable, as the C3 stacks shown in Fig. \ref{MZrelation} follow a consistent trend with the results found by \cite{Cullen2021} using a different stacking procedure.

In order to obtain an independent value for gas metallicity, we also probed  the Si3-O3C3 calibration presented by \cite{Byler2020}. For our stacks, the values obtained with the assumed [O/Fe] are consistent within 0.15dex with the gas-phase metallicities obtained using the Si3-O3C3 calibration. We note that the latter is found to have a median offset of 0.35dex when compared to other well-known metallicity calibrations based on optical indices. Acknowledging these differences and the relatively good agreement between these two methods, we choose to use the re-scaled values with the mean [O/Fe] in the following sections. Clearly, follow up studies probing bright optical lines of CIII] emitters are necessary to provide more reliable gas-phase  metallicity determinations.

On the right panel in Fig. \ref{MZrelation}, our best fit with the assumed mean [O/Fe] is fully consistent with the MZRg at $z\sim$\,2.3 from \cite{Steidel2014}. Our results are also consistent with most data points in the MZRg estimates obtained at $z\sim$\,3 from \cite{Onodera2016} and \cite{Sanders2020}, which are also included for comparison. Our results are thus consistent with the reported redshift evolution of the MZRg, illustrated here using a comparison with the local relation found by  \cite{Andrews2013}. The larger differences are found with respect to the MZRg found by \cite{Troncoso2014}, which could be explained by systematic differences in the excitation conditions and metallicities between the samples, as suggested in \cite{Sanders2020}. Indeed, the \cite{Troncoso2014} sample is likely to be biased towards lower metallicities. But the agreement of our relation with previous MZRg  depends on the $\alpha$-enhanced assumed in our transformation between metallicity phases.   

To better constrain the gas-phase metallicity of CIII] emitters at these redshifts, we need NIR follow-up observations to obtain the rest-optical spectra of these objects. In a future paper, this analysis based on individual galaxies will be performed.
\begin{figure}[t]
    \centering
    \includegraphics[width=\columnwidth]{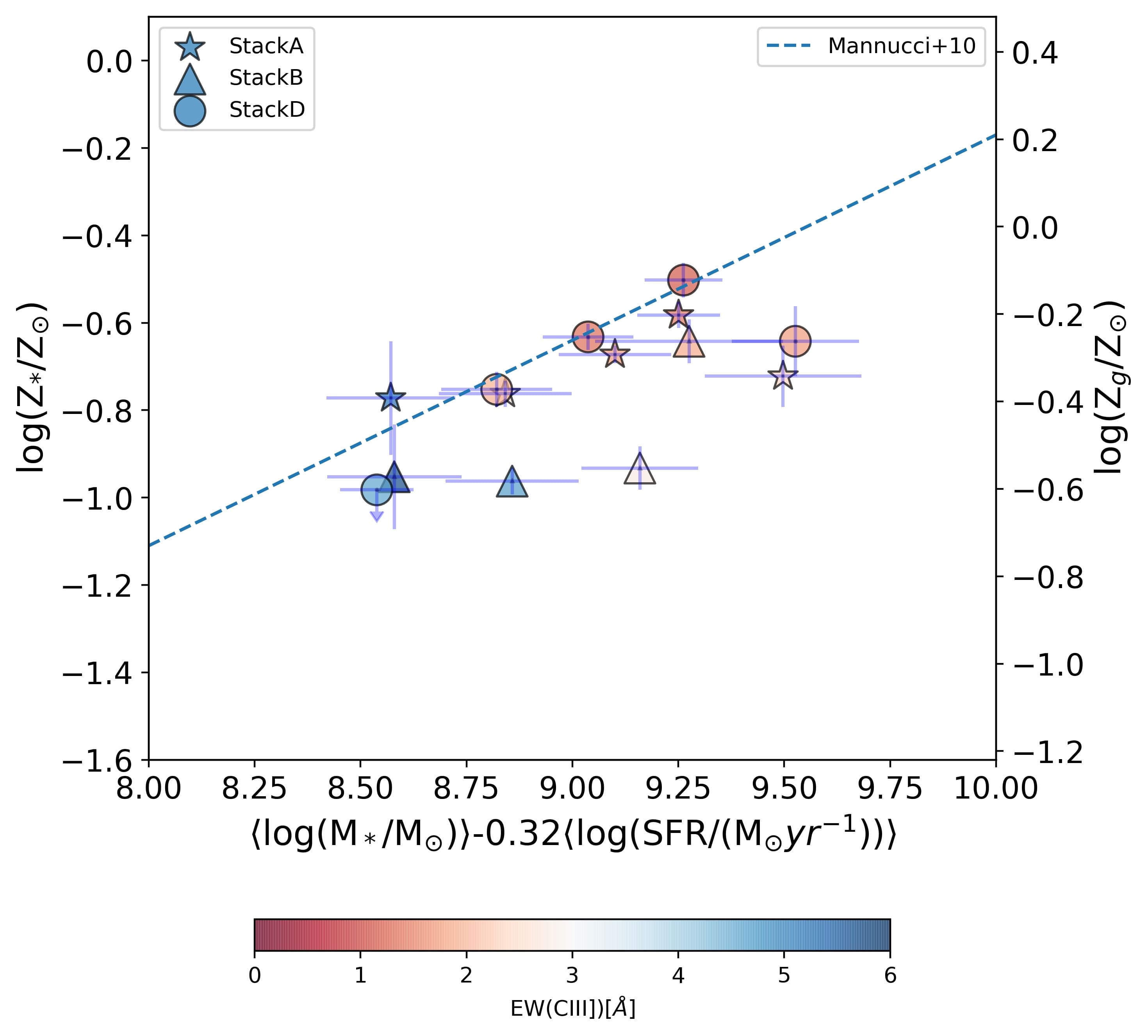}
    \caption{FMR with the stacks A, B, D by stellar mass, color-coded by EW(CIII]). Blue dashed line is the relation shown in \cite{Mannucci2010}. The right scale is the [O/H]$\sim$ log(Z$_g$/Z$_{\odot}$) assuming the $\alpha$-enhanced in Eq. \ref{eq_alpha}.}
    \label{FMRrelation}
\end{figure}
Finally, we use the gas-phase metallicity of the C3 sample to probe the Fundamental Metallicity Relation \citep[FMR, ][]{Mannucci2010}, which describes an invariant  dependence with SFR of the MZRg metallicity of galaxies out to $z\sim 2.5$. Recent work by \citet{Sanders2020} suggest this lack of evolution extends out to $z\sim$\,3.3. In our work, exploring this relation is highly dependent on the adopted Z$_g$. For example, if we assume that Z$_{\star}$=Z$_g$  our results would show an offset to low metallicity of $\sim 0.5$ dex. However, assuming the average $\alpha$-enhancement derived by \citet{Cullen2021} and used in Fig.\ref{MZrelation}, we find a trend that appears in agreement with the slope of the FMR, as shown in Fig. \ref{FMRrelation}. While two stacks in the B scheme (i.e. only galaxies with $z\geq$\,2.9), appear offset towards lower metallicity, stacks D (i.e. only galaxies with $z<$\,2.9) and stacks A (stars, representing all galaxies at 2.4$\lesssim z \lesssim$\,3.9) find a relatively good agreement with the slope of local FMR. 
Considering the typical large uncertainties  involved both in the data measurements and metallicity derivation, especially inherent to the different spectral features and methodologies applied in this and previous works, this result is surprisingly robust. In agreement with recent results \citep{Sanders2020}, this favors the scenario where the FMR does not evolves significantly up to $z\sim$\,3.

\subsection{On the stellar metallicity - C/O relation at \texorpdfstring{$z\sim3$}{z-3} }\label{subsec:discCO}

\begin{figure*}[t]
    \centering
    \includegraphics[width=0.75\textwidth]{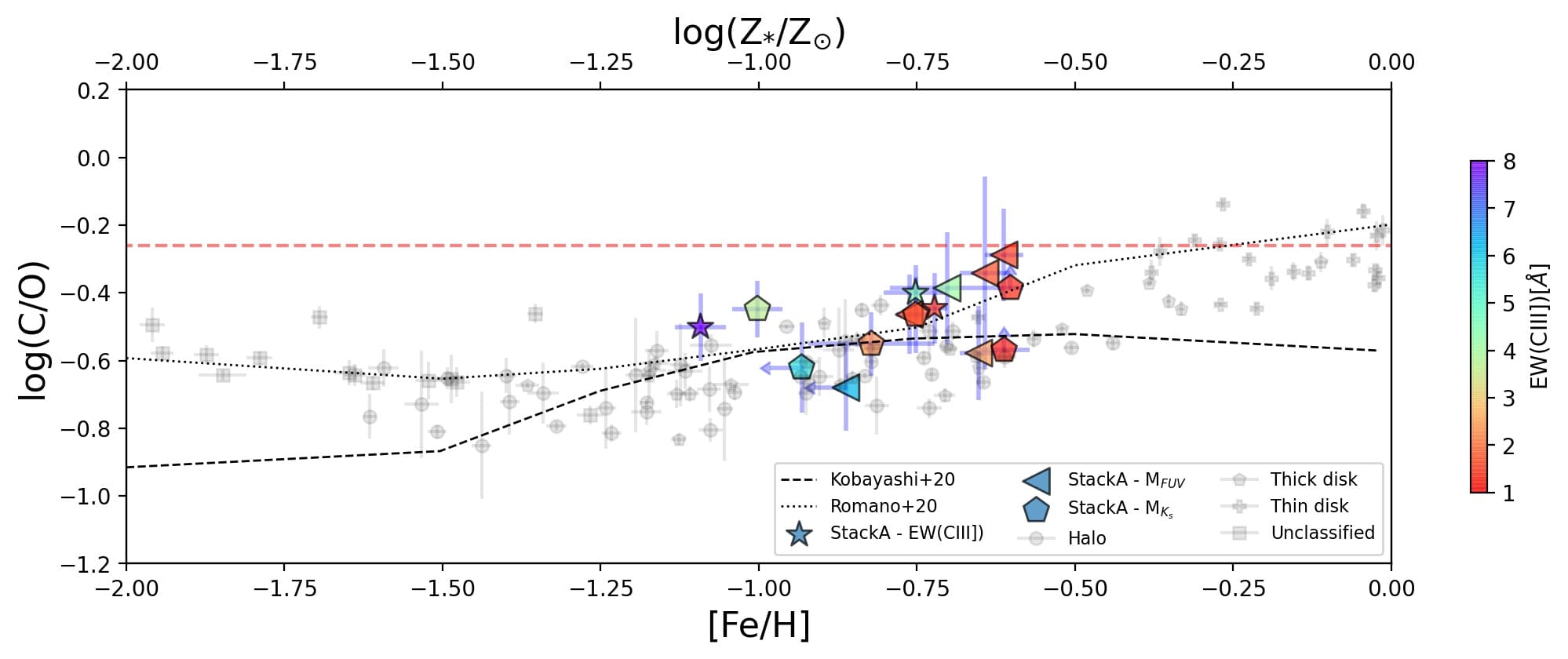}\\ \includegraphics[width=0.75\textwidth]{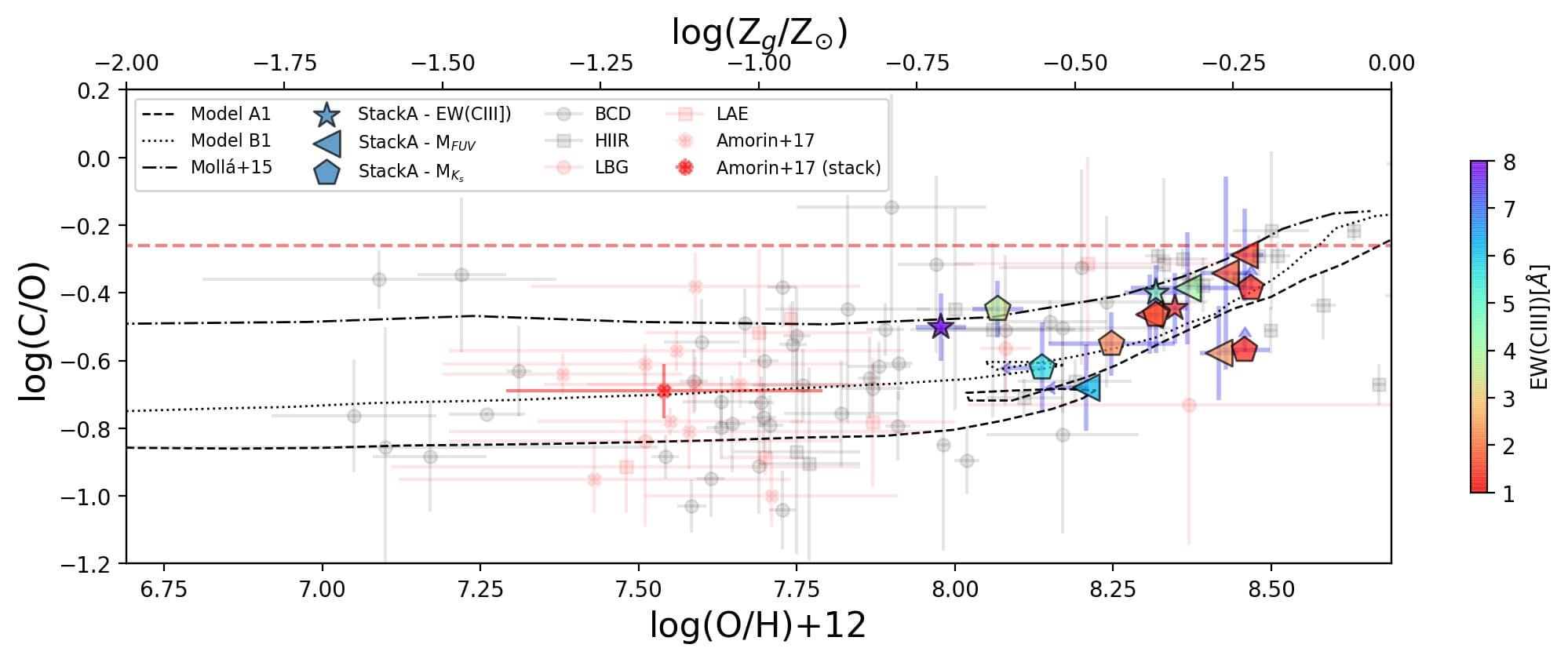}
    \caption{\textit{Left panel:} C/O-Z$_{\star}$ relation. Black symbols are values of stars from the Galactic thin, thick disk, halo and unclassified from the literature \citep{Amarsi2019}. The dashed black line is the K20 model in \cite{Kobayashi2020} and the dotted line is the model MWG-11 in \cite{Romano2020}. \textit{Right panel:} C/O-Z$_{g}$ relation. The red markers are high-redshift galaxies from literature and black markers are local galaxies and HII regions (see text for references). The multi-zone chemical evolution models from \cite{Mattsson2010} are also shown by black lines. The green dotted-dashed line is the chemical model from \cite{Molla2015}. \textit{In both panels:} The red dashed line is the C/O solar value. The stack A by EW(CIII]) and broad band luminosities are shown by markers according to legend and color-coded by EW(CIII]).}
    \label{COZ}
\end{figure*}

The C/O ratio may provide us general trends in the evolutionary state of a galaxy and its ISM. In evolved, metal-enriched galaxies an increase of C/O with increasing metallicity has been observed \citep{Garnett1995,Berg2016,Berg_2019} and also reproduced by models \citep[e.g.][]{Henry2000,Molla2015,Mattsson2010}. This trend can be explained because carbon is primarily produced by the triple-$\alpha$ process in both massive and low- to intermediate-mass stars but, in massive stars, carbon arises almost exclusively from the production due to metallicity-dependent stellar winds, mass loss and ISM enrichment which are greater at higher metallicities \citep{Henry2000}. An evolutionary effect due to the delayed release of carbon (which is mostly produced by low- and intermediate-mass) relative to oxygen (which is produced almost exclusively by massive stars) in younger and less metal-rich systems is an alternative explanation for this trend \citep{Garnett1995}. 
 
 {In this section, we discuss the position of the C3 sample in the Z$_{\star}$-C/O plane, which provide new insights on their chemical enrichment.} On the left panel in Fig. ~\ref{COZ}, we present the Z$_{\star}$-C/O relation for our C3 sample in VANDELS along with the predictions from a chemical evolution model (model MWG-11 in \cite{Romano2020} and model K20 in \cite{Kobayashi2020}) and stellar metallicities $[$Fe/H$]$ estimations of local Milky Way stars in the halo, thick and thin disk \citep{Amarsi2019}. We show results from stack A, which include the entire C3 sample. Symbols are color-coded by EW(CIII]) and their marker depend on the parameter used for stacking, i.e stacks by EW(CIII]), FUV and K$_s$ bands in stars, left-triangle and pentagon, respectively. Our results indicate that C/O increases with stellar metallicity for $Z_{\star}\gtrsim10$\% solar, in agreement with the trends of chemical evolution model and of metal-rich stars in the Galactic thick disk. Our results show higher values that the ones predicted by \cite{Kobayashi2020} for a given stellar metallicity, but this also occurs with most of the thin disk stars. According to these authors, the latter can be partially explained  by an under-prediction of carbon yields by AGB stars on the models. More generally, we note that our results are found in better agreement with the model prediction by \cite{Romano2020}. 

For the gas-phase metallicity, we present results on the right panel of Fig. \ref{COZ}. The stacks are the same on the left panel, but re-scaled assuming Eq. \ref{eq_alpha}. We include four different predictions from chemical evolution models by \cite{Mattsson2010} and \cite{Molla2015}. We also include comparison samples from literature: LAEs \citep{Bayliss2014, Christensen2012, Erb2010, James2014, Villar2004} and LGBs \citep{Barros2016,Vanzella2016, Steidel2016} at $z\sim$2-4, and local analogs such as Blue Compact Dwarfs (BCD) \citep{Garnett1995, Garnett1997, Kobulnicky1997,Kobulnicky1998,Izotov1999,Thuan1999, Berg2016, Senchyna2020} and HII regions \citep{Garnett1995, Kurt1995, Garnett1999, Mattsson2010, Senchyna2020} at $z\sim$0. We note that  metallicities for these objects are derived from nebular lines. Our points in Fig.~\ref{COZ}, instead, assume Eq. \ref{eq_alpha} with a fixed [O/Fe]. However, this assumption might not be accurate and significant (and probably different) levels of $\alpha$-enhancement in individual galaxies may generate systematic offsets between $Z_*$ and $Z_{g}$ \citep{Cullen2021}. Such effect will tend to increase the dispersion of data in the x-axis. With these caveats in mind, the trends shown in Fig.~\ref{COZ} are yet useful to discuss the different levels of chemical enrichment traced by the C/O ratio, particularly for our C3 sample. 

Despite the large scatter shown by observations, Fig.~\ref{COZ} shows a trend of increasing C/O with stellar metallicity for our VANDELS C3 sample. This trend is strongly driven by the low mass (low luminosity) stacks, which show the lower C/O and metallicities.  
Particularly, the stack by EW(CIII]) with the highest EW(CIII]) (purple star in Fig. \ref{COZ}) has the lower stellar metallicity and a lower C/O ratio, which suggest that this population with extreme EW(CIII]) have indeed a young stellar population still dominated by massive stars.  

The comparison of the VANDELS C3 sample with the sample of galaxies at similar redshift (red symbols), shows that our stacks have higher metallicity (even assuming $Z_*=Z_g$) and higher C/O ratios than that of most CIII] emitters in the comparison sample. This suggests that, on average, we are probing galaxies that are chemically more evolved. 

Compared to the local sample, the C/O ratios of our VANDELS C3 sample are comparable to those of local HII regions at similar metallicity, for which the C/O abundances increase toward solar values due to a mix of young and aged stellar populations contributing to C production. Local BCDs, instead, show similar C/O but lower metallicities, i.e. compatible with our results if we assume a lower $\alpha$-enhancement.    

Several mechanisms can affect the position of galaxies in Fig. \ref{COZ}. This includes variations in the star formation histories (i.e. shape, duration of bursts and star formation efficiency), gas fraction and inflow rates that may affect the gas metallicity and even potential changes in the initial mass function, may contribute to the scatter \citep[e.g.][]{Mattsson2010,Molla2015,Berg2016,Vincenzo2018,Kobayashi2020,Palla2020}. High values of C/O at low metallicity could be due to the effects of a massive inflow of pristine gas, which may lower the gas metallicity of an otherwise more  chemically evolved system, without altering C/O \citep{Nakajima_2018}. This is another caveat to consider when there is not a direct measurement of the gas-phase metallicity. 

 In models, C/O is sensitive to the prescriptions for yields, the initial mass function, star formation efficiencies, inflow rates that are not fully understood that make difficult to interpret observations and may also produce variable levels of C/O at low metallicity, as suggested in \cite{Mattsson2010} and \cite{Berg2016}. Although models can reproduce some observations, they do not explain the large scatter observed and the precise values completely. Moreover, the large observational and methodological  uncertainties involved makes the  interpretation of the C/O abundances challenging. 
 
 In this work, we find that our results are consistent with the trend of increasing C/O seen in the models for the range of gas-phase metallicities probed by the stacks (around 20 to 60\% solar), assuming a constant $\alpha$-enhancement. In particular, we compare our data with the multi-zone chemical evolution models from \cite{Mattsson2010}. Case A1 (black dashed line in Fig. \ref{COZ}) represents a case where the low and intermediate mass stars are producing most of the carbon, while case B1 (black dotted line in Fig. \ref{COZ}) represents a case where the carbon is to a large extent produced in high mass stars. 
Other models such as those in \cite{Molla2015}  \citep[see also][]{Berg2016}, which predict higher values of log(C/O) for a given O/H, find a good agreement especially for our stacks at higher EW(CIII]). 

Based on a detailed comparison with models, \citet{Berg_2019} discuss the sensitivity of the C/O ratio to both the detailed star formation history and  supernova (SNe) feedback. In models, lower C/O values are found at lower star formation efficiencies and burst with longer duration. On the other hand, larger C/O ratios are more related with the presence of SNe feedback that ejects oxygen and reduces the effective yields. 
The rapid C/O enrichment found for the C3 sample could be thus related to a more bursty star formation history. Given that our C3 sample are mostly  main-sequence galaxies, we speculate that a recent star formation history with multiple bursts could have enhanced the C/O to their observed levels. 

More detailed observations are needed for both local and high redshift galaxies to fully understand the C/O enrichment history through cosmic time. The HST COS Legacy Archive Spectroscopic SurveY (CLASSY) \citep{Berg2019hst}, which has obtained high S/N UV nebular spectra for a representative sample of local galaxies, will certainly contribute to clarify this scenario in the local universe and future observations with JWST will enable more detailed studies of the C/O  evolution of star-forming galaxies at intermediate and high redshift.

\section{Conclusions}\label{sec:conclu}
{
We study a large representative sample of 217 galaxies with CIII] detection at $2<z<4$ covering a range of $\sim 2$ dex in stellar mass. These CIII] emitters have a broad range of UV luminosities, thus allowing a stacking analysis to characterize their mean physical properties. We consider stacking by different bins of stellar mass, rest-frame FUV and K$_S$ luminosities, and rest-frame EW(CIII]) and  EW(Ly$\alpha$). We derive   stellar metallicity and the C/O abundance from  stack spectra and discuss several relations found between these parameters. We summarize our conclusions as follows:
   \begin{enumerate}[i]
    \item Reliable (S/N$>3$) CIII] emitters represent $\sim$30\% of the VANDELS parent sample at $z \sim$\,2-4. They show EW(CIII]) between 0.3 and 20\r{A} with a mean value of $\sim 4$\r{A}. However, stacked spectra of galaxies with marginal or non detection of CIII] (S/N$<$3) in individual spectra show weak (EW$\lesssim$\,2\r{A}) CIII] emission, suggesting this line is common in normal star-forming galaxies at $z\sim$\,3. On the other hand, extreme emitters (EW(CIII])$\gtrsim$8\r{A}) are exceedingly rare ($\sim$3\%) in VANDELS, which is expected as the C3 sample is drawn from a parent sample of main sequence galaxies.
    \item Stacking reveals that faint UV nebular lines of OIII], SiIII, CIV and HeII, and fine structure emission lines of SiII are ubiquitous in CIII] emitters. We find that the strength of the nebular lines depends on the stellar mass and luminosity. Overall, less massive (and fainter in any band) stacks show  more intense Ly$\alpha$, CIV, HeII, OIII] and CIII] than more massive (and brighter in any band), which tend to have fainter UV emission lines.
    \item According to UV diagnostic diagrams, nebular lines in VANDELS  CIII] emitters are powered by stellar photoionization, suggesting no other ionization source than massive stars. These results add new constraints to models on the flux ratios and EWs of nebular emission lines in main sequence SF regions for the faint CIII] regime at $z\sim 2-4$.
    \item The EW(Ly$\alpha$) and EW(CIII]) appear correlated in our C3 sample but a large scatter is found for individual galaxies. Stacks with larger EW(CIII]) show larger EW(Ly$\alpha$), but not all CIII] emitters are Ly$\alpha$ emitters. 
    \item Galaxies with higher EW(CIII]) show lower stellar metallicities. This result suggests that extremely low metallicities ($<$10\% solar) should be expected for the most extreme galaxies in terms of their EW(CIII]). Our results (Eq. \ref{ZEWc3eq}) show that galaxies with stellar metallicity Z$_{\star}<0.1$Z$_{\odot}$  are typically  strong CIII] emitters   (EW(CIII])$>$7\r{A}).
    \item The stellar metallicities of CIII] emitters are not significantly different from that of the parent sample, increasing from $\sim$10\% to $\sim$40\% solar for stellar masses $\log$(M$_{\star}$/M$_{\odot})$\,$\sim$9-10.5.  The stellar mass-metallicity relation of the CIII] emitters is consistent with previous works showing strong evolution from $z=0$ to $z\sim3$, both using stellar (Fe/H) and nebular (O/H) metallicities inferred from the UV spectra and assuming an average [Fe/O]$=$0.38, recently found for a subsample of VANDELS galaxies at $z\sim$\,3 \citep{Cullen2021}. 
    \item We find the C/O abundances of CIII] emitters ranging 35\%-150\% solar, with a noticeable increase with FUV luminosity, and a smooth decrease with CIII] EWs. Fainter FUV galaxies have lower C/O, higher EW(CIII]), and lower Z$_{\star}$, which suggest a UV spectrum dominated by massive stars and a bright nebular component which is still chemically unevolved. 
    \item We discuss for the first time the C/O-Fe/H and the C/O-O/H relations for star-forming galaxies at $z\sim$\,3. They show stellar and nebular abundances consistent with the trends observed in Milky Way halo and thick disc stars and local HII galaxies, respectively. We find a good agreement with modern chemical evolution models, which suggest that CIII] emitters at $z\sim$\,3 are experiencing an active phase of chemical enrichment.
   \end{enumerate}
Our results provide new insight into the nature of UV line emitters at $z\sim$\,3, paving the way for future studies at higher-$z$ using the James Webb Space Telescope.}

The spectra built and used for the analysis in this paper are publicly available at \url{https://github.com/mfllerena/stacks_C3emittersVANDELS}.

\begin{acknowledgements}
    We thank the anonymous referee for the detailed review and useful suggestions. This work is based on data products from observations made with ESO Telescopes at La Silla Paranal Observatory under ESO programme ID 194.A-2003 (PIs: Laura Pentericci and Ross McLure). MLl acknowledges support from the National Agency for Research and Development (ANID)/Scholarship Program/Doctorado Nacional/2019-21191036. RA acknowledges support from ANID FONDECYT Regular Grant 1202007.
    
    This work has made extensive use of Python packages astropy \citep{astropy:2018}, numpy \citep{harris2020}, and Matplotlib \citep{Hunter:2007}.

\end{acknowledgements}

\bibliographystyle{aa}
\bibliography{main}

\begin{appendix} 
\section{Analysis of non-detected CIII] emitters in the parent sample}\label{appen1}
The results presented in this paper are based on the selection of a sample of galaxies with the CIII] line detected with S/N$>3$. This S/N threshold is motivated by the need of having reliable systematic redshifts, which are obtained fitting  this nebular line --the most intense UV metal line in the parent sample. In this section, we study whether this selection criterion has any effect on our results. We analyse the 521 star-forming galaxies from the VANDELS parent sample with non-detections in the CIII] line, i.e. those objects  excluded from our C3 sample. We perform the same stacking analysis done for the C3 sample, following schemes A and B in bins of FUV luminosity and using the same bin distribution and methodology explained in Section \ref{sec:method}. Because of the lack of accurate systemic redshifts for this sample, stacks are performed using the spectroscopic redshift reported by the VANDELS DR3 catalog, which is mostly based on template fitting of ISM absorption features by the tool PANDORA/EZ \citep{McLure_2018}. We find that the resulting stacks show similar features compared to the stacks in the C3 sample, showing detections of multiple ISM absorption lines and nebular emission lines.  Overall, the effect of using spectroscopic instead of systemic redshifts in the stacking is to broaden emission line profiles in the stacks due to small velocity shifts. As a result, the S/N ratio of the detected lines tend to be lower and their line fluxes more uncertain than that of the C3 sample, despite the number of galaxies averaged in each stack is generally higher. 

\begin{figure}[t]
    \centering
    \includegraphics[width=0.7\linewidth]{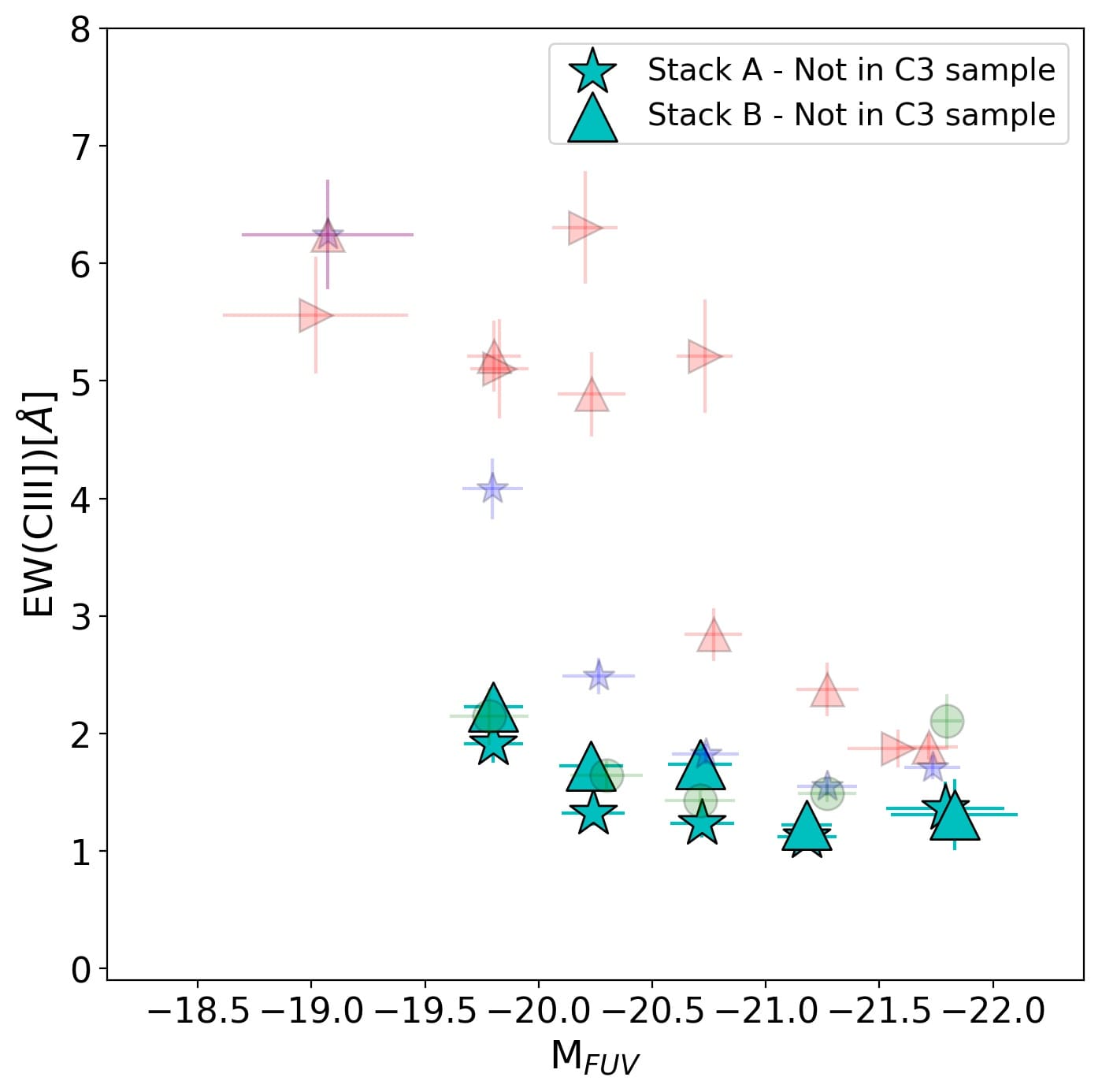}
    \caption{M$_{FUV}$-EW(CIII]) relation for the stacks by FUV luminosity with the non-detected CIII] sample (in cyan symbols). Faint symbols are the same as in Fig. \ref{c3relation}.}
    \label{EW-MFUV-noc3}
\end{figure}

In all the stacks but the one at lower luminosity, we detect CIII] with a S/N$>3$. However, their fluxes are fainter than their counterparts in the C3 sample. In Fig. \ref{EW-MFUV-noc3} we present the resulting EW(CIII]) as a function of the mean FUV luminosity for the stacks A and B with detected CIII]. Their trend is in good  agreement with the same relations found using the C3 sample, showing a very mild increase of EW(CIII]) from $\sim$1\r{A} to $\sim$2\r{A} towards lower FUV  luminosity. This suggest that including these galaxies without individual detection of CIII] in the stacks for the C3 sample will not change the resulting EWs and their relations with luminosity. Similar conclusion is found when comparing line ratios of UV metal lines. Although not all the stacks in this test have good detections (S/N$>$3) in all the relevant emission lines, the position of their line ratios in the  diagnostic diagrams presented in Fig. \ref{diagC}
and \ref{diagO} are consistent with those of the faintest CIII] emitters, i.e. fully consistent with pure stellar photoionization. 

\begin{figure}[t]
    \centering
    \includegraphics[width=0.7\linewidth]{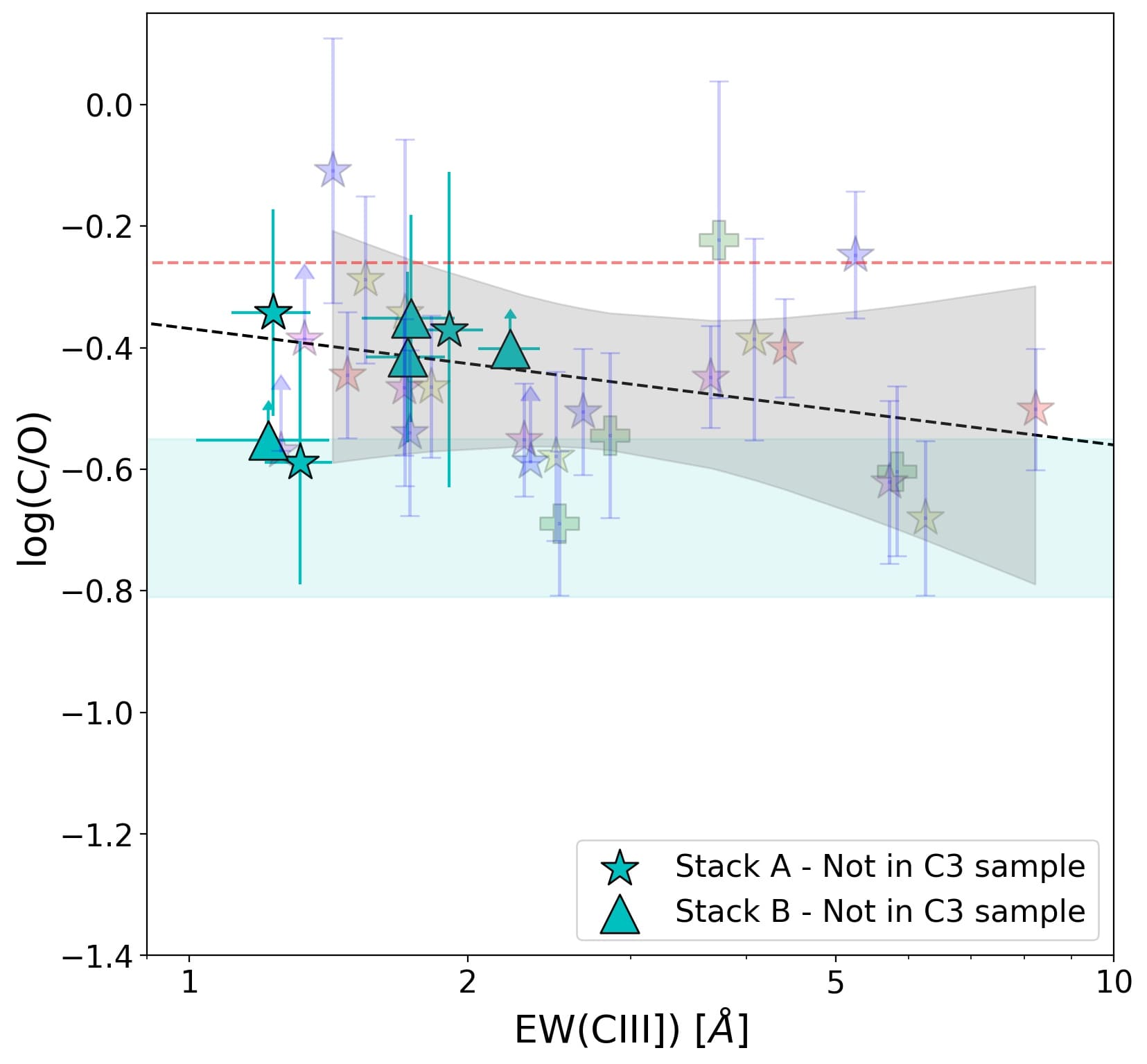}
    \caption{EW(CIII])-C/O relation for the stacks by FUV luminosity with the non-detected CIII] sample (in cyan symbols). Faint symbols, shaded regions and dashed lines are the same as in Fig. \ref{C/O_EW}}
    \label{EW-CO-noc3}
\end{figure}

Finally, we also test the possible effects in the C/O abundances. Following the same methods described in Section~\ref{subsec:co} for the C3 sample, we estimate C/O in the non-C3 stacks where we have at least S/N$\geq$\,2 detections of OIII]. We find consistent C/O values with those of the C3 sample, with $\log$(C/O) ranging from $-0.6$ to $-0.34$ (i.e. 45\% to 80\% solar, respectively). These values show good agreement with the relation between C/O and EW(CIII]) found in Eq.~\ref{COc3} for the C3 sample, as illustrated in Fig. \ref{EW-CO-noc3}. Again, this result suggests that the non-detected CIII] emitters in the parent sample that were excluded from the stacking analysis of the C3 sample include intrinsically faint CIII] emitters, which are well represented by the results found for the C3 sample throughout this paper. We can therefore conclude that CIII] emission should be a common feature in UV faint galaxies at $z\sim$\,3. This result is relevant for reionization studies because the number of UV faint galaxies increase significantly at higher redshift \citep[e.g.][]{Bouwens2016,Yue2018}.  

\clearpage
\newpage

\section{Stacked spectra}
\begin{figure*}[t]
    \centering
    \includegraphics[trim=5 50 5 100, clip, width=0.33\linewidth]{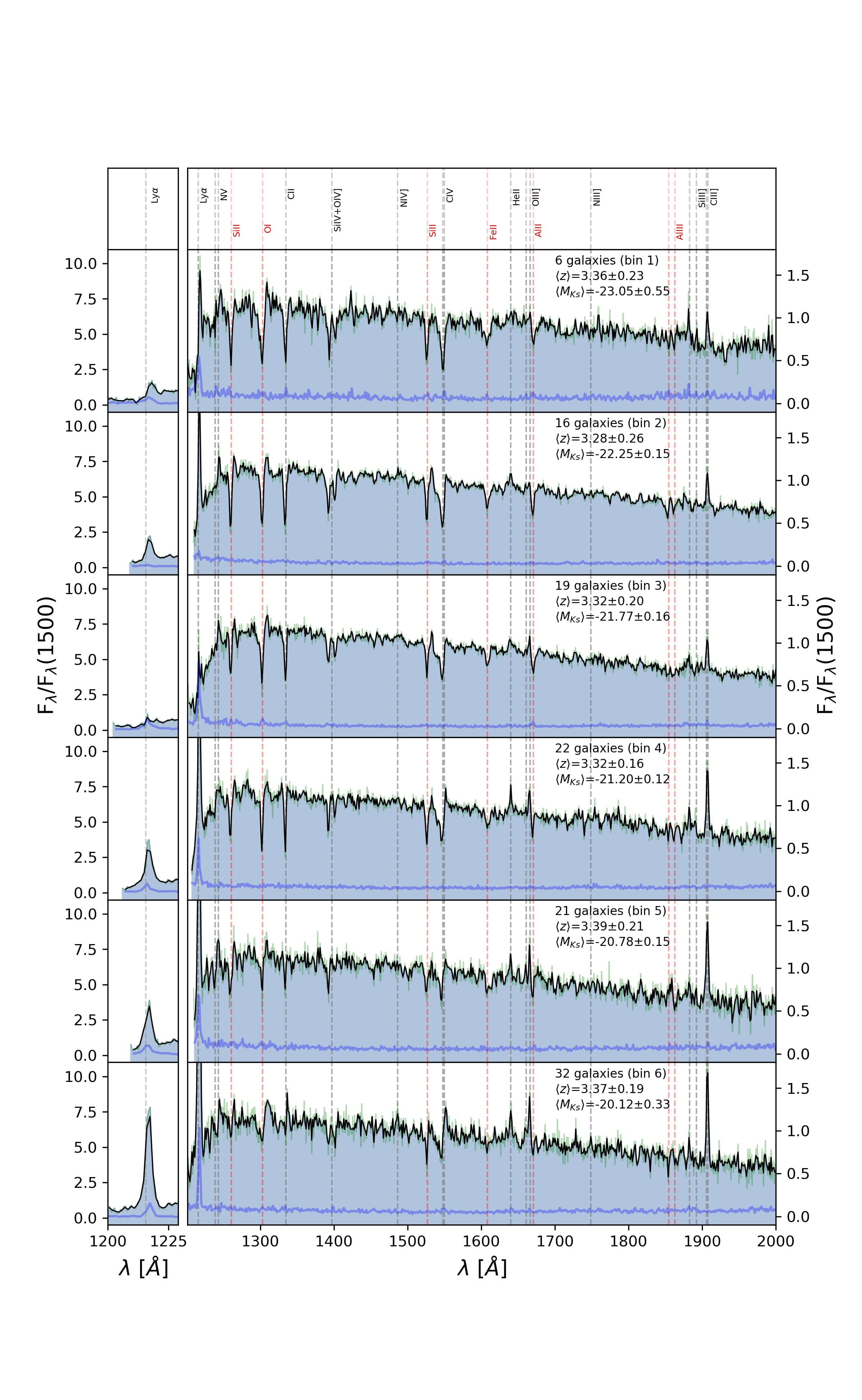}\,\includegraphics[trim=5 50 5 100, clip,width=0.33\linewidth]{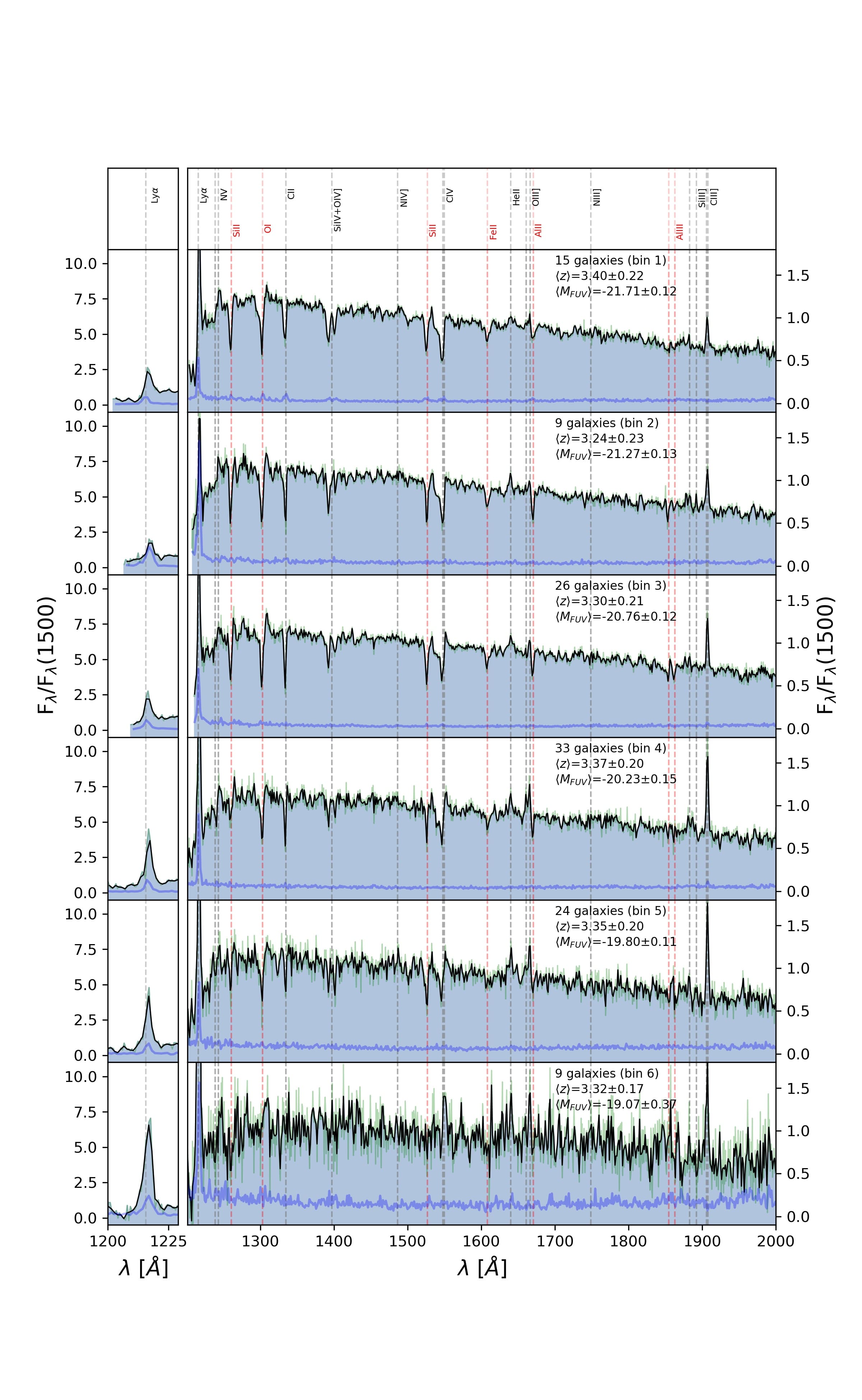}\,\includegraphics[trim=5 50 5 50, clip,width=0.33\linewidth]{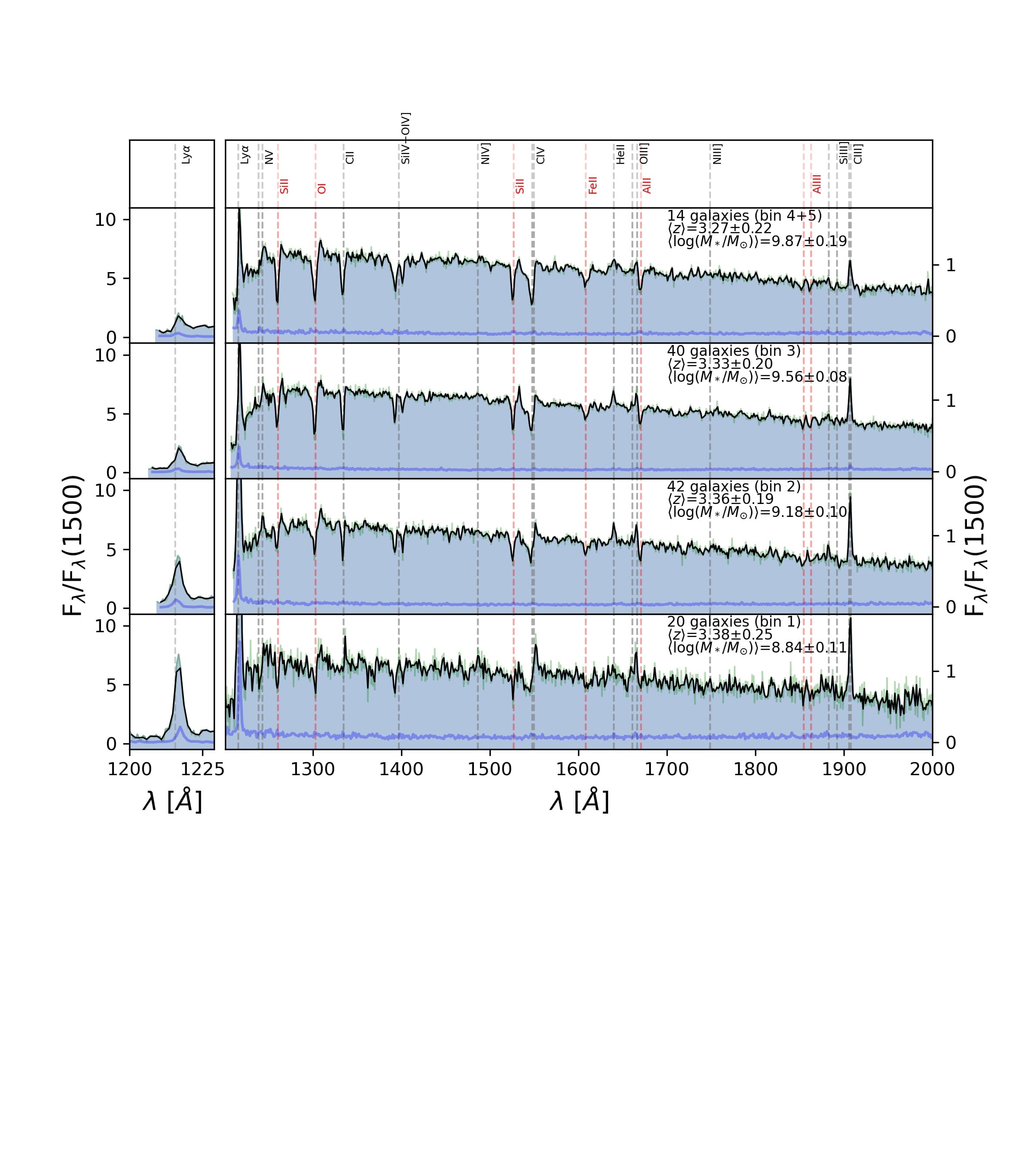}
    \caption{Resulting Stack B for each physical parameter. \textit{From left to right}: M$_{K_s}$, M$_{FUV}$, and stellar mass. In each panel, the green faint line is the stack spectrum with the $\sim$0.6\r{A}/pixel sampling, while the black one is with $\sim$1.2\r{A}/pixel. The blue line in the 1-$\sigma$ error spectrum. The vertical lines mark known UV lines. Information about the number of galaxies, the mean redshift and the mean parameter are included in each panel.}
    \label{StackB}
\end{figure*}

\begin{figure*}[ht]
    \centering
   \includegraphics[trim=5 120 5 75, clip, width=0.33\linewidth]{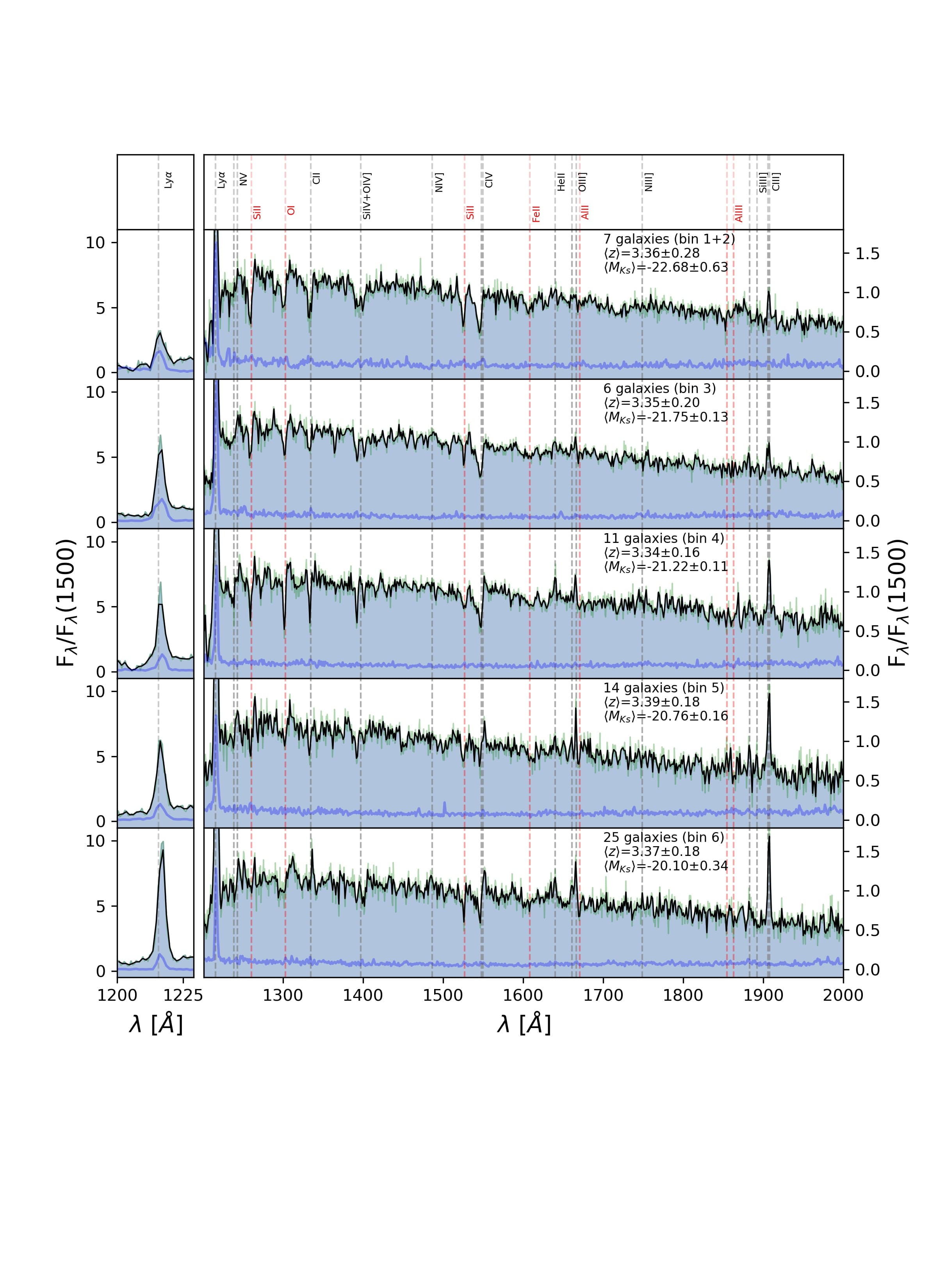}\,\includegraphics[trim=5 120 5 75, clip, width=0.33\linewidth]{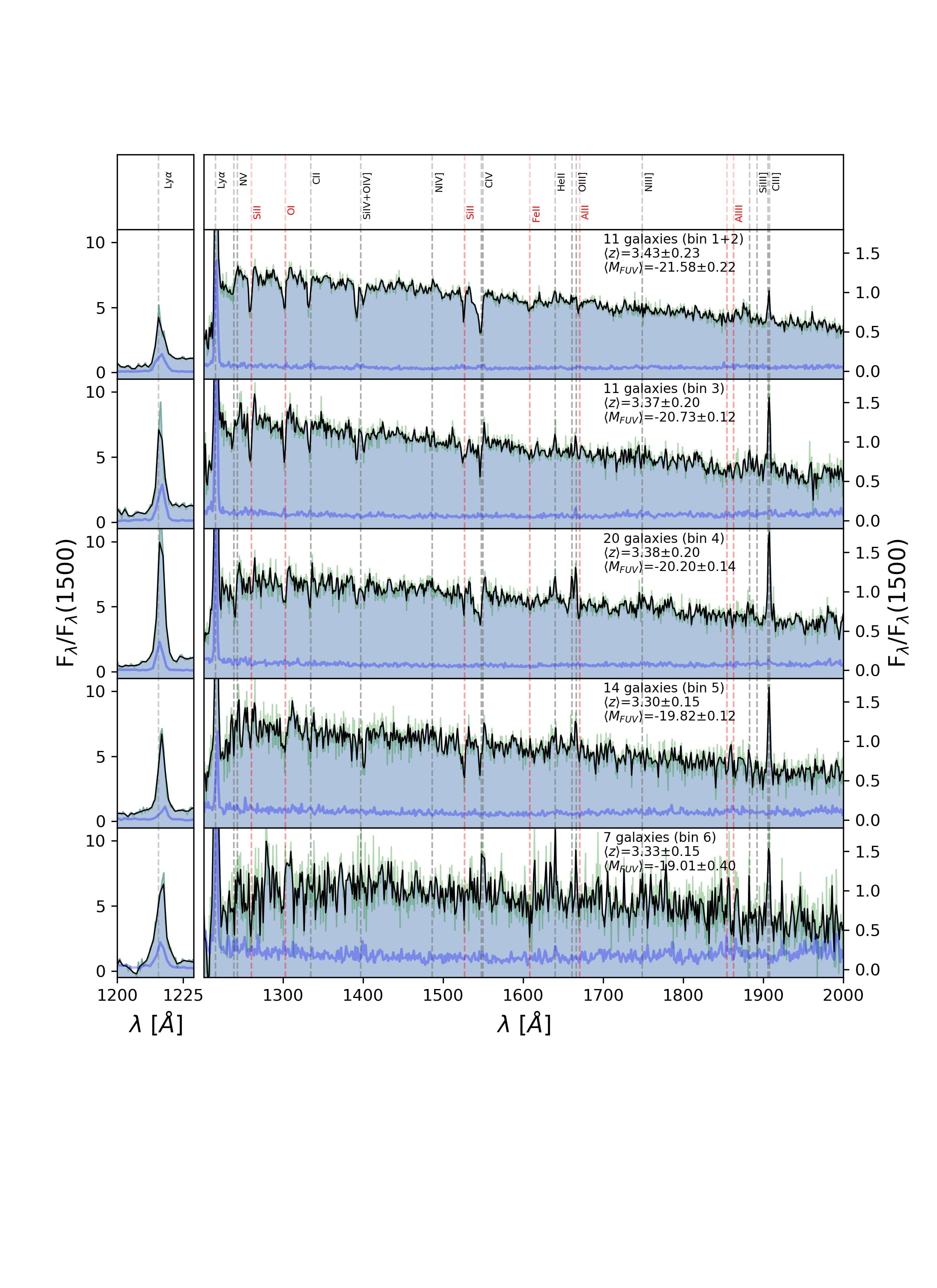}\,\includegraphics[trim=5 120 5 50, clip, width=0.33\linewidth]{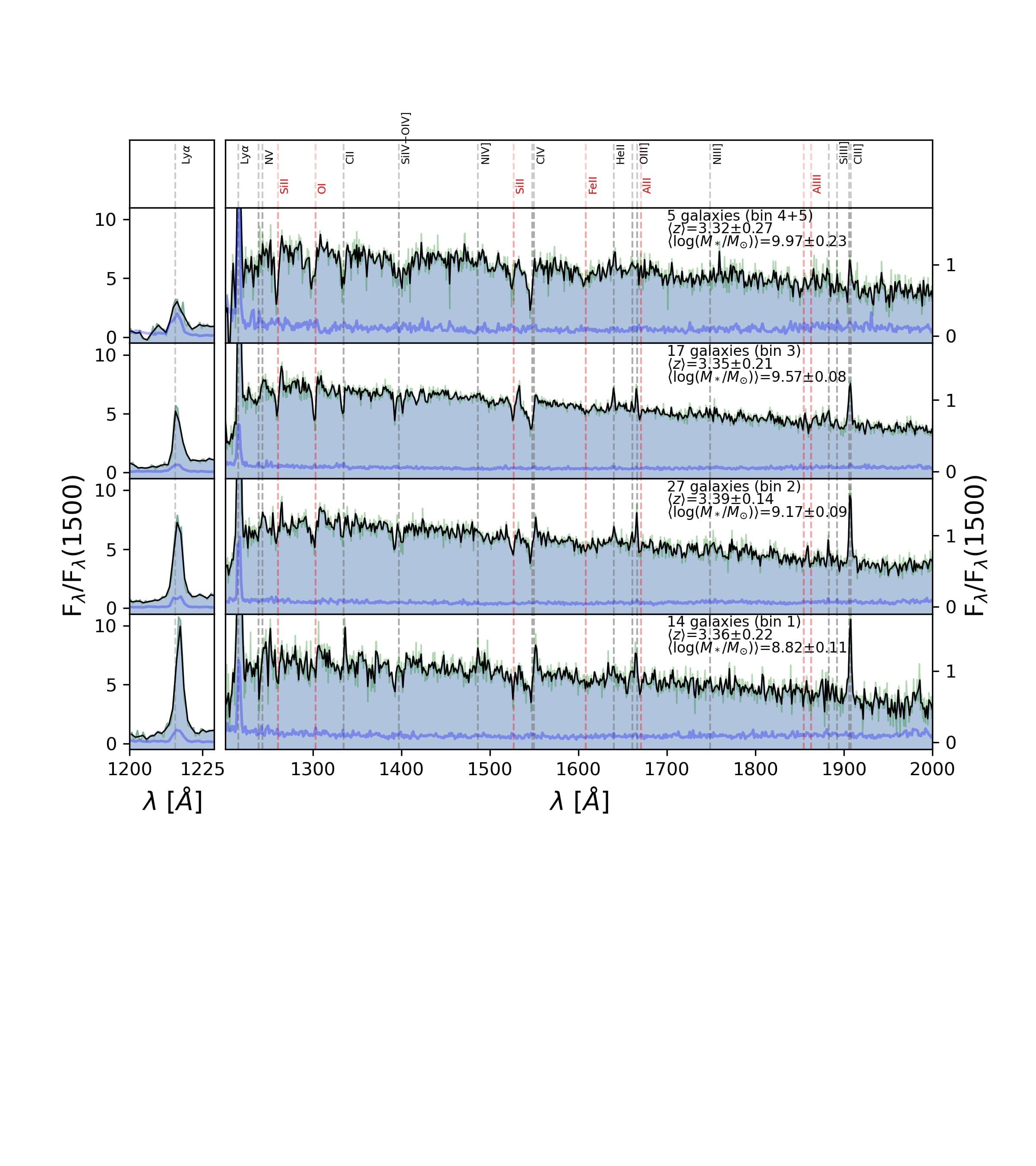}
    \caption{Resulting Stack C for each physical parameter. \textit{From left to right}: M$_{K_s}$, M$_{FUV}$, and stellar mass. In each panel, the green faint line is the stack spectrum with the $\sim$0.6\r{A}/pixel sampling, while the black one is with $\sim$1.2\r{A}/pixel. The blue line in the 1-$\sigma$ error spectrum. The vertical lines mark known UV lines. Information about the number of galaxies, the mean redshift and the mean parameter are included in each panel.}
    \label{StackC}
\end{figure*}

\begin{figure*}[t]
    \centering
    \includegraphics[trim=5 520 5 90, clip,width=0.33\linewidth]{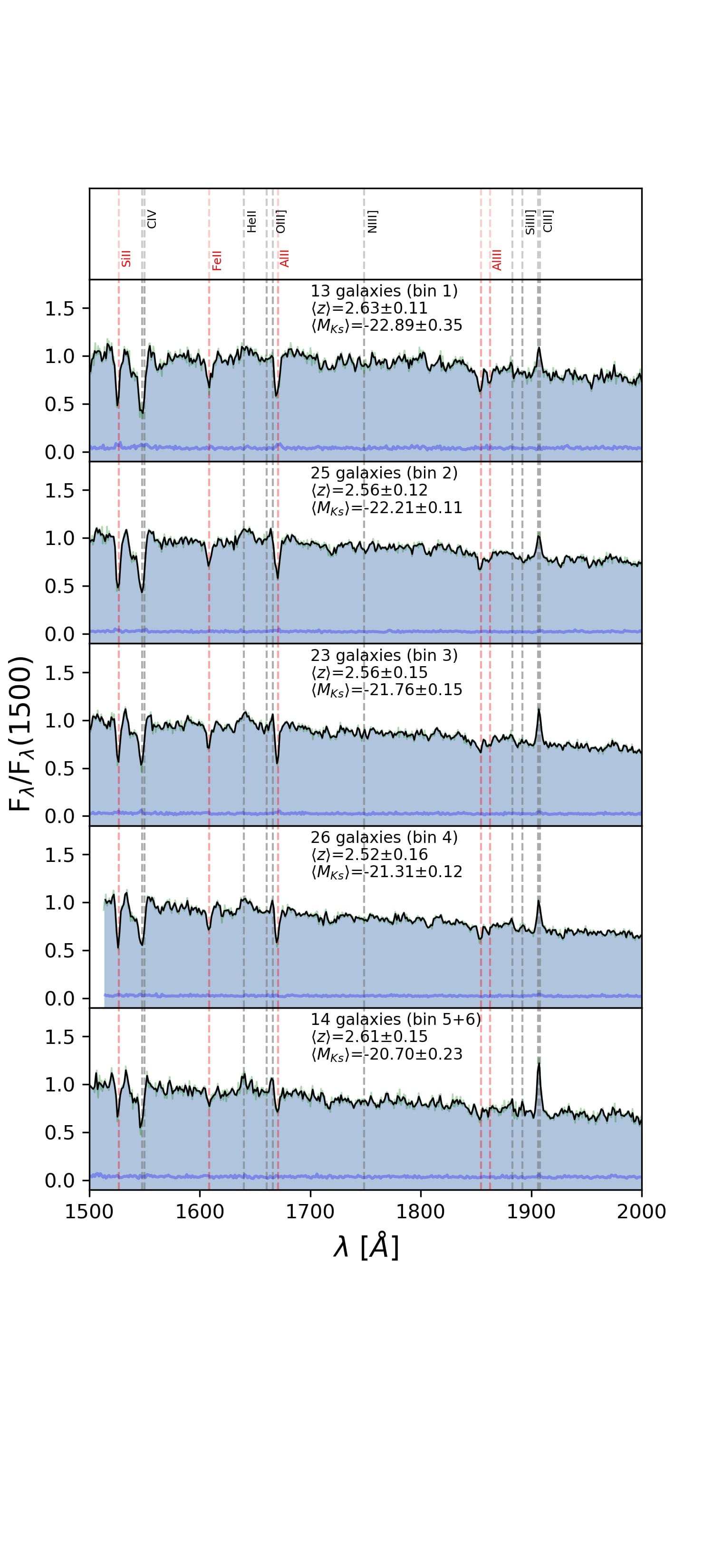}\,\includegraphics[trim=5 520 5 90, clip,width=0.33\linewidth]{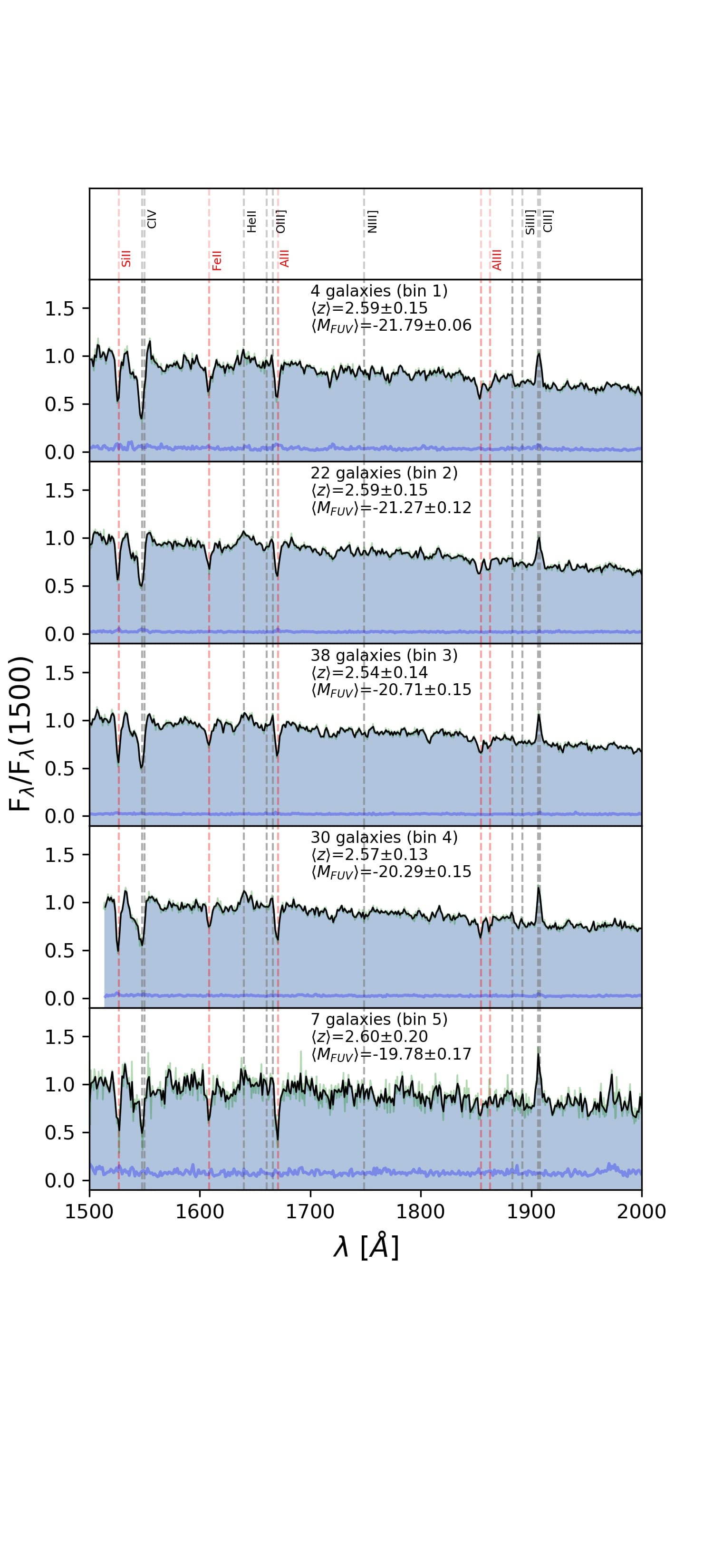}\,\includegraphics[trim=5 520 5 90, clip,width=0.33\linewidth]{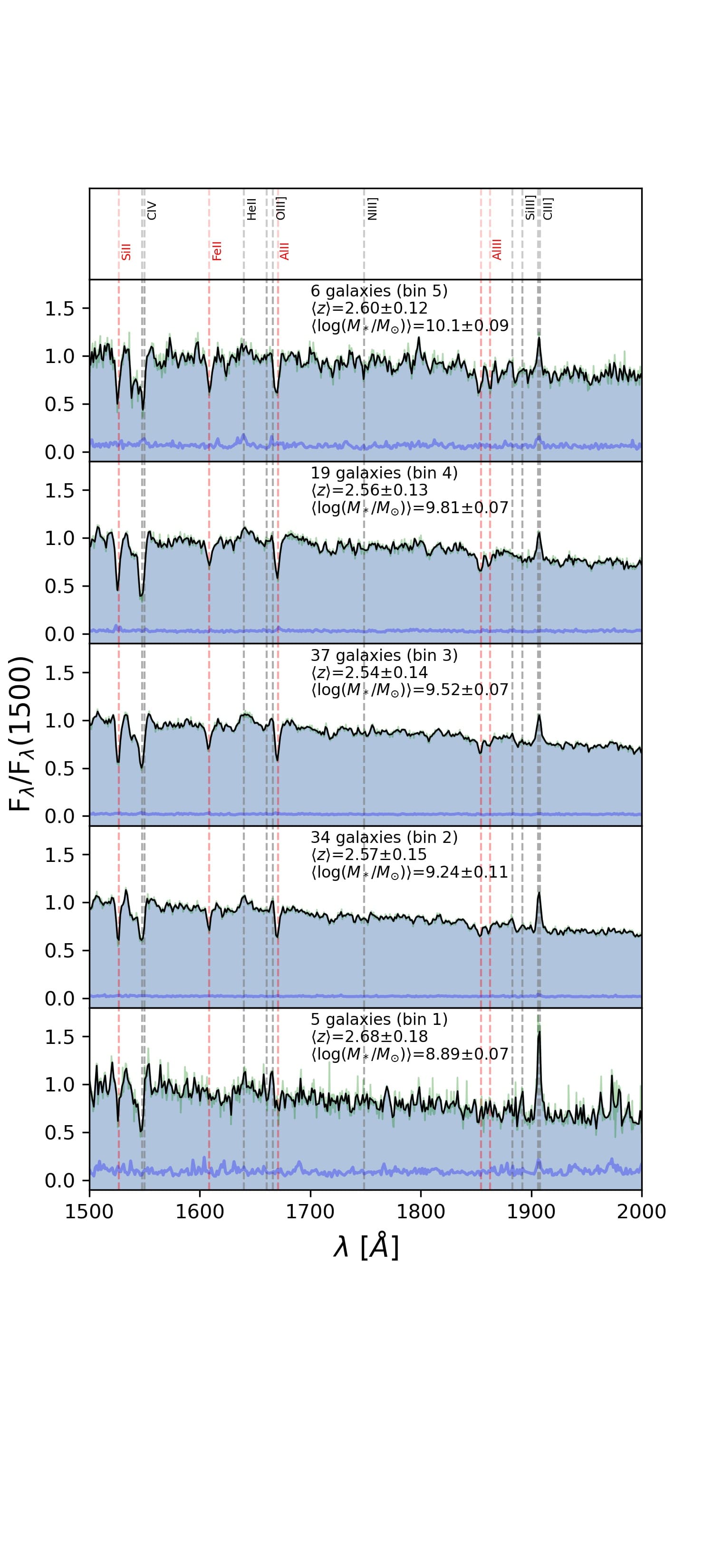}
    \caption{Resulting Stack D for each physical parameter. \textit{From left to right}: M$_{K_s}$, M$_{FUV}$, and stellar mass. In each panel, the green faint line is the stack spectrum with the $\sim$0.6\r{A}/pixel sampling, while the black one is with $\sim$1.2\r{A}/pixel. The blue line in the 1-$\sigma$ error spectrum. The vertical lines mark known UV lines. Information about the number of galaxies, the mean redshift and the mean parameter are included in each panel.}
    \label{StackD}
\end{figure*}

\begin{figure*}[t]
    \centering
    \includegraphics[trim=5 1400 5 290, clip,width=0.45\linewidth]{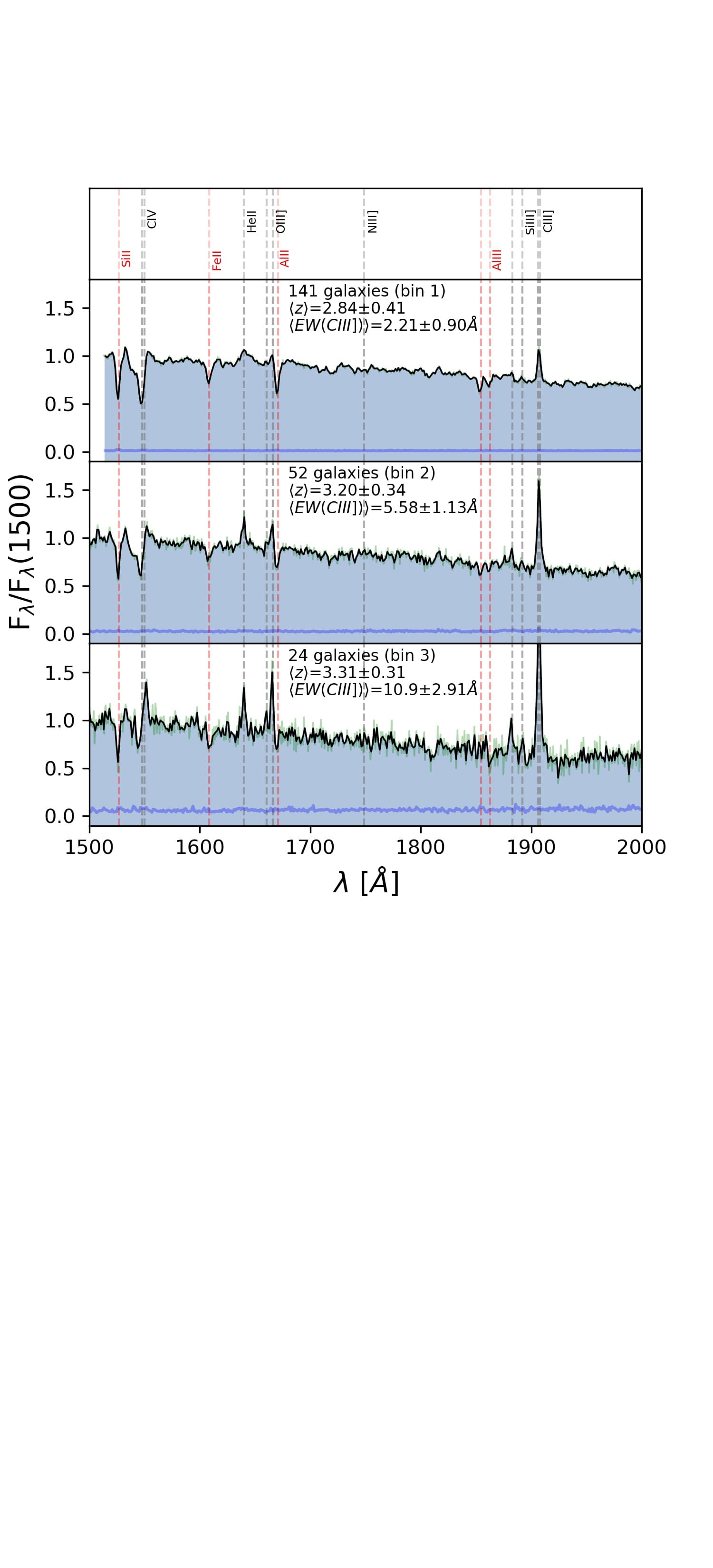}\,\includegraphics[trim=5 1450 5 290, clip,width=0.45\linewidth]{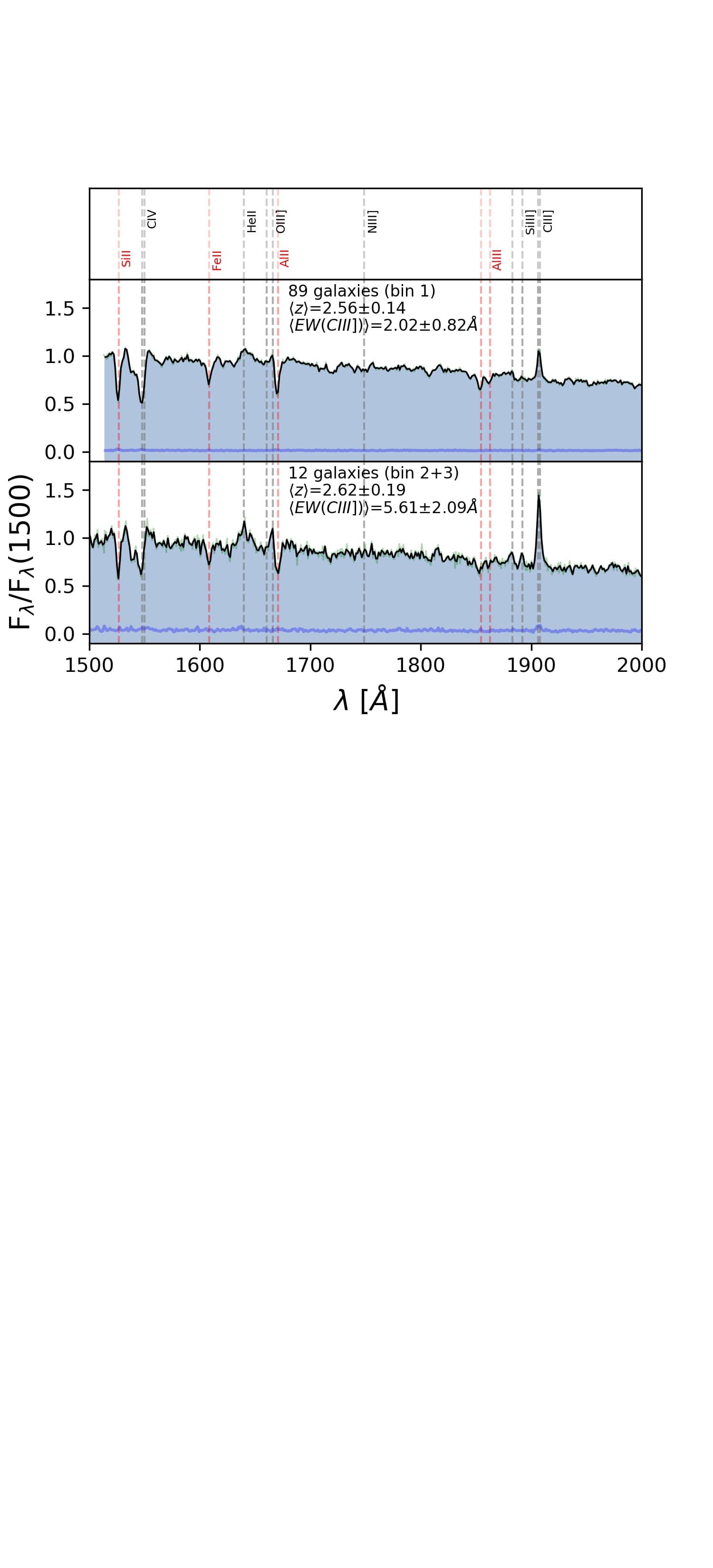}\\\includegraphics[trim=5 1100 5 270, clip,width=0.7\linewidth]{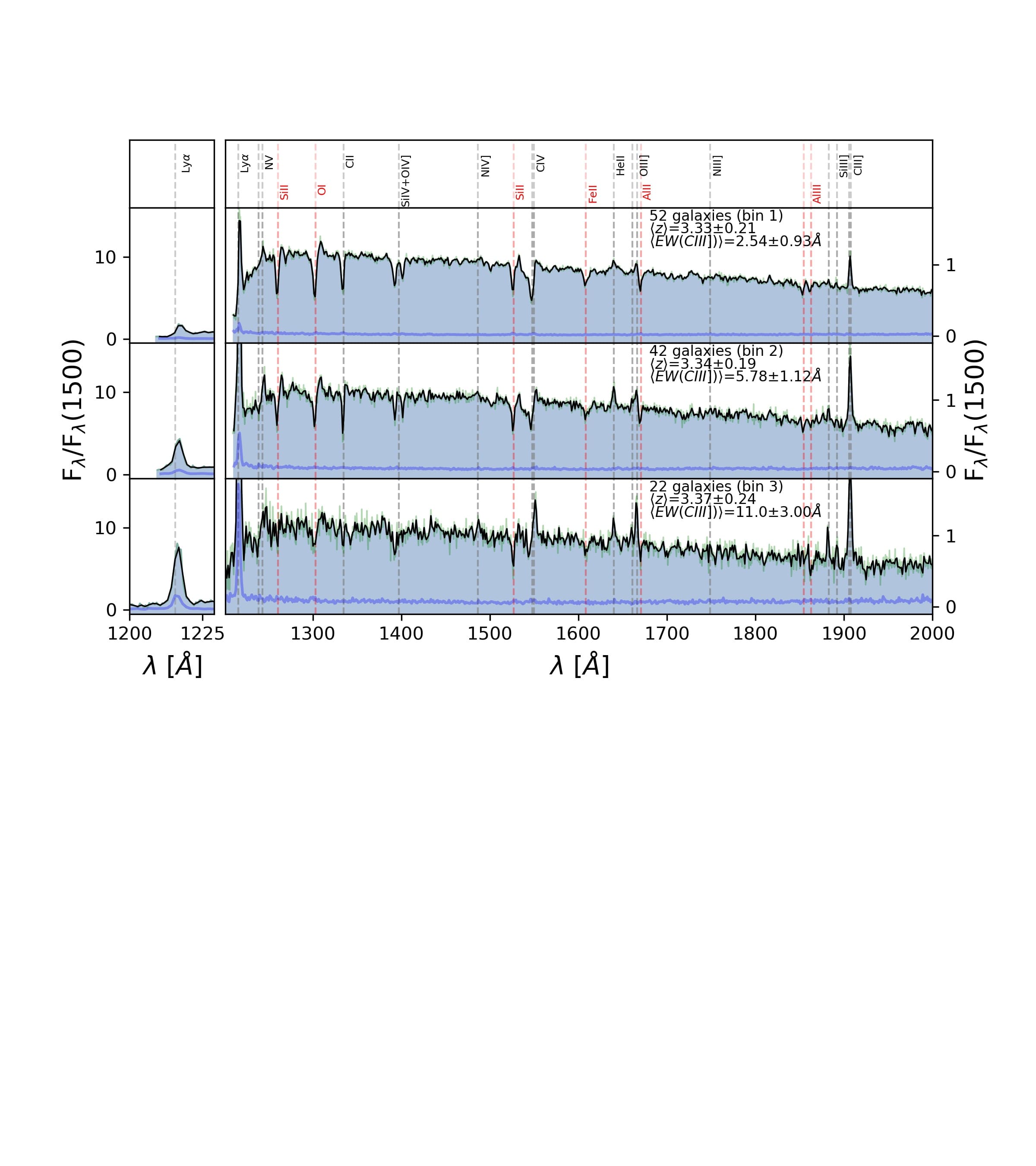}\\\includegraphics[trim=5 1100 5 270, clip,width=0.7\linewidth]{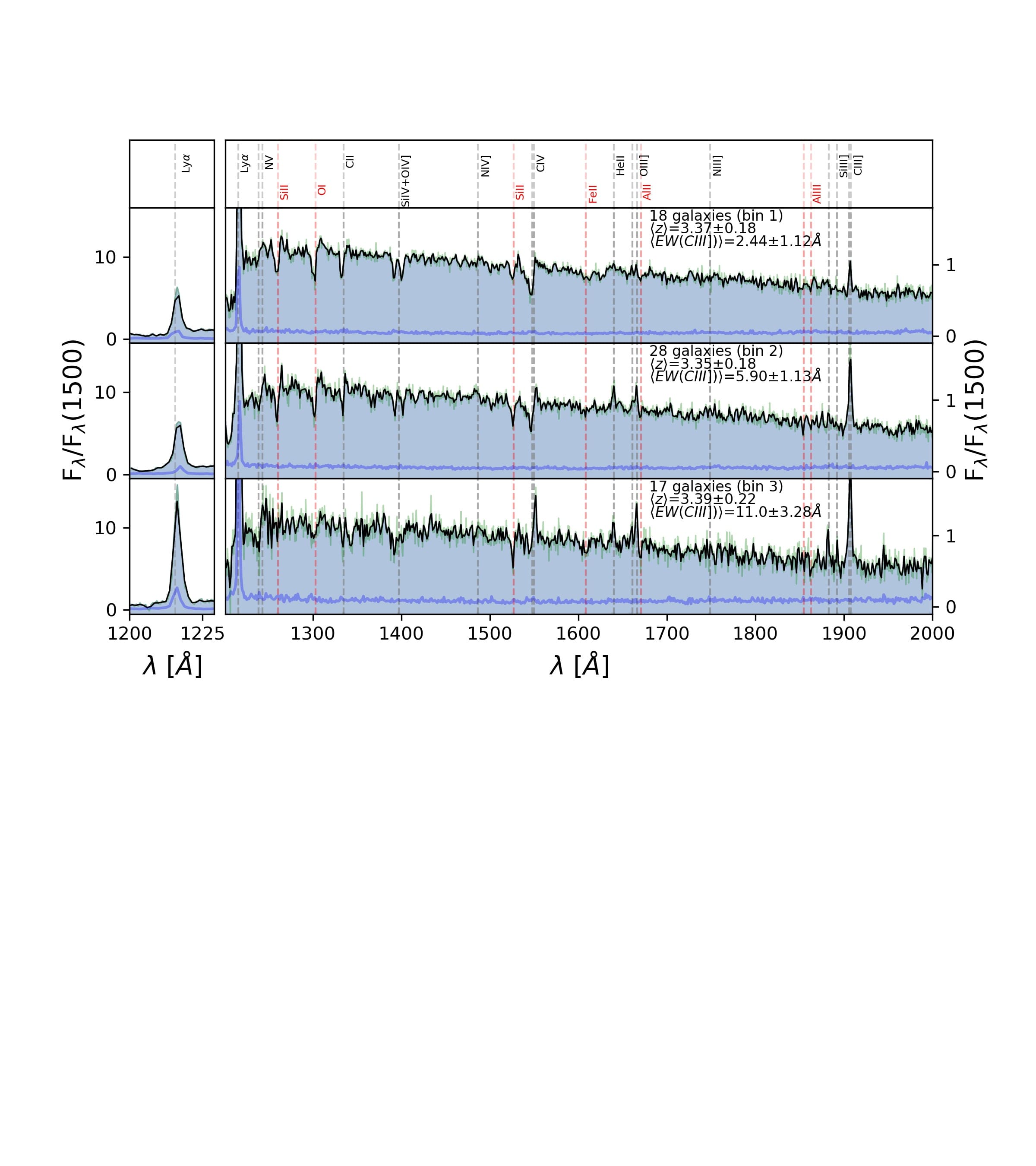}
    \caption{Stacks by EW(CIII]). \textit{From the left to the right on top row}: Stack A, Stack D, \textit{Middle row:} Stack B, and \textit{Bottom row:} Stack C. In each panel, the green faint line is the stack spectrum with the $\sim$0.6\r{A}/pixel sampling, while the black one is with $\sim$1.2\r{A}/pixel. The blue line in the 1-$\sigma$ error spectrum. The vertical lines mark known UV lines. Information about the number of galaxies, the mean redshift and the mean parameter are included in each panel.
    }
    \label{StackEWC3}
\end{figure*}

In this section, we present the resulting stacks B (Fig. \ref{StackB}), C  (Fig. \ref{StackC}), and D  (Fig. \ref{StackD}) grouping galaxies by FUV and $K_s$ luminosities, and by stellar mass, as well as stacks by EW(CIII]) (Fig. \ref{StackEWC3}), which are used in the different analyses of this work.  

\clearpage
\newpage
\section{Caveats related with stacking methods}\label{apen0}
\begin{figure}[t]
    \centering
    \includegraphics[width=0.25\textwidth]{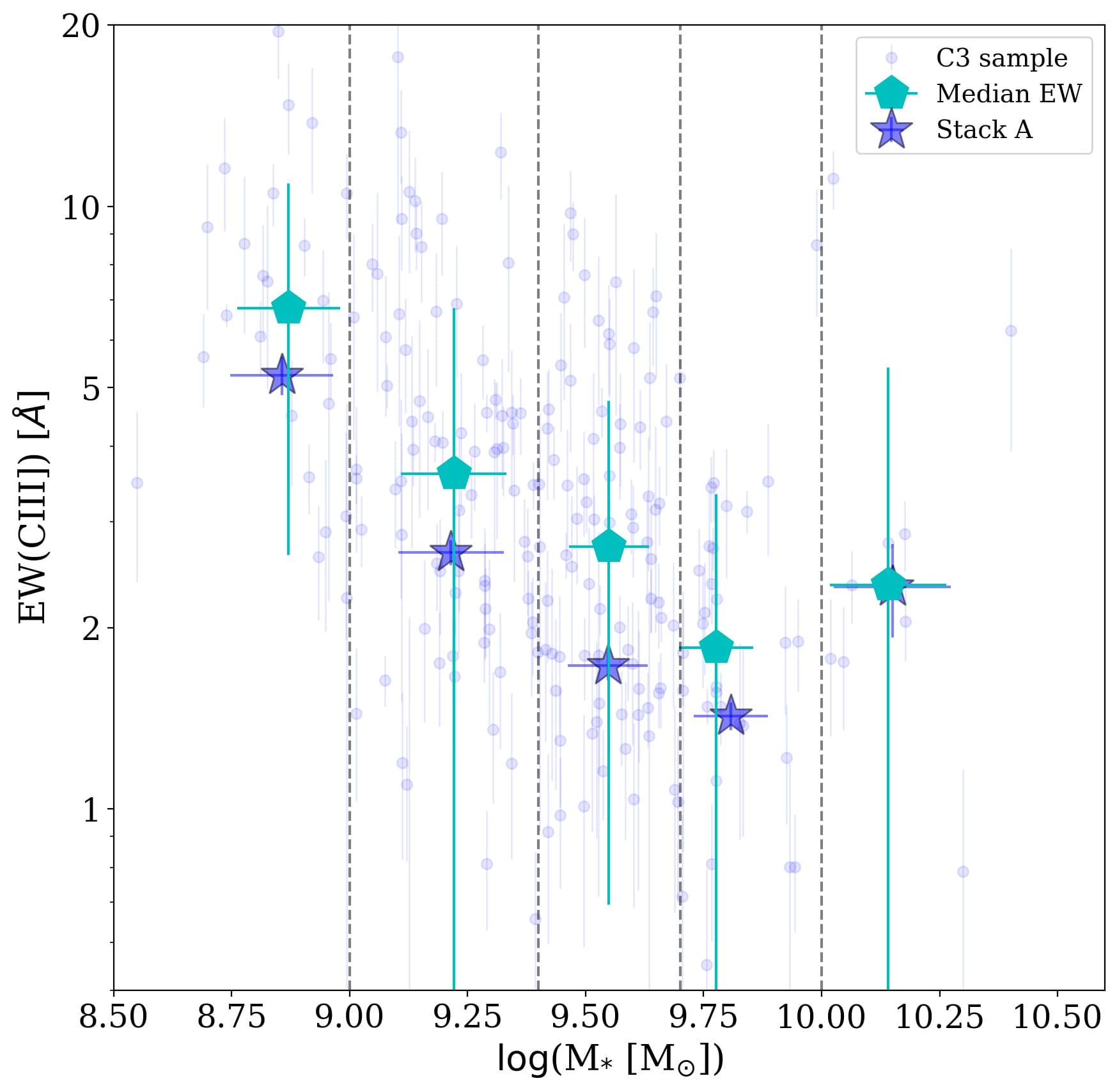}\,\includegraphics[width=0.25\textwidth]{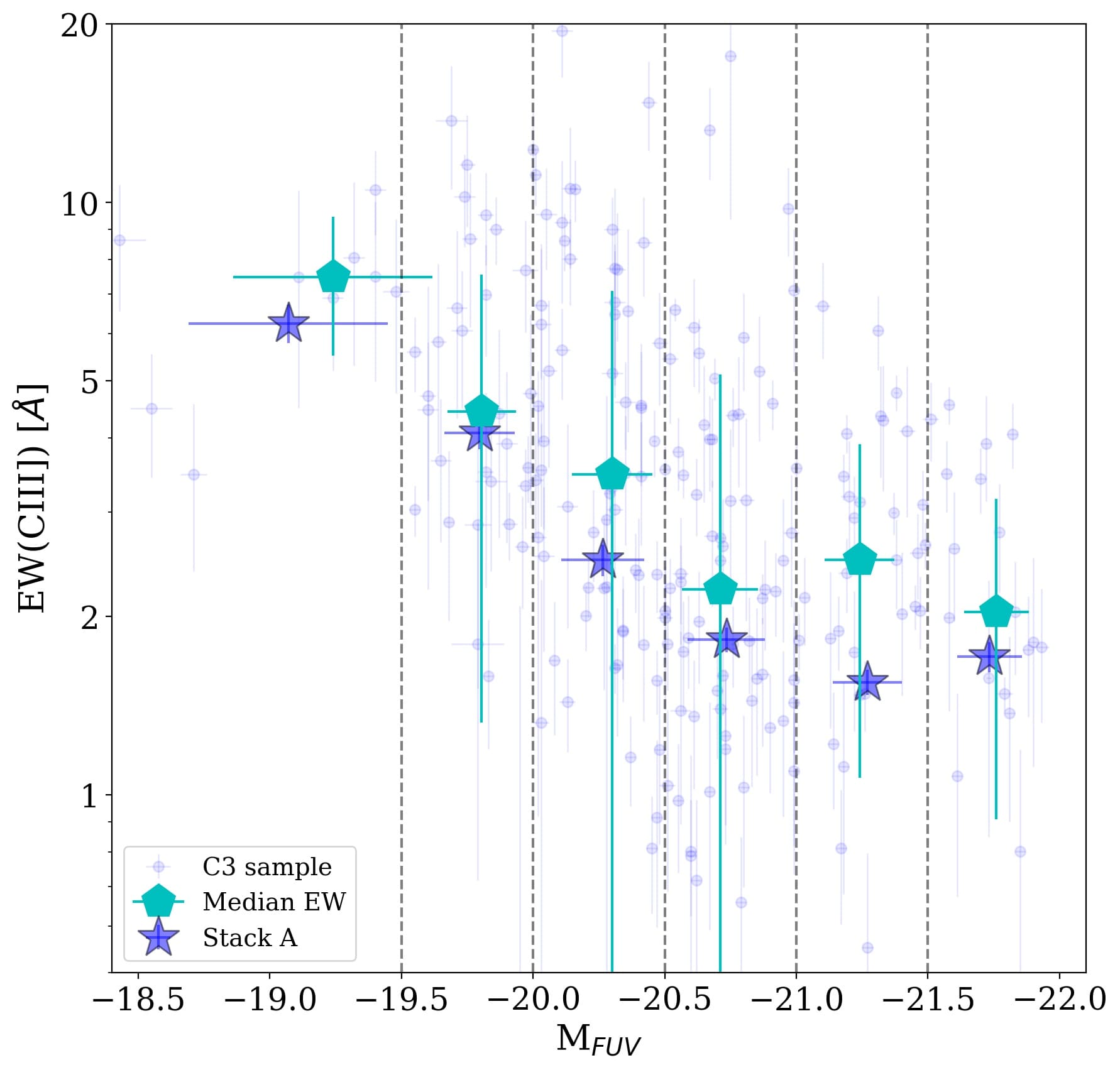}
    \caption{Relation between EW(CIII]) and stellar mass (\textit{left panel}) and FUV (\textit{right panel}) luminosity, similar as shown in Fig. \ref{c3relation} for stack A (blue stars) and C3 sample (small blue circles). The cyan pentagons are the median values for the individual galaxies in the bin and their errorbars correspond to the standard deviation in the bin (boundaries marked by vertical dashed line). }
    \label{medianprop}
\end{figure}

In this section, we discuss possible effects of stacking and the binning choice in the relations shown in Fig. \ref{c3relation}. For this aim, in Fig. \ref{medianprop} we compare the EW(CIII]) measured on the stacked spectra with the median EW(CIII]) of the galaxies in each bin as measured in individual spectra. We do this comparison with stacks A, which are obtained after binning by stellar mass and FUV luminosity. The error bars for the median EWs represent the standard deviation in the bin. We notice that the trends are consistent between both methods, with some differences in the actual value of EW(CIII]), but consistent within the large errors. 
Therefore, measurements of the stacks are representative of the median properties of the galaxies in each bin and we do not expect strong biases due to bin size or the particular choose of the limits.

\begin{figure}[h]
    \centering
    \includegraphics[width=0.25\textwidth]{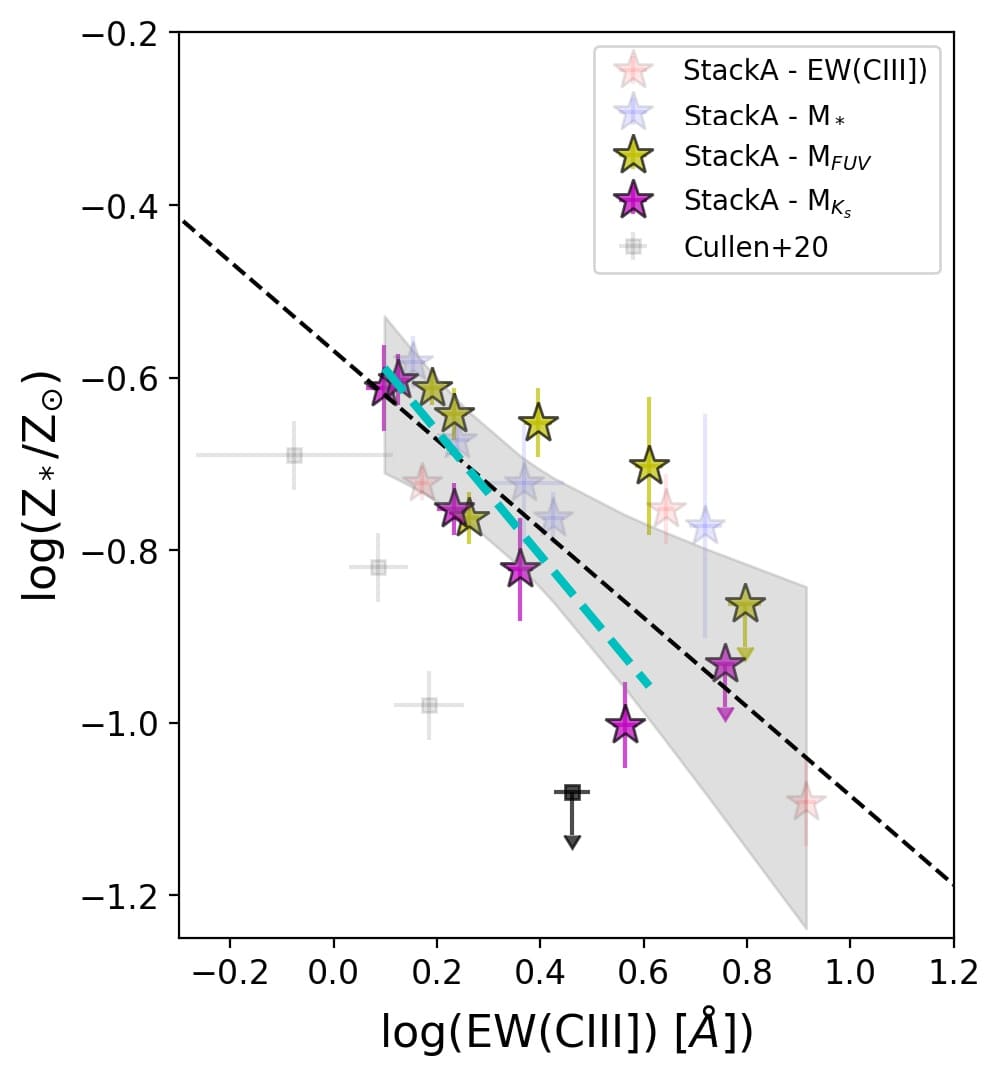}\,\includegraphics[width=0.25\textwidth]{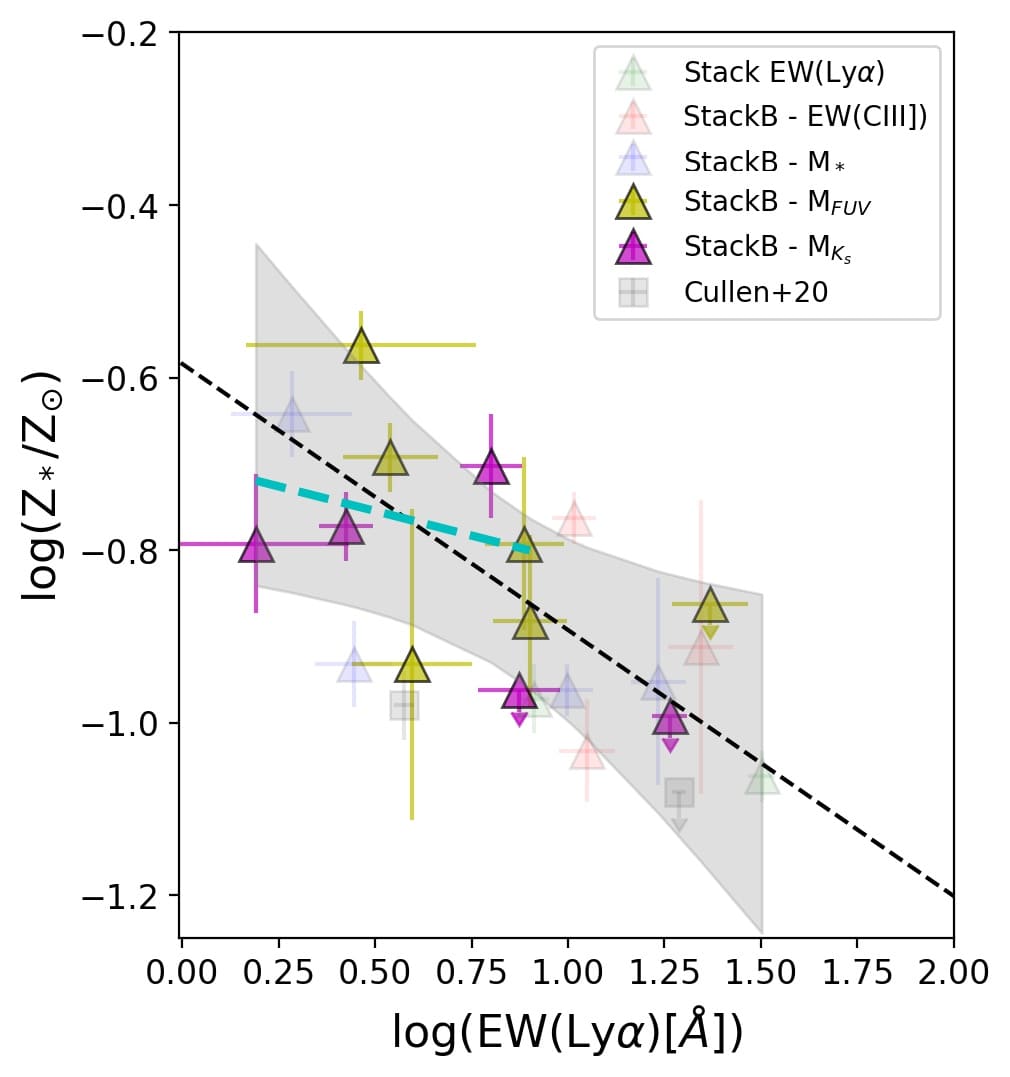}\\
    \includegraphics[width=0.25\textwidth]{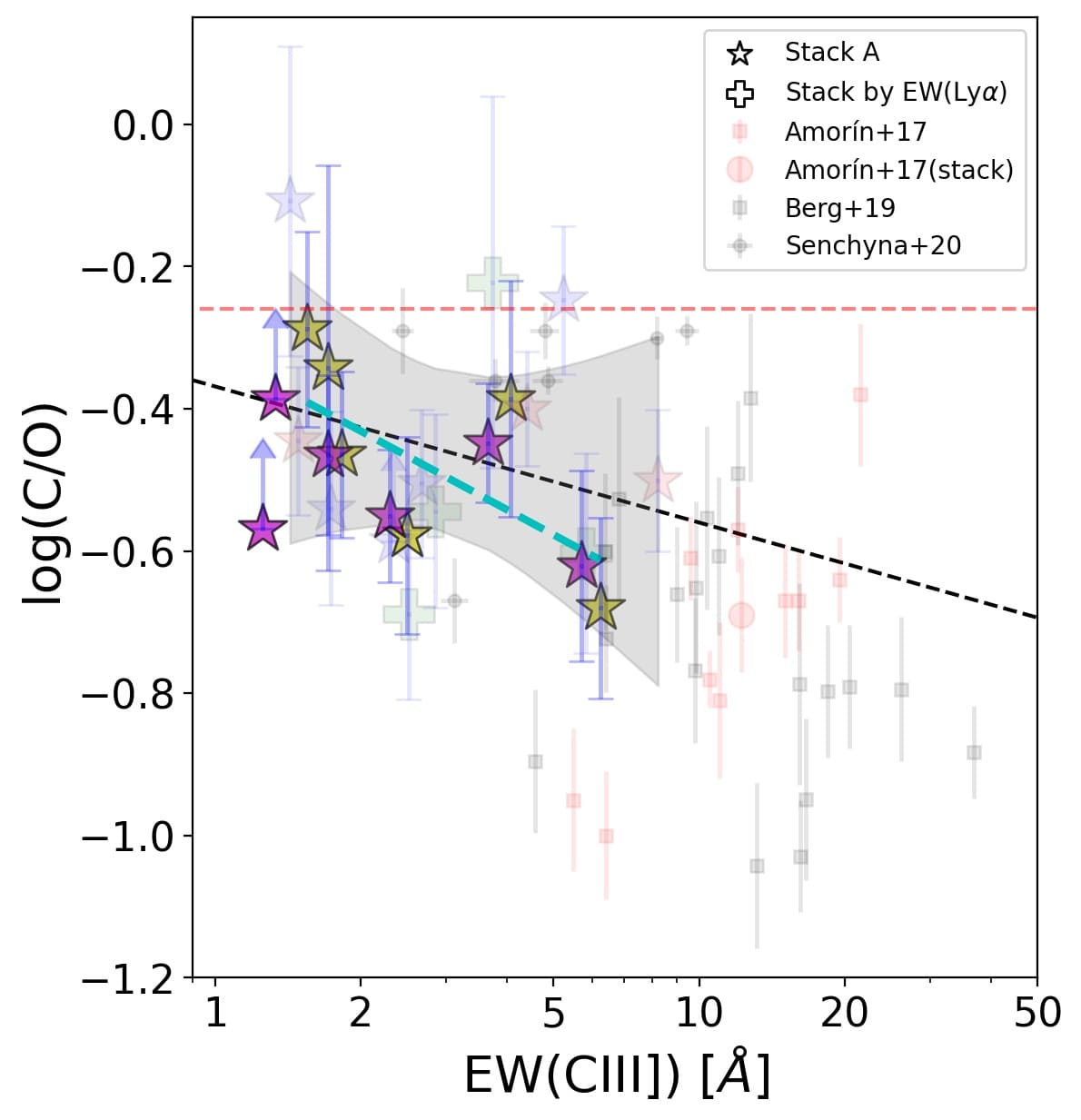}\,\includegraphics[width=0.25\textwidth]{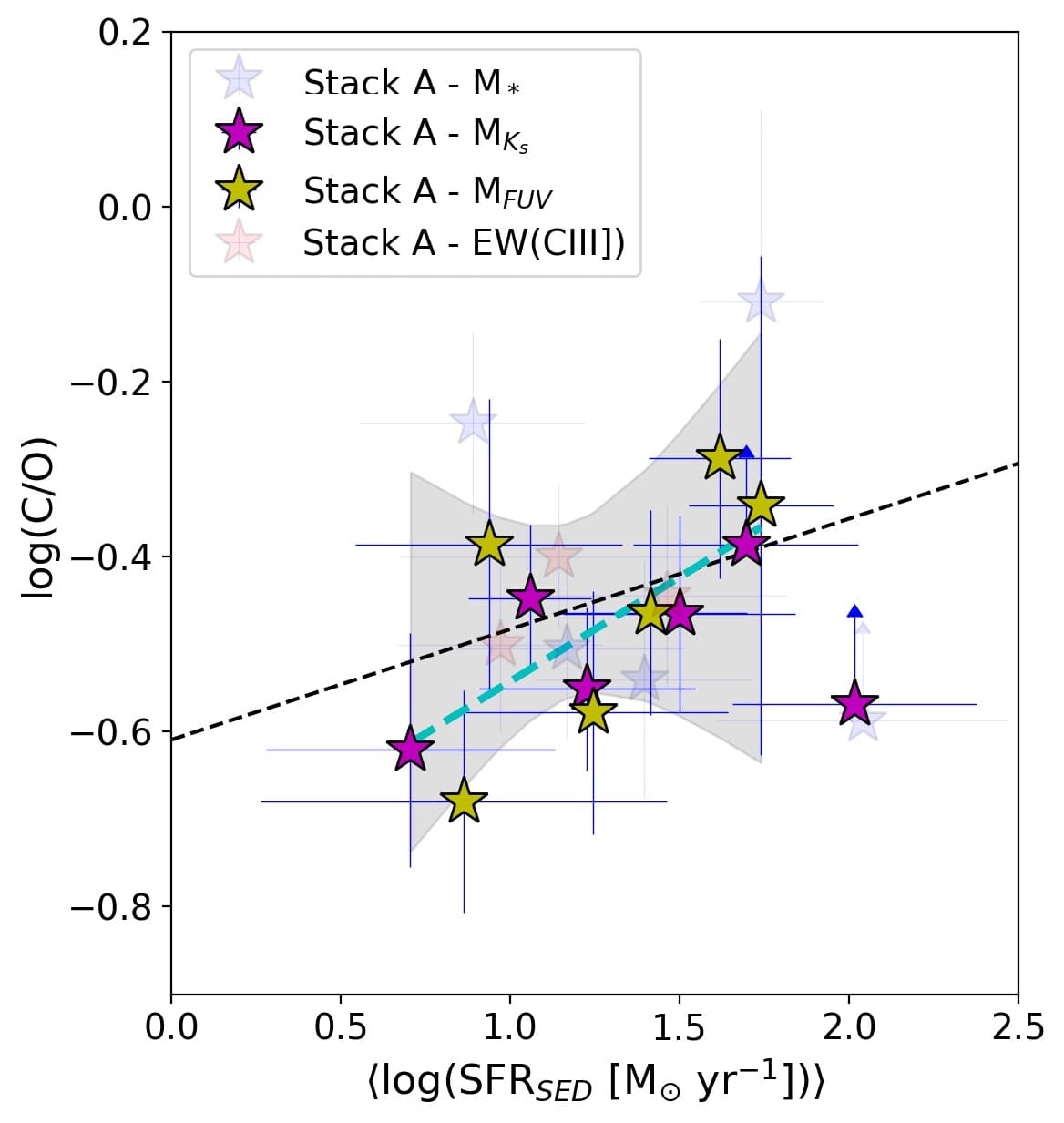}
    \caption{Same as in Fig. \ref{corrZEW}, \ref{corrZEWlyalog}, \ref{C/O_EW}, and \ref{SFR-relation}. The cyan dashed line shows the best fit including only the stacks A binned by luminosity. The stacks excluded in the fit are shown using transparent symbols.}
    \label{rel_lum}
\end{figure}

On the other hand, we note that the parameter used for stacking may result in different ranges in stellar metallicity and C/O (e.g. Fig. \ref{corrZEW}, \ref{corrZEWlyalog}, \ref{C/O_EW}, and \ref{SFR-relation}). This is the result of the distribution of the galaxies in the bins depending on their physical properties. As shown in Fig. \ref{sampleUMassbag}, the relations between stellar mass and luminosities show scatter and then the same galaxies are not in the corresponding same bin for each parameter. However, the trends are still found if we use only one set of stacking. For instance, in Fig. \ref{rel_lum} we show by the cyan dashed line the relations found if we only considered the stacks by luminosity. We note the the cyan lines follow the same trend that the black lines, which are the relations reported in the paper including the other set of stacks in the legends. They also are within the 3-$\sigma$ uncertainty of the reported relations. In the EW(Ly$\alpha$)-Z$_{\star}$ relation (top-right panel in Fig. \ref{rel_lum}), we note that the dynamical range in EW(Ly$\alpha$) is shorter compared with the entire set of stacks, because only upper limits in stellar metallicity are determined due to the lower S/N of the stacks with higher EWs in this set of stacks. 
The inclusion of additional sets of stacks binned by different parameters allows exploration of a larger dynamical space, generally at the cost of a larger variance, but without affecting the trends found in the relations.

\section{Tables}

In this section, we present Tables \ref{tablemassbin},  \ref{tableKsmagbin}, \ref{tableFUVmagbin}, \ref{tableEWLyabin}, \ref{tableC3bin} with the corresponding properties of each stack by stellar mass, K$_s$, and FUV luminosities, EW(Ly$\alpha$), and EW(CIII]), respectively. In particular, the observed fluxes measured for the lines used in the analysis, and the C/O and Z$_*$ estimated are reported.    

\begin{table*}
\caption{\label{tablemassbin}Parameters estimated for the stacks by stellar mass}
\centering
\begin{scriptsize}
\begin{tabular}{cccccccccccc}
\hline\hline
Bin&log(M$_{\star}$/M$_{\odot}$)\tablefootmark{a}&E(B-V)\tablefootmark{b}& CIV/CIII]\tablefootmark{c} & HeII/CIII]\tablefootmark{c}& OIII]/CIII]\tablefootmark{c} &EW(CIV)\tablefootmark{d}&EW(CIII])\tablefootmark{d}&log(C/O)\tablefootmark{e}&log(Z$_{\star}$/Z$_{\odot}$)\tablefootmark{f}\\
\hline\multicolumn{9}{c}{Stack A }\\ \hline
1 & 8.85 $\pm$ 0.10 & 0.06 & 0.32 $\pm$ 0.05 & 0.14 $\pm$ 0.05 & 0.11 $\pm$ 0.02 & 1.18 $\pm$ 0.17 & 5.24 $\pm$ 0.38 & -0.24 $\pm$ 0.10 & -0.77 $\pm$ 0.20 \\
2 & 9.21 $\pm$ 0.11 & 0.07 & 0.20 $\pm$ 0.02 & 0.38 $\pm$ 0.03 & 0.18 $\pm$ 0.01 & 0.40 $\pm$ 0.05 & 2.66 $\pm$ 0.12 & -0.50 $\pm$ 0.10 & -0.76 $\pm$ 0.03 \\
3 & 9.54 $\pm$ 0.08 & 0.10 & 0.18 $\pm$ 0.03 & 0.45 $\pm$ 0.04 & 0.19 $\pm$ 0.03 & 0.23 $\pm$ 0.03 & 1.73 $\pm$ 0.09 & -0.54 $\pm$ 0.13 & -0.67 $\pm$ 0.02 \\
4 & 9.80 $\pm$ 0.07 & 0.13 & 0.26 $\pm$ 0.05 & 0.54 $\pm$ 0.06 & 0.08 $\pm$ 0.03 & 0.28 $\pm$ 0.05 & 1.42 $\pm$ 0.07 & -0.10 $\pm$ 0.21 & -0.58 $\pm$ 0.02 \\
5 & 10.1 $\pm$ 0.12 & 0.21 & $<$ 0.03 & $<$ 0.12 & $<$ 0.15 & $<$ 0.06 & 2.33 $\pm$ 0.41 & $>$ -0.58 & -0.72 $\pm$ 0.06 \\
\hline\multicolumn{9}{c}{Stack B }\\ \hline
1 & 8.84 $\pm$ 0.11 & 0.05 & 0.38 $\pm$ 0.06 & $<$ 0.12 & 0.20 $\pm$ 0.06 & 1.45 $\pm$ 0.21 & 5.69 $\pm$ 0.39 & -0.52 $\pm$ 0.19 & -0.95 $\pm$ 0.07 \\
2 & 9.18 $\pm$ 0.10 & 0.06 & 0.19 $\pm$ 0.03 & 0.25 $\pm$ 0.04 & 0.13 $\pm$ 0.02 & 0.59 $\pm$ 0.10 & 4.65 $\pm$ 0.24 & -0.33 $\pm$ 0.10 & -0.96 $\pm$ 0.04 \\
3 & 9.56 $\pm$ 0.08 & 0.08 & 0.23 $\pm$ 0.04 & 0.33 $\pm$ 0.05 & 0.20 $\pm$ 0.04 & 0.45 $\pm$ 0.07 & 2.76 $\pm$ 0.19 & -0.57 $\pm$ 0.13 & -0.93 $\pm$ 0.03 \\
4 & 9.87 $\pm$ 0.19 & 0.16 & 0.15 $\pm$ 0.05 & 0.40 $\pm$ 0.07 & 0.18 $\pm$ 0.05 & 0.19 $\pm$ 0.07 & 1.86 $\pm$ 0.14 & -0.51 $\pm$ 0.20 & -0.64 $\pm$ 0.03 \\
\hline\multicolumn{9}{c}{Stack C }\\ \hline
1 & 8.82 $\pm$ 0.11 & 0.04 & 0.34 $\pm$ 0.07 & $<$ 0.09 & 0.22 $\pm$ 0.06 & 1.22 $\pm$ 0.22 & 5.75 $\pm$ 0.46 & -0.52 $\pm$ 0.16 & -0.74 $\pm$ 0.18 \\
2 & 9.17 $\pm$ 0.09 & 0.05 & 0.22 $\pm$ 0.05 & 0.20 $\pm$ 0.05 & 0.22 $\pm$ 0.04 & 0.77 $\pm$ 0.18 & 5.55 $\pm$ 0.31 & -0.61 $\pm$ 0.13 & -1.05 $\pm$ 0.05 \\
3 & 9.57 $\pm$ 0.08 & 0.05 & 0.24 $\pm$ 0.06 & 0.23 $\pm$ 0.05 & 0.22 $\pm$ 0.04 & 0.56 $\pm$ 0.12 & 3.39 $\pm$ 0.34 & -0.54 $\pm$ 0.1 & -0.96 $\pm$ 0.02 \\
4 & 9.97 $\pm$ 0.23 & 0.22 & $<$ 0.10 & 0.41 $\pm$ 0.20 & $<$ 0.21 & $<$ 0.13 & 1.96 $\pm$ 0.69 & $>$ -0.66 & -0.48 $\pm$ 0.08 \\
\hline\multicolumn{9}{c}{Stack D }\\ \hline
1 & 8.89 $\pm$ 0.07 & 0.07 & 0.28 $\pm$ 0.08 & 0.22 $\pm$ 0.10 & 0.12 $\pm$ 0.03 & 0.98 $\pm$ 0.26 & 4.59 $\pm$ 0.69 & -0.30 $\pm$ 0.13 & $<$ -0.982 \\
2 & 9.24 $\pm$ 0.11 & 0.08 & 0.30 $\pm$ 0.05 & 0.46 $\pm$ 0.05 & 0.20 $\pm$ 0.03 & 0.43 $\pm$ 0.07 & 1.94 $\pm$ 0.11 & -0.55 $\pm$ 0.09 & -0.75 $\pm$ 0.04 \\
3 & 9.52 $\pm$ 0.07 & 0.12 & 0.18 $\pm$ 0.03 & 0.49 $\pm$ 0.06 & 0.09 $\pm$ 0.04 & 0.18 $\pm$ 0.03 & 1.35 $\pm$ 0.08 & -0.18 $\pm$ 0.19 & -0.63 $\pm$ 0.02 \\
4 & 9.81 $\pm$ 0.07 & 0.14 & 0.24 $\pm$ 0.07 & 0.59 $\pm$ 0.09 & $<$ 0.20 & 0.22 $\pm$ 0.06 & 1.17 $\pm$ 0.09 & $>$ -0.58 & -0.50 $\pm$ 0.04 \\
5 & 10.1 $\pm$ 0.09 & 0.19 & $<$ 0.06 & $<$ 0.60 & $<$ 0.46 & $<$ 0.07 & 1.71 $\pm$ 0.43 & $>$ -1.02 & -0.64 $\pm$ 0.06 \\
\hline\hline
\end{tabular}
\end{scriptsize}
\tablefoot{Panels depending on the kind of stack according to Subsection \ref{sectionstack}\\
\tablefoottext{a}{Bin mean log(M$_*$/M$_{\odot}$) according to Table \ref{tab:binsDR3}}
\tablefoottext{b}{Mean E(B-V) in mag}
\tablefoottext{c}{Flux normalized to CIII] not corrected by extinction.}
\tablefoottext{d}{Equivalent width in units of \r{A}.}
\tablefoottext{e}{Estimated using \citet{Perez-Montero_2017}}
\tablefoottext{f}{Estimated using \citet{Cullen_2019}}
}
\end{table*}

\begin{table*}
\caption{\label{tableKsmagbin}Parameters estimated for the stacks by K$_s$-band luminosity}
\centering
\begin{scriptsize}
\begin{tabular}{ccccccccccc}
\hline\hline
Bin&Mean M$_{Ks}$\tablefootmark{a}&E(B-V)\tablefootmark{b}& CIV/CIII]\tablefootmark{c} & HeII/CIII]\tablefootmark{c}& OIII]/CIII]\tablefootmark{c} &EW(CIV)\tablefootmark{d}&EW(CIII])\tablefootmark{d}&log(C/O)\tablefootmark{e}&log(Z$_{\star}$/Z$_{\odot}$)\tablefootmark{f}\\
\hline\multicolumn{9}{c}{Stack A }\\ \hline
1 & -22.94 $\pm$ 0.43 & 0.18 & $<$ 0.03 & 0.37 $\pm$ 0.07 & 0.08 $\pm$ 0.03 & $<$ 0.03 & 1.25 $\pm$ 0.10 & $>$-0.56 & -0.61 $\pm$ 0.04 \\
2 & -22.23 $\pm$ 0.13 & 0.14 & 0.26 $\pm$ 0.04 & 0.53 $\pm$ 0.05 & $<$ 0.14 & 0.26 $\pm$ 0.04 & 1.33 $\pm$ 0.06 & $>$ -0.38 & -0.60 $\pm$ 0.02 \\
3 & -21.76 $\pm$ 0.15 & 0.09 & 0.22 $\pm$ 0.04 & 0.38 $\pm$ 0.05 & 0.17 $\pm$ 0.03 & 0.28 $\pm$ 0.05 & 1.70 $\pm$ 0.12 & -0.46 $\pm$ 0.11 & -0.75 $\pm$ 0.02 \\
4 & -21.26 $\pm$ 0.13 & 0.08 & 0.25 $\pm$ 0.04 & 0.44 $\pm$ 0.04 & 0.20 $\pm$ 0.02 & 0.42 $\pm$ 0.06 & 2.29 $\pm$ 0.12 & -0.55 $\pm$ 0.09 & -0.82 $\pm$ 0.10 \\
5 & -20.76 $\pm$ 0.16 & 0.06 & 0.20 $\pm$ 0.03 & 0.24 $\pm$ 0.04 & 0.16 $\pm$ 0.02 & 0.53 $\pm$ 0.08 & 3.66 $\pm$ 0.19 & -0.44 $\pm$ 0.08 & -1.00 $\pm$ 0.04 \\
6 & -20.12 $\pm$ 0.32 & 0.04 & 0.27 $\pm$ 0.04 & 0.26 $\pm$ 0.05 & 0.24 $\pm$ 0.03 & 1.04 $\pm$ 0.14 & 5.71 $\pm$ 0.26 & -0.62 $\pm$ 0.13 & $<$ -0.932 \\
\hline\multicolumn{9}{c}{Stack B }\\ \hline
1 & -23.05 $\pm$ 0.55 & 0.18 & $<$ 0.06 & 0.46 $\pm$ 0.22 & $<$ 0.30 & $<$ 0.04 & 1.09 $\pm$ 0.37 & $>$ -0.85 & -0.79 $\pm$ 0.08 \\
2 & -22.25 $\pm$ 0.15 & 0.15 & 0.14 $\pm$ 0.03 & 0.35 $\pm$ 0.06 & $<$ 0.09 & 0.21 $\pm$ 0.05 & 2.06 $\pm$ 0.10 & $>$ -0.28 & -0.77 $\pm$ 0.10 \\
3 & -21.77 $\pm$ 0.16 & 0.08 & 0.21 $\pm$ 0.07 & 0.27 $\pm$ 0.08 & 0.18 $\pm$ 0.06 & 0.31 $\pm$ 0.09 & 2.13 $\pm$ 0.23 & -0.53 $\pm$ 0.18 & -0.91 $\pm$ 0.04 \\
4 & -21.20 $\pm$ 0.12 & 0.07 & 0.26 $\pm$ 0.04 & 0.19 $\pm$ 0.03 & 0.20 $\pm$ 0.03 & 0.73 $\pm$ 0.10 & 4.07 $\pm$ 0.24 & -0.56 $\pm$ 0.11 & -0.70 $\pm$ 0.07 \\
5 & -20.78 $\pm$ 0.15 & 0.06 & 0.14 $\pm$ 0.05 & 0.14 $\pm$ 0.04 & 0.23 $\pm$ 0.04 & 0.48 $\pm$ 0.16 & 5.11 $\pm$ 0.36 & -0.59 $\pm$ 0.10 & $<$ -0.962 \\
6 & -20.12 $\pm$ 0.33 & 0.04 & 0.29 $\pm$ 0.04 & 0.25 $\pm$ 0.05 & 0.23 $\pm$ 0.04 & 1.14 $\pm$ 0.18 & 5.87 $\pm$ 0.26 & -0.57 $\pm$ 0.12 & $<$ -0.992 \\
\hline\multicolumn{9}{c}{Stack C }\\ \hline
1 & -22.68 $\pm$ 0.63 & 0.19 & $<$ 0.12 & $<$ 0.28 & $<$ 0.17 & $<$ 0.15 & 1.88 $\pm$ 0.43 & $>$ -0.57 & -0.42 $\pm$ 0.06 \\
2 & -21.75 $\pm$ 0.13 & 0.05 & 0.23 $\pm$ 0.07 & 0.25 $\pm$ 0.11 & 0.19 $\pm$ 0.09 & 0.36 $\pm$ 0.10 & 2.35 $\pm$ 0.43 & -0.45 $\pm$ 0.24 & -0.75 $\pm$ 0.06 \\
3 & -21.22 $\pm$ 0.11 & 0.07 & 0.24 $\pm$ 0.05 & 0.18 $\pm$ 0.03 & 0.19 $\pm$ 0.06 & 0.70 $\pm$ 0.15 & 4.24 $\pm$ 0.37 & -0.52 $\pm$ 0.15 & -0.73 $\pm$ 0.20 \\
4 & -20.76 $\pm$ 0.16 & 0.06 & 0.21 $\pm$ 0.06 & 0.10 $\pm$ 0.04 & 0.26 $\pm$ 0.03 & 0.73 $\pm$ 0.21 & 5.48 $\pm$ 0.36 & -0.64 $\pm$ 0.12 & -1.05 $\pm$ 0.05 \\
5 & -20.10 $\pm$ 0.34 & 0.03 & 0.25 $\pm$ 0.04 & 0.14 $\pm$ 0.04 & 0.22 $\pm$ 0.04 & 0.95 $\pm$ 0.17 & 5.97 $\pm$ 0.34 & -0.59 $\pm$ 0.11 & $<$ -0.982 \\
\hline\multicolumn{9}{c}{Stack D }\\ \hline
1 & -22.89 $\pm$ 0.35 & 0.19 & $<$ 0.07 & 0.28 $\pm$ 0.08 & $<$ 0.21 & $<$ 0.07 & 1.39 $\pm$ 0.13 & $>$ -0.68 & -0.66 $\pm$ 0.05 \\
2 & -22.21 $\pm$ 0.11 & 0.13 & 0.24 $\pm$ 0.07 & 0.58 $\pm$ 0.09 & $<$ 0.20 & 0.21 $\pm$ 0.06 & 1.12 $\pm$ 0.09 & $>$ -0.56 & -0.54 $\pm$ 0.04 \\
3 & -21.76 $\pm$ 0.15 & 0.11 & 0.15 $\pm$ 0.05 & 0.41 $\pm$ 0.06 & 0.19 $\pm$ 0.03 & 0.20 $\pm$ 0.07 & 1.78 $\pm$ 0.14 & -0.54 $\pm$ 0.13 & -0.66 $\pm$ 0.03 \\
4 & -21.31 $\pm$ 0.12 & 0.08 & 0.40 $\pm$ 0.08 & 0.50 $\pm$ 0.07 & 0.29 $\pm$ 0.05 & 0.41 $\pm$ 0.07 & 1.44 $\pm$ 0.13 & -0.64 $\pm$ 0.11 & -0.94 $\pm$ 0.02 \\
5 & -20.70 $\pm$ 0.23 & 0.07 & 0.15 $\pm$ 0.06 & 0.28 $\pm$ 0.05 & 0.14 $\pm$ 0.02 & 0.29 $\pm$ 0.12 & 2.66 $\pm$ 0.18 & -0.43 $\pm$ 0.12 & -1.01 $\pm$ 0.05 \\
\hline\hline
\end{tabular}
\end{scriptsize}
\tablefoot{Same as in \ref{tablemassbin} but \tablefoottext{a}{Bin mean Ks-band luminosity according to Table \ref{tab:binsDR3}.}}
\end{table*}

\begin{table*}
\caption{\label{tableFUVmagbin}Parameters estimated for the stacks by FUV luminosity}
\centering
\begin{scriptsize}
\begin{tabular}{cccccccccc}
\hline\hline
Bin&Mean M$_{FUV}$\tablefootmark{a}&E(B-V)\tablefootmark{b}& CIV/CIII]\tablefootmark{c} & HeII/CIII]\tablefootmark{c}& OIII]/CIII]\tablefootmark{c} &EW(CIV)\tablefootmark{d}&EW(CIII])\tablefootmark{d}&log(C/O)\tablefootmark{e}&log(Z$_{\star}$/Z$_{\odot}$)\tablefootmark{f}\\
\hline\multicolumn{9}{c}{Stack A }\\ \hline
1 & -21.73 $\pm$ 0.12 & 0.08 & 0.24 $\pm$ 0.05 & 0.31 $\pm$ 0.07 & 0.13 $\pm$ 0.06 & 0.28 $\pm$ 0.06 & 1.70 $\pm$ 0.10 & -0.34 $\pm$ 0.28 & -0.64 $\pm$ 0.03 \\
2 & -21.27 $\pm$ 0.13 & 0.09 & 0.25 $\pm$ 0.04 & 0.52 $\pm$ 0.05 & 0.11 $\pm$ 0.02 & 0.28 $\pm$ 0.05 & 1.54 $\pm$ 0.07 & -0.28 $\pm$ 0.13 & -0.61 $\pm$ 0.03 \\
3 & -20.73 $\pm$ 0.14 & 0.10 & 0.12 $\pm$ 0.03 & 0.35 $\pm$ 0.03 & 0.14 $\pm$ 0.02 & 0.17 $\pm$ 0.04 & 1.82 $\pm$ 0.08 & -0.46 $\pm$ 0.11 & -0.76 $\pm$ 0.02 \\
4 & -20.26 $\pm$ 0.15 & 0.10 & 0.21 $\pm$ 0.04 & 0.42 $\pm$ 0.04 & 0.20 $\pm$ 0.02 & 0.40 $\pm$ 0.07 & 2.49 $\pm$ 0.15 & -0.57 $\pm$ 0.13 & -0.65 $\pm$ 0.02 \\
5 & -19.79 $\pm$ 0.13 & 0.08 & 0.14 $\pm$ 0.05 & 0.26 $\pm$ 0.06 & 0.14 $\pm$ 0.03 & 0.43 $\pm$ 0.17 & 4.08 $\pm$ 0.25 & -0.38 $\pm$ 0.16 & -0.70 $\pm$ 0.09 \\
6 & -19.07 $\pm$ 0.37 & 0.13 & 0.52 $\pm$ 0.10 & 0.27 $\pm$ 0.08 & 0.30 $\pm$ 0.06 & 2.13 $\pm$ 0.40 & 6.24 $\pm$ 0.46 & -0.68 $\pm$ 0.12 & $<$ -0.862 \\
\hline\multicolumn{9}{c}{Stack B }\\ \hline
1 & -21.71 $\pm$ 0.12 & 0.07 & 0.18 $\pm$ 0.06 & 0.25 $\pm$ 0.05 & 0.16 $\pm$ 0.06 & 0.23 $\pm$ 0.07 & 1.88 $\pm$ 0.10 & -0.50 $\pm$ 0.17 & -0.69 $\pm$ 0.02 \\
2 & -21.27 $\pm$ 0.13 & 0.08 & 0.25 $\pm$ 0.08 & 0.21 $\pm$ 0.05 & 0.15 $\pm$ 0.05 & 0.41 $\pm$ 0.13 & 2.37 $\pm$ 0.22 & $>$-0.73  & -0.56 $\pm$ 0.04 \\
3 & -20.76 $\pm$ 0.12 & 0.08 & 0.20 $\pm$ 0.03 & 0.24 $\pm$ 0.04 & 0.15 $\pm$ 0.03 & 0.41 $\pm$ 0.06 & 2.84 $\pm$ 0.22 & -0.40 $\pm$ 0.13 & -0.93 $\pm$ 0.04 \\
4 & -20.23 $\pm$ 0.15 & 0.07 & 0.20 $\pm$ 0.04 & 0.21 $\pm$ 0.04 & 0.18 $\pm$ 0.03 & 0.69 $\pm$ 0.12 & 4.88 $\pm$ 0.35 & -0.53 $\pm$ 0.12 & -0.79 $\pm$ 0.12 \\
5 & -19.80 $\pm$ 0.11 & 0.06 & 0.25 $\pm$ 0.06 & 0.26 $\pm$ 0.06 & 0.23 $\pm$ 0.04 & 0.91 $\pm$ 0.21 & 5.21 $\pm$ 0.30 & -0.60 $\pm$ 0.11 & -0.88 $\pm$ 0.08 \\
6 & -19.07 $\pm$ 0.37 & 0.13 & 0.52 $\pm$ 0.10 & 0.27 $\pm$ 0.08 & 0.30 $\pm$ 0.06 & 2.13 $\pm$ 0.40 & 6.24 $\pm$ 0.46 & -0.66 $\pm$ 0.13 & $<$ -0.862 \\
\hline\multicolumn{9}{c}{Stack C }\\ \hline
1 & -21.58 $\pm$ 0.22 & 0.07 & 0.21 $\pm$ 0.07 & 0.17 $\pm$ 0.07 & $<$ 0.13 & 0.27 $\pm$ 0.08 & 1.87 $\pm$ 0.16 & $>$ -0.35 & -0.76 $\pm$ 0.09 \\
2 & -20.73 $\pm$ 0.12 & 0.05 & 0.21 $\pm$ 0.05 & 0.11 $\pm$ 0.04 & $<$ 0.13 & 0.74 $\pm$ 0.16 & 5.21 $\pm$ 0.48 & $>$ -0.37 & $<$ -1.002 \\
3 & -20.20 $\pm$ 0.14 & 0.06 & 0.16 $\pm$ 0.04 & 0.21 $\pm$ 0.04 & 0.18 $\pm$ 0.03 & 0.68 $\pm$ 0.20 & 6.30 $\pm$ 0.47 & -0.54 $\pm$ 0.10 & -1.00 $\pm$ 0.06 \\
4 & -19.82 $\pm$ 0.12 & 0.05 & 0.32 $\pm$ 0.07 & 0.13 $\pm$ 0.05 & 0.23 $\pm$ 0.05 & 1.14 $\pm$ 0.23 & 5.10 $\pm$ 0.42 & -0.63 $\pm$ 0.15 & -0.91 $\pm$ 0.09 \\
5 & -19.01 $\pm$ 0.40 & 0.11 & 0.54 $\pm$ 0.14 & 0.45 $\pm$ 0.14 & 0.31 $\pm$ 0.08 & 1.85 $\pm$ 0.47 & 5.56 $\pm$ 0.49 & -0.57 $\pm$ 0.21 & $<$ -0.792 \\
\hline\multicolumn{9}{c}{Stack D }\\ \hline
1 & -21.79 $\pm$ 0.06 & 0.09 & 0.29 $\pm$ 0.10 & 0.36 $\pm$ 0.08 & 0.08 $\pm$ 0.02 & 0.42 $\pm$ 0.15 & 2.10 $\pm$ 0.22 & -0.06 $\pm$ 0.24 & -0.38 $\pm$ 0.04 \\
2 & -21.27 $\pm$ 0.12 & 0.09 & 0.26 $\pm$ 0.06 & 0.49 $\pm$ 0.06 & 0.08 $\pm$ 0.03 & 0.27 $\pm$ 0.06 & 1.48 $\pm$ 0.07 & -0.09 $\pm$ 0.19 & -0.59 $\pm$ 0.03 \\
3 & -20.71 $\pm$ 0.15 & 0.10 & 0.15 $\pm$ 0.04 & 0.42 $\pm$ 0.06 & 0.12 $\pm$ 0.04 & 0.15 $\pm$ 0.05 & 1.42 $\pm$ 0.10 & -0.33 $\pm$ 0.13 & -0.69 $\pm$ 0.04 \\
4 & -20.29 $\pm$ 0.15 & 0.13 & 0.20 $\pm$ 0.06 & 0.57 $\pm$ 0.08 & 0.13 $\pm$ 0.03 & 0.25 $\pm$ 0.07 & 1.64 $\pm$ 0.13 & -0.37 $\pm$ 0.14 & -0.64 $\pm$ 0.03 \\
5 & -19.78 $\pm$ 0.17 & 0.16 & $<$ 0.16 & 0.33 $\pm$ 0.11 & 0.17 $\pm$ 0.06 & $<$ 0.28 & 2.14 $\pm$ 0.24 & -0.44 $\pm$ 0.18 & -0.68 $\pm$ 0.10 \\
\hline
\hline
\end{tabular}
\end{scriptsize}
\tablefoot{Same as in \ref{tablemassbin} but \tablefoottext{a}{Bin mean FUV luminosity according to Table \ref{tab:binsDR3}.}}
\end{table*}

\begin{table*}
\caption{\label{tableEWLyabin}Parameters estimated for the stacks by EW(Ly$\alpha$)}
\centering
\begin{scriptsize}
\begin{tabular}{cccccccccc}
\hline\hline
Bin&EW(Ly$\alpha$)\tablefootmark{a}&E(B-V)\tablefootmark{b}& CIV/CIII]\tablefootmark{c} & HeII/CIII]\tablefootmark{c}& OIII]/CIII]\tablefootmark{c} &EW(CIV)\tablefootmark{d}&EW(CIII])\tablefootmark{d}&log(C/O)\tablefootmark{e}&log(Z$_{\star}$/Z$_{\odot}$)\tablefootmark{f}\\
\hline
1 & -29.1 $\pm$ 8.18 & 0.09 & $<$ 0.14 & 0.34 $\pm$ 0.08 & 0.22 $\pm$ 0.05 & $<$ 0.24 & 2.51 $\pm$ 0.33 & -0.68 $\pm$ 0.11 & -0.87 $\pm$ 0.02 \\
2 & -8.93 $\pm$ 4.10 & 0.10 & 0.15 $\pm$ 0.04 & 0.23 $\pm$ 0.04 & 0.19 $\pm$ 0.04 & 0.31 $\pm$ 0.08 & 2.84 $\pm$ 0.20 & -0.54 $\pm$ 0.13 & -0.71 $\pm$ 0.04 \\
3 & 10.87 $\pm$ 5.52 & 0.06 & 0.12 $\pm$ 0.05 & 0.12 $\pm$ 0.03 & 0.09 $\pm$ 0.03 & 0.30 $\pm$ 0.12 & 3.73 $\pm$ 0.33 & -0.22 $\pm$ 0.26 & -0.97 $\pm$ 0.04 \\
4 & 44.90 $\pm$ 20.4 & 0.06 & 0.25 $\pm$ 0.03 & 0.26 $\pm$ 0.04 & 0.22 $\pm$ 0.03 & 0.95 $\pm$ 0.12 & 5.82 $\pm$ 0.24 & -0.60 $\pm$ 0.14 & -1.06 $\pm$ 0.04 \\
\hline
\hline
\end{tabular}
\end{scriptsize}
\tablefoot{Same as in \ref{tablemassbin}but \tablefoottext{a}{Bin mean EW(Ly$\alpha$) according to Table \ref{tab:binsDR3EW}}.
}
\end{table*}

\begin{table*}
\caption{\label{tableC3bin}Parameters estimated for the stacks by EW(CIII])}
\centering
\begin{scriptsize}
\begin{tabular}{cccccccccc}
\hline\hline
Bin&EW(CIII])\tablefootmark{a}&E(B-V)\tablefootmark{b}& CIV/CIII]\tablefootmark{c} & HeII/CIII]\tablefootmark{c}& OIII]/CIII]\tablefootmark{c} &EW(CIV)\tablefootmark{d}&EW(CIII])\tablefootmark{d}&log(C/O)\tablefootmark{e}&log(Z$_{\star}$/Z$_{\odot}$)\tablefootmark{f}\\
\hline\multicolumn{9}{c}{Stack A }\\ \hline
1 & 2.21 $\pm$ 0.90 & 0.10 & 0.21 $\pm$ 0.03 & 0.58 $\pm$ 0.03 & 0.15 $\pm$ 0.02 & 0.23 $\pm$ 0.03 & 1.48 $\pm$ 0.04 & -0.44 $\pm$ 0.10 & -0.72 $\pm$ 0.02 \\
2 & 5.58 $\pm$ 1.13 & 0.07 & 0.17 $\pm$ 0.02 & 0.29 $\pm$ 0.03 & 0.14 $\pm$ 0.01 & 0.52 $\pm$ 0.08 & 4.40 $\pm$ 0.15 & -0.4 $\pm$ 0.08 & -0.75 $\pm$ 0.04 \\
3 & 10.9 $\pm$ 2.91 & 0.07 & 0.26 $\pm$ 0.03 & 0.13 $\pm$ 0.02 & 0.19 $\pm$ 0.02 & 1.40 $\pm$ 0.19 & 8.21 $\pm$ 0.35 & -0.50 $\pm$ 0.1 & -1.09 $\pm$ 0.04 \\
\hline\multicolumn{9}{c}{Stack B }\\ \hline
1 & 2.54 $\pm$ 0.93 & 0.09 & 0.22 $\pm$ 0.04 & 0.28 $\pm$ 0.04 & 0.13 $\pm$ 0.03 & 0.32 $\pm$ 0.06 & 2.08 $\pm$ 0.09 & -0.38 $\pm$ 0.10 & -0.76 $\pm$ 0.04 \\
2 & 5.78 $\pm$ 1.12 & 0.07 & 0.16 $\pm$ 0.03 & 0.26 $\pm$ 0.03 & 0.13 $\pm$ 0.02 & 0.58 $\pm$ 0.11 & 5.05 $\pm$ 0.20 & -0.30 $\pm$ 0.12 & -0.91 $\pm$ 0.05 \\
3 & 11.0 $\pm$ 3.00 & 0.07 & 0.36 $\pm$ 0.04 & 0.12 $\pm$ 0.02 & 0.22 $\pm$ 0.02 & 1.88 $\pm$ 0.21 & 7.98 $\pm$ 0.39 & -0.56 $\pm$ 0.10 & -1.03 $\pm$ 0.06 \\
\hline\multicolumn{9}{c}{Stack C }\\ \hline
1 & 2.44 $\pm$ 1.12 & 0.06 & 0.31 $\pm$ 0.09 & 0.27 $\pm$ 0.09 & $<$ 0.13 & 0.42 $\pm$ 0.11 & 2.03 $\pm$ 0.18 & $>$ -0.33 & -0.99 $\pm$ 0.03 \\
2 & 5.90 $\pm$ 1.13 & 0.06 & 0.19 $\pm$ 0.03 & 0.24 $\pm$ 0.04 & 0.15 $\pm$ 0.03 & 0.69 $\pm$ 0.13 & 5.41 $\pm$ 0.23 & -0.40 $\pm$ 0.12 & -1.02 $\pm$ 0.04 \\
3 & 11.0 $\pm$ 3.28 & 0.07 & 0.33 $\pm$ 0.04 & 0.12 $\pm$ 0.03 & 0.19 $\pm$ 0.02 & 1.56 $\pm$ 0.18 & 7.64 $\pm$ 0.43 & -0.49 $\pm$ 0.1 & -1.01 $\pm$ 0.07 \\
\hline\multicolumn{9}{c}{Stack D }\\ \hline
1 & 2.02 $\pm$ 0.82 & 0.12 & 0.20 $\pm$ 0.03 & 0.56 $\pm$ 0.04 & 0.17 $\pm$ 0.02 & 0.22 $\pm$ 0.04 & 1.47 $\pm$ 0.06 & -0.48 $\pm$ 0.10 & -0.69 $\pm$ 0.02 \\
2 & 5.61 $\pm$ 2.09 & 0.08 & 0.14 $\pm$ 0.05 & 0.31 $\pm$ 0.05 & 0.13 $\pm$ 0.02 & 0.39 $\pm$ 0.13 & 4.03 $\pm$ 0.30 & -0.36 $\pm$ 0.10 & -0.64 $\pm$ 0.05 \\
\hline
\hline
\end{tabular}
\end{scriptsize}
\tablefoot{Same as in \ref{tablemassbin} but \tablefoottext{a}{Bin mean EW(CIII]) according to Table \ref{tab:binsDR3EW}}.
}
\end{table*}

\end{appendix} 

\end{document}